\documentclass[%
 reprint,
 superscriptaddress,
 amsmath,amssymb,
 aps,
 pra,
 floatfix,
 longbibliography,
]{revtex4-2}

\usepackage{graphicx}
\usepackage{dcolumn}
\usepackage{bm}
\usepackage{fp}
\usepackage{siunitx}
\usepackage{booktabs}
\usepackage[dvipsnames]{xcolor}
\usepackage{mathtools}
\usepackage{textcomp}
\usepackage[overload]{empheq}
\usepackage{mleftright}
\usepackage{xparse}
\usepackage{cancel}
\usepackage[super]{nth}
\usepackage{stackengine}
\usepackage{fixfoot}
\usepackage{chngcntr}  
\usepackage[hidelinks,colorlinks=true,linkcolor=blue,citecolor=blue,urlcolor=blue]{hyperref}

\DeclareFixedFootnote{\AppSwenson}{See Supplemental Material S3} 
\DeclareFixedFootnote{\AppGth}{See Supplemental Material S1} 

\newcommand{\nbar}[0]{\overline{n}}
\newcommand{\dv}[2]{\frac{\textrm{d}#1}{\textrm{d}#2}} 
\newcommand{\dvstar}[2]{\textrm{d}#1/\textrm{d}#2}
\DeclareMathOperator{\sech}{sech}

\begin{document}

\preprint{APS/123-QED}

\title{TLS-induced thermal nonlinearity in a micro-mechanical resonator}

\author{C. Metzger}
\email{cyril.metzger@yale.edu}
\affiliation{Department of Physics, Yale University, New Haven, Connecticut 06511, USA}\textbf{}
\affiliation{JILA, National Institute of Standards and Technology and the University of Colorado, Boulder, Colorado 80309, USA}
\affiliation{Department of Physics, University of Colorado, Boulder, Colorado 80309, USA}

\author{A. L. Emser}
\author{B. C. Rose}
\affiliation{JILA, National Institute of Standards and Technology and the University of Colorado, Boulder, Colorado 80309, USA}
\affiliation{Department of Physics, University of Colorado, Boulder, Colorado 80309, USA}

\author{K. W. Lehnert}
\affiliation{Department of Physics, Yale University, New Haven, Connecticut 06511, USA}
\affiliation{JILA, National Institute of Standards and Technology and the University of Colorado, Boulder, Colorado 80309, USA}
\affiliation{Department of Physics, University of Colorado, Boulder, Colorado 80309, USA}

\date{\today}

\begin{abstract}
We present experimental evidence of a thermally-driven amplitude-frequency nonlinearity in a thin-film quartz phononic crystal resonator at millikelvin temperatures. 
The nonlinear response arises from the coupling of the mechanical mode to an ensemble of microscopic two-level system defects driven out of equilibrium by a microwave drive. In contrast to the conventional Duffing oscillator, the observed nonlinearity exhibits a mixed reactive-dissipative character. Notably, the reactive effect can manifest as either a softening or hardening of the mechanical resonance, depending on the ratio of thermal to phonon energy. By combining the standard TLS theory with a thermal conductance model, the measured power-dependent response is quantitatively reproduced and readout-enhanced relaxation damping from off-resonant TLSs is identified as the primary mechanism limiting mechanical coherence. Within this framework, we delineate the conditions under which similar systems will realize this nonlinearity.
\end{abstract}

\maketitle

\setlength{\abovedisplayskip}{5pt}
\setlength{\belowdisplayskip}{5pt}

\section{Introduction}\label{sec:Intro}

Micro-mechanical resonators for quantum processing are particularly prone to readout-power heating, due to the competing demands of isolating the mechanical mode of interest from its environment to preserve high mechanical coherence, while also ensuring strong thermal anchoring to a low-temperature bath to keep the mechanical resonator cold \cite{Meenehan2015}. This trade-off is known to limit the performance of both piezoelectric phononic crystal resonators (PCR), currently explored as a platform for circuit quantum acoustodynamics \cite{Arrangoiz2016, Arrangoiz2018, Bozkurt2025}, and their optical counterpart, optomechanical crystal cavities (OMC) \cite{Eichenfield2009, Chan2012, Bochmann2013}. Both systems exploit periodically-patterned suspended structures to generate phononic bandgaps that spatially confine a mechanical mode without introducing deleterious clamping losses \cite{SafaviNaeini2010, Tsaturyan2017}. 
Although such phononic engineering can substantially enhance phonon lifetimes \cite{MacCabe2020}, it adversely affects device thermalization by lowering the thermal conductivity of the resonator supports, since heat removal can occur only through phonon modes outside the frequency bandgap \cite{Cleland2001, Tian2019}. 
In OMC cavities, parasitic optical absorption substantially degrades the achievable quantum cooperativity due to heating and damping of the mechanical-mode caused by the absorption-induced hot phonon bath \cite{Meenehan2014, Meenehan2015}. This has been a significant roadblock to their use for quantum applications and different strategies to improve thermal anchoring have been investigated, such as moving to two-dimensional phononic crystal arrays \cite{Ren2020,Povey2024} or switching to a release-free design \cite{Sarabalis2017, Burger2025}.

Mechanical damping is also known to occur from coupling to a dissipative bath of microscopic two-level-system material defects (TLS) that can drain energy from the resonant acoustic mode and disperse it to substrate phonons \cite{Muller2019, Bachtold2022}. TLSs represent a persistent source of decoherence and noise in various solid-state devices at millikelvin temperature. Therefore, efforts are underway to decrease their density through improved fabrication techniques and surface treatments \cite{Gruenke2023, Crowley2023, Gruenke2025}, and to mitigate their detrimental impact through phononic engineering of the resonator material \cite{Chen2024}. Due to their high surface participation, PCRs not only exhibit enhanced susceptibility to TLS loss, but may also host long-lived TLSs whose relaxation is suppressed by the phononic bandgap \cite{Odeh2025}.
PCRs have thus appeared as one valuable platform for exploring TLS physics, yielding new insights into dephasing mechanisms in nanomechanical systems \cite{Cleland2023, Maksymowych2025, Hitchcock2025, Fiaschi2025}, and enabling the control and engineering of TLS–phonon interactions at the single-phonon level \cite{Yuksel2025}. 
Recently, investigations of thin-film PCRs at millikelvin temperature also revealed an anomalous nonlinear behavior at high probe power (mean phonon number $\nbar \gtrsim 5000$), suggesting a thermal origin arising from readout-induced heating of a TLS ensemble \cite{Emser2024}.

Thermal nonlinearities in mechanical oscillators have been widely studied across diverse platforms, including levitated nanoparticles \cite{Gieseler2013}, quantum dots in suspended nanotubes \cite{Samanta2023}, and micro-electromechanical systems (MEMS) such as contour-mode resonators \cite{Tazzoli2012, Segovia2013, Miller2014}.
At ambient temperatures, thermal nonlinearities in MEMS typically originate from material softening upon Joule heating and manifest as a readout-induced negative frequency shift \cite{Lu2015}. In the millikelvin regime, however, phonon freeze-out suppresses lattice effects, leaving coupling to TLS defects as the dominant source of nonlinearity in surface-dominated devices such as thin-film resonators \cite{Muller2019, Bachtold2022}.
Although TLSs are commonly associated with saturable loss and dissipative nonlinearities, they can also produce significant dispersive shifts leading to reactive nonlinear behavior. Yet, such shifts are usually unobserved in single-tone spectroscopy, where symmetric saturation of TLSs around the resonance frequency causes only linewidth narrowing \cite{Kirsh2017}. Detecting TLS-induced frequency shifts thus requires either an additional off-resonant pump tone \cite{Capelle2020, Andersson2021} or broadband heating of the ensemble to produce a net spectral imbalance in the TLS population about the resonator frequency.
In fact, varying the device temperature and monitoring the resulting frequency shift has become an increasingly common technique for extracting the loss tangent associated to TLSs and to characterize their ensemble response \cite{Gao2008, Wollack2021, Gruenke2023, Emser2024}. 

With sufficient power, spectroscopy itself can also induce such device heating and therefore a subsequent frequency shift. In PCRs, the combination of small mode volume and strong thermal isolation due to the suspended structure and the engineered acoustic bandgap makes them especially sensitive to readout-power heating. As a result, simply probing the resonance can elevate the TLS temperature, altering their collective dispersive response and rendering the resonator frequency effectively power-dependent. We argue that for resonators with a frequency $f_r$ such that $T \sim h f_r / k_B$, TLS heating due to the readout tone can thus generate a strong reactive-dissipative nonlinearity.
For typical dilution refrigerator temperatures ($T \sim 25$-$100$~mK), this translates to $f_r \sim 0.5$-$2$~GHz, with this work focusing on the lower end of the range.
Although arising from different mechanisms, similar heating-induced nonlinearities with mixed reactive-dissipative character have been observed in superconducting resonators, notably in kinetic inductance detectors where the power dissipated by the readout signal can lead to resonance distortion, hysteresis, and switching \cite{deVisser2010, Thompson2013}. In addition to the intrinsic equilibrium ``supercurrent'' nonlinearity, power-induced quasiparticle generation through Cooper-pair breaking gives rise to a power-dependent kinetic inductance resulting in a non-equilibrium ``heating'' nonlinearity \cite{Zhao2022, Thomas2022}. While both effects contribute to soft-spring Duffing dynamics at high drive powers \cite{Swenson2013, Joshi2022}, a few studies have reported anomalous positive frequency shifts at low powers. These shifts were tentatively attributed to resonant TLS heating, but a definitive theoretical explanation and quantitative modeling have so far remained elusive \cite{Sage2011, Wei2020, Kirsh2021}.

In this Article, we demonstrate how readout-induced heating and coupling to TLSs can conspire to endow a micro-mechanical resonator at millikelvin temperature with strong dissipative and reactive nonlinearities, and we develop a model to describe this effect.
In Section~\ref{sec:TLSnonlinPCR}, we present experimental evidence of this thermal nonlinearity in the swept-frequency response of a thin-film quartz PCR. 
Unlike ``spectral hole burning'' two-tone experiments that probe the cavity pull from a subset of TLSs resonant with a detuned pump \cite{Kirsh2017, Capelle2020, Andersson2021}, the nonlinearity reported here emerges from local heating of a spectrally-broad TLS ensemble, generated and detected with a single probe tone.  
Although the TLS-driven dissipative nonlinearity is a well-known effect in quantum circuits, its reactive counterpart has not been discussed extensively in the literature.
Here, owing to the weak external coupling and the high degree of thermal isolation provided by the acoustic bandgap, this reactive effect becomes apparent at relatively low readout powers ($\nbar\sim 10^{3}$). Depending on the ratio $k_B T/h f_r$, the resonance either softens or hardens, giving rise to an asymmetric and hysteretic frequency response. Across all measured powers and temperatures ($T<1$~K), the resonator quality factor is found to be consistently limited by TLS dissipation.
In Section~\ref{sec:ReacDissip}, we formulate a model to account for the measured behavior and show that, when combined with a thermal conductance model linking the TLS effective temperature to the power dissipated in the resonator, the standard tunneling model (STM) \cite{Anderson1972, Phillips1987} can successfully reproduce the strong power-dependent resonator response provided that both reactive and dissipative TLS responses are included. 
We accurately fit the steady-state resonator response in both the frequency (Section~\ref{sec:PowerFreqShift}) and time domains (Section~\ref{sec:Ringdowns}) using an iterative numerical procedure to obtain the power-dependent self-consistent solution. This analysis identifies TLS relaxation damping activated by readout-induced heating as the primary mechanism limiting the mechanical quality factor to $\sim10^7$.
In the low-power limit, we show that the model predicts a generalized Duffing-oscillator physics with a power-dependent and complex-valued Kerr ``constant''. Experimentally we depart from this prediction as we resolve the influence of individual TLSs that appear strongly coupled with the mechanical oscillator. Notably, this observation of a discrete density of TLSs implies that the common procedure of inferring the TLS-induced mechanical loss from the temperature-dependent frequency shift of the resonator can be inaccurate.
Finally, in Section~\ref{sec:phaseDiag}, we simulate a ``phase diagram'' for the reactive nonlinearity and use the model to identify a critical TLS density below which these power-dependent effects could be neglected.

\section{Thermal nonlinearity in phononic crystal resonators}\label{sec:TLSnonlinPCR}

\begin{figure*}[!ht]
\centering
\includegraphics[width=1.0\linewidth]{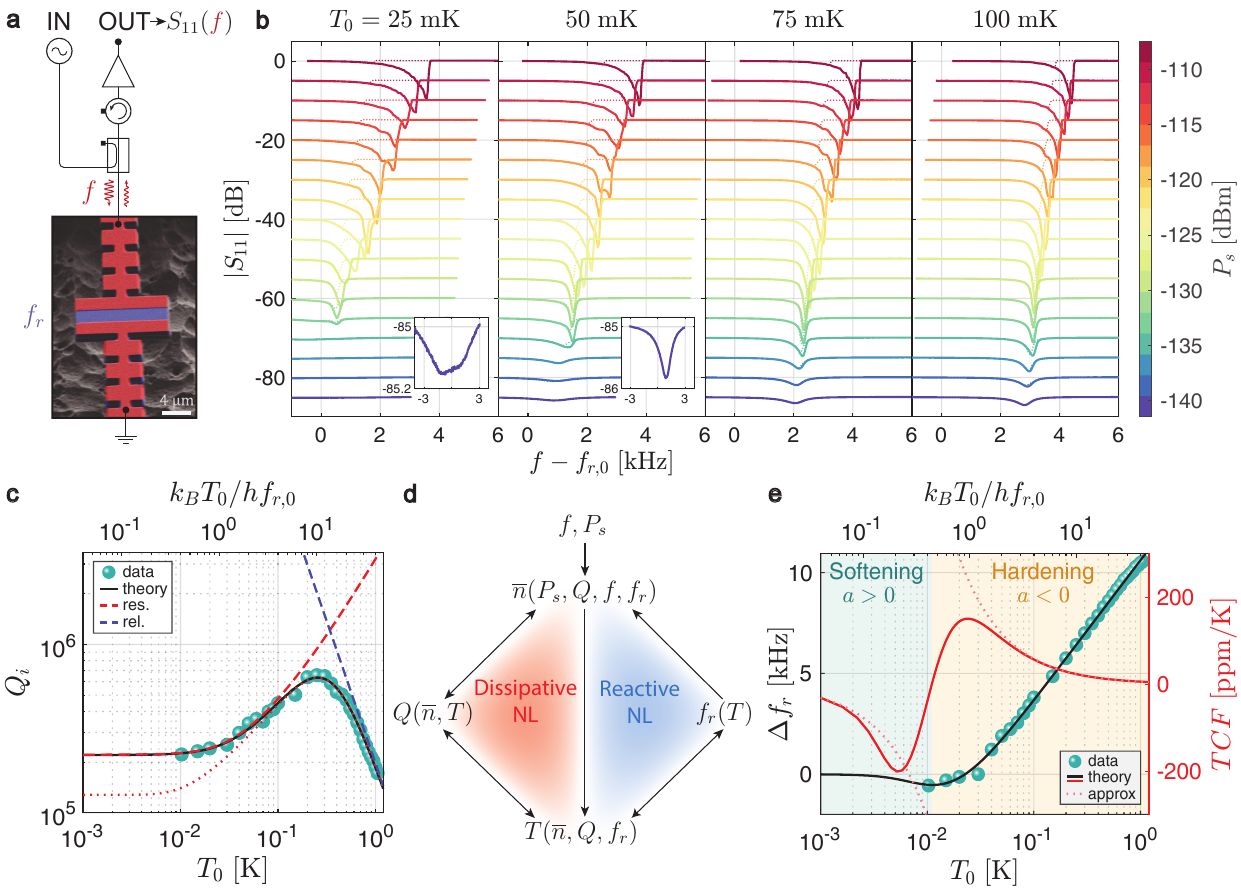}
\caption{A mixed reactive-dissipative thermal nonlinearity in a thin-film quartz PCR. (a)  False-colored scanning electron micrograph of a PCR: a 1-{\textmu}m-thick freestanding quartz beam (blue) patterned with aluminum electrodes (red) to excite the fundamental extensional mode (resonance frequency $f_r$) of the central block. A continuous microwave tone ($f\approx f_r$) is applied, and the reflected signal is analyzed to extract $S_{11}$. (b) Measured $|S_{11}|(f)$ for increasing drive power $P_s$ at select baseplate temperatures $T_0=25$-$100$~mK (Dataset~\#1). Traces were acquired by sweeping $f$ up (solid) and down (dotted) over a 10~kHz window centered at $f_r \approx 520.81$~MHz. With increasing $P_s$, the resonance dip becomes asymmetric and shifts to higher frequency. Curves are vertically offset in 5~dB steps for clarity; insets compare the lineshape at $P_s=-141$~dBm for $T_0=25$ and 50~mK. (c, e) The temperature-dependent internal quality factor $Q_i$ and resonance frequency shift $\Delta f_r$ measured in the low-power limit with $\nbar\approx 300$ (Dataset~\#2, teal disks). Black solid lines show fits to Eqs.~\ref{dfTLS}–\ref{Qimodel}; in (c), resonant-TLS (red dashed) and relaxation-TLS (blue dashed) contributions are indicated. Resonant TLSs within the phononic bandgap are assumed to thermalize at an elevated temperature $T_0'=\sqrt{T_0^2+T_\mathrm{sat}^2}$ with $T_\mathrm{sat}=30$~mK; the red dotted line shows the ideal case $T_0'=T_0$. (d) Schematic of the coupling between external ($f, P_s$), internal ($\nbar, T$), and resonator ($f_r, Q$) parameters giving rise to reactive and dissipative nonlinearities (NL). (e) Right axis: the temperature coefficient of frequency, $TCF \equiv f_r^{-1} \dvstar{\Delta f_r}{T}$ (red solid line), and its high- and low-temperature approximations (Eqs.~\ref{TCFhighT}, \ref{TCFlowT}, red dotted lines) as a function of $T_0$. The sign change at $T_c \approx 0.44 h f_{r,0}/k_B \approx 11$~mK marks a crossover from softening ($TCF<0$) to hardening ($TCF>0$).}
\label{fig1}
\end{figure*}

In this section, we present experimental data from thin-film quartz piezoelectric PCRs showing evidence of a 
thermally-driven amplitude-frequency nonlinearity.
These measurements were performed on the same device reported in Ref.~\cite{Emser2024}, with a focus on the resonance at $f_r = 520.81$~MHz, which combines a strong nonlinear response with relatively low loss. Other resonators on the same device exhibit qualitatively similar nonlinear behavior, differing only in magnitude.  
In Fig.~\ref{fig1}(a), we show a picture of one such PCR, which consists of a 1-{\textmu}m thin periodically-structured beam of quartz hosting in its center a defect site, whose fundamental extensional mode can be excited electrically by means of local aluminum electrodes patterned on its top surface.   
These quartz micro-resonators were extensively characterized in their linear regime in Ref.~\cite{Emser2024} over a broad range of microwave power and temperature ($25<T<800$~mK) using single-tone spectroscopy. 
Although this technique was successfully applied to extract the mechanical damping rate at low power when the response is Lorentzian, the spectral lineshapes of the measured resonators were observed to become strongly asymmetric at moderate drive strengths ($\nbar\gtrsim 2000$). In this regime, the response could not be modeled with the standard Duffing-oscillator formalism for nonlinear resonances, motivating the development of a dedicated theoretical description. Here we extend the device characterization to the high-power regime and model data measured at an average phonon occupancy spanning six orders of magnitude, $10^2<\nbar<10^8$. 

In Fig.~\ref{fig1}(b), we show explicitly the onset of the nonlinear regime and carefully characterize it as a function of power and temperature using single-tone spectroscopy. We refer to this power- and temperature-sweep measurements as ``Dataset~\#1''.
The baseplate temperature of the refrigerator to which the device is thermalized is initially maintained at $T_0=25$~mK and the microwave probe power applied to the sample, $P_s$, is swept over a 34-dB range by increments of 2~dB (Fig.~\ref{fig1}(b) \nth{1} column). 
At the lowest drive power, $P_s=-141$~dBm, the sub-dB resonance depth indicates that the PCR is under-coupled by a factor of about $40$ \footnote{The low external coupling rate $\kappa_e$ from the $Z_0=50\,\Omega$ input line to the PCR devices stems from the weak electromechanical coupling of ST quartz ($K^2\sim 0.14\%$). Using the BAW model of Ref.~\cite{Wollack2021}, we can crudely estimate $\kappa_e$ for the $\omega_r/2\pi=520$~MHz resonator. Taking $\epsilon_r\approx4.6$ for ST quartz and an electrode length $\ell=12$~{\textmu}m gives a static capacitance $C_0\approx \epsilon_0(\epsilon_r+1)\ell/2 \approx 0.3$~fF, leading to $\kappa_e = (8K^2/\pi^2)\omega_r^2 C_0 Z_0 \approx 2\pi\times 30$~Hz. At $T_0=25$~mK, the internal loss rate is $\kappa_i = \omega_r F\delta_0 \tanh{(\hbar\omega_r/2k_B T_0)} \approx 2\pi\times 2.4$~kHz for a typical TLS loss tangent $F\delta_0=1.0\times10^{-5}$, yielding $\kappa_i/\kappa_e \approx 80$.}.
A distinct ``double-resonance'' structure is also visible (see inset), pointing to interaction with a strongly-coupled low-frequency fluctuator, which will be later discussed in Section~\ref{sec:PowerFreqShift}.
From the extracted internal quality factor at that power, $Q_i=1.9\times 10^5$, one can estimate the average phonon number to be $\nbar\approx 120$. 
As $P_s$ is increased, the mechanical resonance dip becomes skewed and shifts to higher frequency, manifesting the emergence of a reactive nonlinearity.
This hardening behavior is at odds with usual Kerr nonlinearities that result in negative shifts of the resonance frequency \footnote{In superconducting circuits, the most common Kerr-like nonlinearity arises from the power-dependent kinetic inductance of the thin superconducting films \cite{Swenson2013, Yurke2006, Joshi2022} and is characterized by a softening response (negative resonance shift with power). Mechanical systems such as MEMS based on doubly-clamped beams can feature nonlinearities of purely geometric origin with both softening and hardening behaviors \cite{Li2023, Zhang2021, Collin2010}. We estimate this geometric effect to be negligible for our PCRs.}. 
Above $P_s\approx-130$~dBm, the resonance enters a bifurcation regime characterized by a hysteretic frequency-domain response, which depends on the sweep direction of the probe tone frequency \cite{Swenson2013}. 
At the highest power, $P_s=-107.5$~dBm, the resonance frequency -- defined as the frequency at which $|S_{11}|$ is minimal \footnote{This is true for upward frequency sweeps, however for downwards sweeps, the probe tone is only \textit{approximately} resonant at the applied frequency where $|S_{11}|$ exhibits its minimum (see Supplemental Material S3).} -- 
has shifted by $3$~kHz from its bare value, more than its $2.7$~kHz linewidth at vanishing power.
Moreover, the resonance experiences a similar shift to higher frequency as the baseplate temperature of the refrigerator is increased. These observations suggest that the power-dependent resonance shift is temperature-driven, with the associated reactive nonlinearity arising from readout-power heating of the device.

To better understand the origin of this thermal nonlinearity, we measured the same set of curves at increased refrigerator temperature and observed a suppression of the reactive effect, both a smaller shift and a less asymmetric resonance at $100$~mK compared to $25$~mK, as evidenced in Fig.~\ref{fig1}(b).
In addition to a power-dependent resonance shift, the spectroscopy data also shows a strong change in the mechanical quality factor, characterized by a drastic increase in the resonance depth across the range of applied power. With $P_s=-141$~dBm, the resonance depth is sub-dB and barely visible, while at $-125$~dBm, it spans almost 15~dB. Similarly, a $2$~dB dip is visible at $-141$~dBm in the $100$~mK series, while at the same power it was less than $0.2$~dB at $25$~mK.
This ``switch-on'' behavior both with power and temperature is very suggestive of TLS saturation \cite{Thomas2020}. Accordingly, Ref.~\cite{Emser2024} identified resonant TLS absorption as the dominant limitation of the quartz PCR response at low powers.
In the next section, we formalize the dissipative nonlinearity arising from the coupling to the TLS ensemble and show how, in the presence of device heating, an additional reactive nonlinearity can occur from the temperature-dependent frequency shift inherited from TLSs. 

\section{TLS-induced nonlinearities}\label{sec:ReacDissip}

The nonlinear response of a resonator manifests as variations of the reflection coefficient of incident microwave power, $S_{11}$, as the amplitude of the readout tone is increased. As reviewed in Ref.~\cite{Thomas2020}, this can occur through a dependence of the resonance frequency and/or quality factor on the power $P_d$ dissipated internally. 
We refer to changes in the resonant frequency with the dissipated power, $f_r(P_d)=\omega_r(P_d)/2\pi$, as \textit{reactive nonlinearities} and to changes in the quality factor, $Q(P_d)$, as \textit{dissipative nonlinearities}. The realized values of $S_{11}$ can then be found for a given applied readout tone (with frequency $f$ and power $P_s$) by looking for the self-consistent solution to the following coupled equations:
\begin{align}[left=\empheqlbrace]
    &S_{11}(f,P_d) = 1 - \frac{2Q(P_d)/Q_e}{1+2 j Q(P_d)x(f, P_d)}\label{S11self}\\
    &P_d = (1 - |S_{11}(f,P_d)|^2)P_s,\label{PdS11}
\end{align}
where $x(f,P_d)=(f-f_r(P_d))/f_r(P_d)$ is the realized reduced detuning. 
Eq.~\ref{PdS11} relating $P_d$ to the incident power $P_s$ results from energy conservation applied to a one-port network \footnote{A similar expression can be worked out for a 2-port network. For the common case of a short-circuited $\lambda/4$ resonator notch-port coupled to a transmission line (the so-called ``hanger'' style), Ref.~\cite{Chen2022, Thomas2020} derive the following expressions: $S_{11}=S_{22}=-(Q/Q_e)/(1+2 j Q x)$ and $S_{12}=S_{21}=1+S_{11}=1-(Q/Q_e)/(1+2 j Q x)$. The dissipated power is then $P_d=(1-|S_{11}|^2-|S_{21}|^2)P_s = 2(Q^2/Q_e Q_i)P_s/(1+4Q^2 x^2)$, a factor of 2 smaller than the expression for a one-port resonator in Eq.~\ref{Pdnbar}.}. 
The total (loaded) quality factor is given by $Q(P_d)=\omega_r/\kappa=(Q_i^{-1}(P_d) + Q_e^{-1})^{-1}$ and has two contributions: the external (coupling) quality factor, $Q_e=\omega_r/\kappa_e$, associated with the power lost from the resonator to the readout circuit that is assumed to be independent of $P_d$ \footnote{Eqs.~\ref{S11self} and \ref{PdS11} describe a one-port resonator measured in reflection via a lossless circuit. As in Ref.~\cite{Thomas2020}, we assume a power-independent coupling quality factor $Q_e$ and that nonlinearities arise solely through $x(f,P_d)$ and $Q(P_d)$, corresponding to a fixed circuit topology with power-dependent component values.}, and the internal quality factor, $Q_i=\omega_r/\kappa_i$, which characterizes losses within the resonator. 

We now wish to elucidate the $f_r(P_d)$ and $Q(P_d)$ dependences that arise from the coupling of a micro-mechanical resonator to TLSs.
Due to dispersive interactions with TLSs, the resonator frequency is shifted from its bare frequency $f_r^{(0)}$ by an amount $\delta f_r=\delta\omega_r/(2\pi)$ and acquires some finite linewidth $\kappa_{i,\textrm{res}}$, assumed to exceed the intrinsic mechanical linewidth $\kappa_i^{(0)}$. 
Assuming coupling to a TLS continuum with a uniform energy density of states, the standard tunneling model of TLS provides expressions for these two quantities in terms of the resonator's internal variables $\nbar$ and $T$, respectively the average number of phonons in the resonator and the TLS-bath temperature \cite{Phillips1987}:

\begin{equation}\label{dfTLS}\hspace*{-0.1cm}
        \frac{\Delta f_r(T)}{f_r^{(0)}}=\frac{F\delta_0^\textrm{reac}}{\pi}\Bigl[\Re \Bigl\{\Psi\Bigl(\frac{1}{2}+\frac{h f_r^{(0)}}{2\pi i k_B T}\Bigr)\Bigr\}-\textrm{ln}\Bigl(\frac{h f_r^{(0)}}{2\pi k_B T}\Bigr)\Bigr]
\end{equation}\vspace*{-0.7cm}
\begin{equation}\label{QTLS}
    Q_\textrm{res}^{-1}(T,\nbar)= \frac{\kappa_{i,\textrm{res}}}{\omega_r} = \frac{F\delta_0^\textrm{diss}}{\sqrt{1+\nbar/n_s}}\tanh{\left(\frac{h f_r}{2 k_B T}\right)}.
\end{equation}
The resonance frequency shift is expressed here relatively to the shift at zero temperature, $\Delta f_r(T) \equiv \delta f_r(T)-\delta f_r(0)=f_r(T)-f_{r,0}$ \footnote{Using that $\Psi(z)\underset{z\to\infty}{\sim} \ln{(z) + \mathcal{O}(1/z)}$, one can directly verify from Eq.~\ref{dfTLS} that $\lim_{T \rightarrow 0} \Delta f_r(T) = 0$.}, where $f_{r,0}\equiv f_r(T=0)$ denotes the resonator frequency at zero temperature when all the TLSs are in their ground state. $\delta_0$ is the intrinsic TLS loss tangent at zero temperature, $F$ is the filling fraction of TLS in the resonator material, $n_s$ is the critical phonon number for TLS saturation and $\Psi$ denotes the complex digamma function.
Only near-resonant TLSs result in the mechanical loss captured by Eq.~\ref{QTLS}, while Eq.~\ref{dfTLS} describes the net shift arising from a continuum of dispersively coupled TLSs. Hence, both quantities sample a different population of TLSs. Following Ref.~\cite{Emser2024}, we therefore distinguish the average loss tangents contributed by near-resonant ($\delta_0^\textrm{diss}$) and far-detuned TLSs ($\delta_0^\textrm{reac}$) and notice that the two quantities may differ if the TLS density of states is not uniform in energy.
For a detailed derivation of these equations, we refer the reader to Appendix H of Ref.~\cite{Emser2024}.

Eqs.~\ref{dfTLS} and \ref{QTLS} are known as the \textit{resonant} TLS contribution to the mechanical susceptibility and dominate the low-temperature/low-power response. At higher temperature, when the TLS relaxation rate $\Gamma_{1,\mathrm{TLS}}$ becomes comparable to the resonator frequency $f_r$, another mechanism originating from the longitudinal interaction between the TLS and the resonator's strain field comes into play \cite{Phillips1987, Meissner2021}. Although it generally does not contribute to any appreciable frequency shift when $f_r \Gamma_{1,\mathrm{TLS}}^{-1}\gg 1$, this \textit{relaxation} TLS contribution does introduce an additional term in the mechanical loss, $Q_\textrm{rel}^{-1}\propto T^d$ with $d$ the dimensionality of the phonon bath interacting with the TLSs, that competes with the Q-enhancement from saturation of resonant TLSs \cite{MacCabe2020, Mittal2024}. 
Following \cite{Wollack2021, Emser2024}, we therefore model the total internal loss as the sum of the following three contributions:
\begin{equation}
    Q_i^{-1}(T,\nbar)=Q_\textrm{res}^{-1}(T,\nbar)+Q_\textrm{rel}^{-1}(T)+Q_\textrm{bkg}^{-1},\label{Qimodel}
\end{equation}
where the power- and temperature-independent $Q_\textrm{bkg}=\omega_r/\kappa_i^{(0)}$ denotes the intrinsic (background) quality factor that would be achieved in absence of any TLS. 

The applied input power $P_s$ determines the average number of phonons $\nbar$ in the resonator \cite{Aspelmeyer2014}:
\begin{equation}\hspace*{-0.1cm}
    \nbar(P_d) = \frac{Q_i\bigl(T(P_d),\nbar(P_d)\bigr)}{\hbar\omega_r\bigl(T(P_d)\bigr)^2}P_d=\frac{\nbar_r(P_d)}{1+4Q(P_d)^2 x(f,P_d)^2},\label{Pdnbar}
\end{equation}
where we introduced the average phonon number at resonance $\nbar_r(P_d)=4[Q(P_d)^2/Q_e](P_s/\hbar\omega_r^2)$ and used Eqs.~\ref{S11self} and \ref{PdS11} to express $P_d$ in terms of the resonator parameters and obtain the second expression. 
As described by Eq.~\ref{Pdnbar} and illustrated in Fig.~\ref{fig1}(d), the achieved $\nbar$ value depends on the resonance quality factor $Q$, which is entirely determined by TLS-induced internal losses since $\kappa_i \gg \kappa_e,\kappa_i^{(0)}$. Because $\kappa_i$ itself depends on $\nbar$, this feedback leads to a dissipative nonlinearity. In addition, $\nbar$ depends on the reduced detuning $x$ between the probe tone and the resonator frequency, which is shifted from its bare value by the the sum of the dispersive shifts from each individual TLS. This overall shift, $\delta f_r$, is governed by the population imbalance between the TLS ground and excited states, set by the bath temperature $T$ \footnote{The state of near-resonant TLSs is also affected by the average phonon number $\nbar$; however, their contribution to the overall resonance shift is negligible compared to the one from the thermally populated, off-resonant TLS continuum.}.
Under sufficient microwave drive, device heating renders the TLS temperature $T$ a dynamic quantity that depends on the dissipated power $P_d$, itself a function of $\nbar$, $Q$, and $f_r$, such that the TLS ensemble can not be treated as a passive thermal bath. Crucially, as the TLS ensemble heats up under microwave probing, the temperature-dependent frequency shift it imparts to the mechanical resonator evolves. $T$ inherits an $\nbar$-dependence, introducing a reactive nonlinearity via Eq.~\ref{dfTLS}, and causing the resonator frequency to shift dynamically with probe power.

In summary, dissipative and reactive nonlinearities arise from the $Q(P_d)$ and $f_r(P_d)$ dependences, respectively. Within the standard TLS framework, these may be expressed in terms of the internal variables $\nbar$ and $T$, allowing $f_r$ and $Q$ to be treated as functionals of $T(P_d)$ and $\nbar(P_d)$. Consequently, evaluating $S_{11}$ requires knowledge of both quantities: $\nbar(P_d)$ is given by Eq.~\ref{Pdnbar}, while a generic expression for $T(P_d)$, based on a thermal conductance model, is provided in Appendix~\ref{app:AppGth}.

\section{Power-dependent resonance shift}\label{sec:PowerFreqShift}

\begin{figure*}[!ht]
\centering
\includegraphics[width=0.9\linewidth]{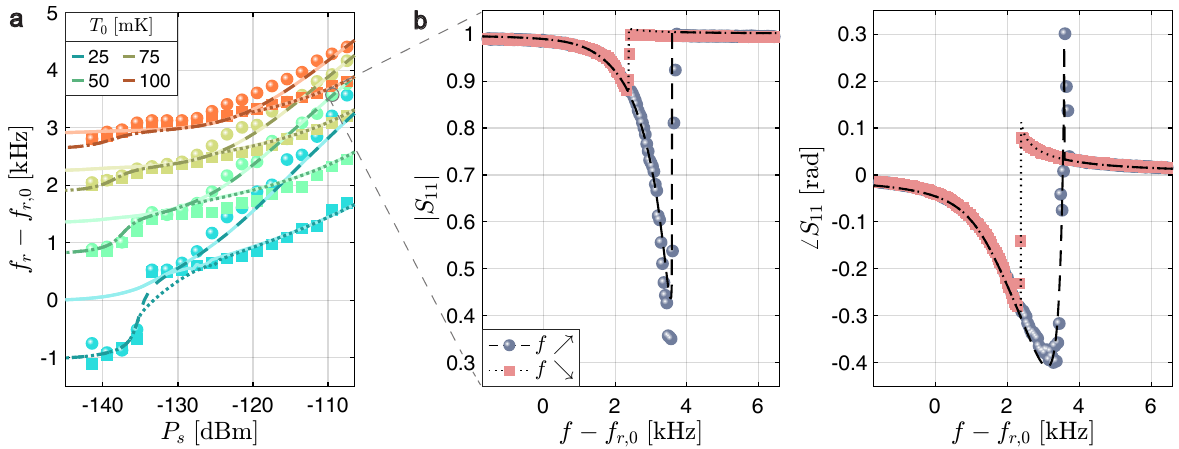}
\caption{Nonlinear resonance frequency shift with power. (a) Resonance frequency $f_r$ versus applied probe power $P_s$ at temperatures $T_0 = 25, 50, 75, 100$~mK, extracted from the $S_{11}$ data in Fig.~\ref{fig1}(b). $f_r$ is shown relative to its extrapolated zero-temperature value $f_{r,0} = 520.8083(1)$~MHz. Disk (square) markers correspond to upward (downward) frequency sweeps, with dashed (dotted) lines showing fits to the numerical model described in Appendix~\ref{app:frPssolver}. A smooth step near $P_s \approx -140$~dBm is captured by including a discrete TLS at $546$~MHz; lighter solid lines show the model without this additional TLS.
(b) The asymmetry and hysteresis in the magnitude and phase of $S_{11}$ at $P_s=-109$~dBm and $T_0=50$~mK are well-reproduced by the model.}
\label{fig2}
\end{figure*}

As a first step towards modeling the power-dependent scattering data of Fig.~\ref{fig1}(b), we measured the temperature dependence of the $520.81$~MHz resonance and verified that it can be fully captured by the TLS theory. Both $Q_i(T_0)$ and $f_r(T_0)$ (Fig.~\ref{fig1}(c) and (e)) were extracted using single-tone spectroscopy performed in the low-power limit ($\nbar\approx 300$) such that $T\approx T_0$, so as to eschew the nonlinear regime where their achieved values depend on the applied probe power. We refer to these baseplate-temperature-sweep measurements as ``Dataset~\#2''; they were acquired in a different cooldown compared to the power-sweep measurements labeled ``Dataset~\#1'', approximately 5 months later. 
The measured $Q_i(T_0)$ is well captured by Eq.~\ref{Qimodel} over the whole range of temperature, $T_0=20$~mK up to $1$~K. At low temperature, the rise in $Q_i$ reflects the saturation of near-resonant TLSs that contribute to mechanical dissipation via resonant exchange. For $T_0>200$~mK, relaxation damping from off-resonant TLSs becomes the dominant loss mechanism, as detailed in Ref.~\cite{Emser2024}. In spite of the device electrodes being superconducting, quasiparticle dissipation remains negligible below $1$~K because the mode participation is almost entirely mechanical. 

Similarly, we extracted the resonance frequency as a function of temperature and fit it with Eq.~\ref{dfTLS}. The shift $\Delta f_r$ from its extrapolated zero-temperature value, $f_{r,0}$, is shown in Fig.~\ref{fig1}(e) on a logarithmic temperature axis to emphasize its asymptotic $\log T_0$ behavior.
At high $T_0$, the observed logarithmic increase is well captured by Eq.~\ref{dfTLS}, which predicts this form from the thermally weighted sum of dispersive shifts across a flat TLS distribution. The slope of this curve reflects how $f_r$ responds to a small TLS temperature increase $\delta T$ due to microwave-induced heating. We introduce a ``temperature coefficient of frequency'', $TCF(T_0)\equiv f_r^{-1}\dvstar{\Delta f_r}{T}|_{T_0}$, as the figure of merit to quantify the strength of the reactive response.  
As shown on the right axis of Fig.\ref{fig1}(e), the $TCF$ changes sign at a characteristic temperature $T_c$ where $k_B T_c \approx h f_{r,0}$. This marks a crossover from frequency softening to hardening. For our $520.8$~MHz resonance, $T_c \approx 11$~mK, near the base temperature of our dilution refrigerator, so only the hardening regime is clearly observed. Importantly, $TCF$ vanishes at high $T$ (since $\dv{}{T}\log T=1/T$), explaining the smaller resonance shift measured at $100$~mK compared to $25$~mK in Fig.~\ref{fig1}(b).

In Appendix~\ref{app:TLSreac}, we derive a first-order model for the reactive nonlinearity based on the linearization of $\Delta f_r(T)$ around $T_0$. The result is a cubic equation for the realized detuning, akin to Swenson's equation that was derived in Ref.~\cite{Swenson2013} to model the kinetic inductance nonlinearity in superconducting micro-resonators. Although such a first-order theory sheds some light on the general mechanism behind the TLS-induced thermal nonlinearity, for our PCRs its validity is restricted to the range $P_s\lesssim -135$~dBm where the device heating remains small enough that $\delta T\ll T_0$. 
Reproducing the asymmetric lineshapes at high power in Fig.~\ref{fig1}(b) requires a non-perturbative approach that accounts for the full $\Delta f_r(T)$ and $Q(T, \nbar)$ dependence. This is done numerically via an iterative method that computes the self-consistent values ${f_r(T), Q_i(T, \nbar)}$ for each probe condition ${f, P_s}$ (Appendix~\ref{app:TLSIterativeSolver}). We used this approach to model the power-dependent frequency $f_r(P_s)$ from Dataset~\#1.

The results are summarized in Fig.~\ref{fig2}(a) and reveal several key features. First, we observe the emergence at high $P_s$ of the bifurcation regime that was discussed in Section~\ref{sec:TLSnonlinPCR}, with distinct resonance frequencies for upward and downward sweeps. The threshold power for the onset of hysteresis increases with $T_0$, indicating a weakening with temperature of the reactive nonlinearity as expected from the $1/T_0$ decrease of the $TCF$. Indeed, compared to its value at vanishing probe power, $f_r$ shifts by as much as $3$~kHz for $P_s=-110$~dBm at $T_0=25$~mK, while it does by only $\approx 1.5$~kHz at $T_0=100$~mK. 
Second, the resonance frequency for upward sweeps grows as $\sim\log{P_s}$ at high power, unlike the linear scaling for a Duffing oscillator \AppSwenson. This weaker dependence reflects the asymptotic $\log T$ scaling of $f_r(T)$ in Eq.~\ref{dfTLS} and is a signature of the readout-induced TLS heating.
Third, for $P_s<-130$~dBm, $f_r(P_s)$ steps down rather than saturating at the value expected from Eq.~\ref{dfTLS}, $f_{r,0} + \Delta f_r(T_0)$. This can be modeled as an additional dispersive shift from a strongly-coupled TLS slightly detuned above the mechanical mode (Appendix~\ref{app:discreteTLS}). As $T_0$ is raised, this TLS becomes increasingly saturated and its contribution to the resonance shift ultimately vanishes, hence the smaller step at $T_0=100$~mK compared to $25$~mK.
A related mechanism could explain the ``double-dip'' lineshape at $25$~mK seen below $-135$~dBm  (inset, Fig.\ref{fig1}(b)), where the frequency of the strongly-coupled TLS -- and \textit{a fortiori} its dispersive shift on $f_r$ -- itself fluctuates due to an interaction with a low-frequency TLS undergoing random thermal switching \cite{Faoro2015, Muller2019}. At $50$~mK, broadening of the strongly-coupled TLS likely surpasses its energy drift caused by the fluctuator, restoring the Lorentzian lineshape of the resonator.

The colored lines in Fig.~\ref{fig2}(a) show the result of a global fit of Dataset~\#1 to the numerical model presented in Appendix~\ref{app:TLSIterativeSolver}. This model is seen to reproduce well all three features described above.
The TLS temperature inferred from the fit suggests significant heating from the probe: at the highest measured power, $P_s=-107$~dBm, $T$ reaches about $120$~mK for an initial base temperature $T_0=25$~mK and around $200$~mK for $T_0=100$~mK (Fig.~\ref{figSim}). 
Additionally, we show in Fig.~\ref{fig2}(b) how the model successfully reproduces the asymmetric and hysteretic resonance lineshape in the bifurcation regime. 

We emphasize that, unlike typical nonlinear effects in superconducting resonators, the measured $S_{11}$ data cannot generally be captured by a Kerr-type model with a complex frequency shift linear in $\nbar$, that is, a resonance shift of the form $\delta\omega_r = K_\textrm{nl} \nbar$ and a total loss rate $\kappa=\kappa_e+\kappa_i+\gamma_\textrm{nl}\nbar$, where $K_\textrm{nl}$ and $\gamma_\textrm{nl}$ represent Kerr shift and two-photon loss, respectively \cite{Yurke2006, Lozano2025}. TLS effects are an uncommon example of a process in which $Q_i$ increases with stored energy, implying  $\gamma_\textrm{nl}<0$ under TLS saturation \cite{Thomas2022}. 
Such a first-order model treating the reactive and dissipative nonlinearities on equal footing is presented in Appendix~\ref{app:genDuffing}. Based on a linearization of Eq.~\ref{QTLS} about $\nbar=0$, it is formally valid only in the weak-drive regime $\nbar\ll n_s\approx 10^2$ ($P_s\ll -140$~dBm).  Although this confines its applicability to a small subset of Fig.~\ref{fig2}, this framework provides a natural starting point and is expected to extend to higher $\nbar$ in systems with improved thermal anchoring, including 2D PCRs and OMCs.

The product $F\delta_0$, which we refer to altogether as the ``TLS loss tangent'' (extrapolated to zero temperature), is the central parameter in our TLS nonlinearity model as it sets the overall nonlinearity strength. 
As detailed in Appendix~\ref{app:extractTLSFd}, fitting Eq.~\ref{dfTLS} to the four $f_r(T_0)$ points from Dataset~\#1 measured at $P_s=-142$~dBm ($\nbar\sim10^2$) yields $F\delta^{(1)}_0=(1.88\pm0.10)\times 10^{-5}$. This value agrees well with the extracted $F\delta_0^{(2)}=(1.80\pm0.02)\times 10^{-5}$ from Dataset~\#2 (Fig.~\ref{fig1}(e)), measured at a similar phonon occupancy but in a later cooldown.
While Dataset~\#2 was primarily used to independently constrain the parameters of the TLS ensemble for modeling Fig.~\ref{fig2}(a), we found that using $F\delta_0^{(2)}$ to simulate Dataset~\#1 systematically overestimated the frequency shift and failed to reproduce the full power dependence.
This discrepancy arises from the presence of the discrete, strongly-coupled TLS inferred from Dataset~\#1, which contributes a large additional dispersive shift at low power. Fitting $f_r(T_0)$ at a higher power ($P_s=-130$~dBm), where this discrete TLS is saturated and its shift suppressed, yields a reduced loss tangent $\widetilde{F\delta_0}=(1.42\pm0.23)\times10^{-5}$.
Using this adjusted value as the background TLS contribution allows accurate modeling of the measured frequency shift.
Figure~\ref{fig2}(a) shows the expected shift from the TLS continuum (light lines) and the full model including the additional discrete TLS (darker lines). This analysis highlights how smooth steps in the measured frequency shift vs. power can reveal individual strongly coupled TLSs and how, in such a regime, the common procedure of inferring the zero-temperature TLS loss from temperature sweeps at high phonon occupancy may underestimate $F\delta_0$.

\section{Time-domain resonator response}\label{sec:Ringdowns}

\begin{figure*}[!ht]
\centering
\includegraphics[width=0.85\linewidth]{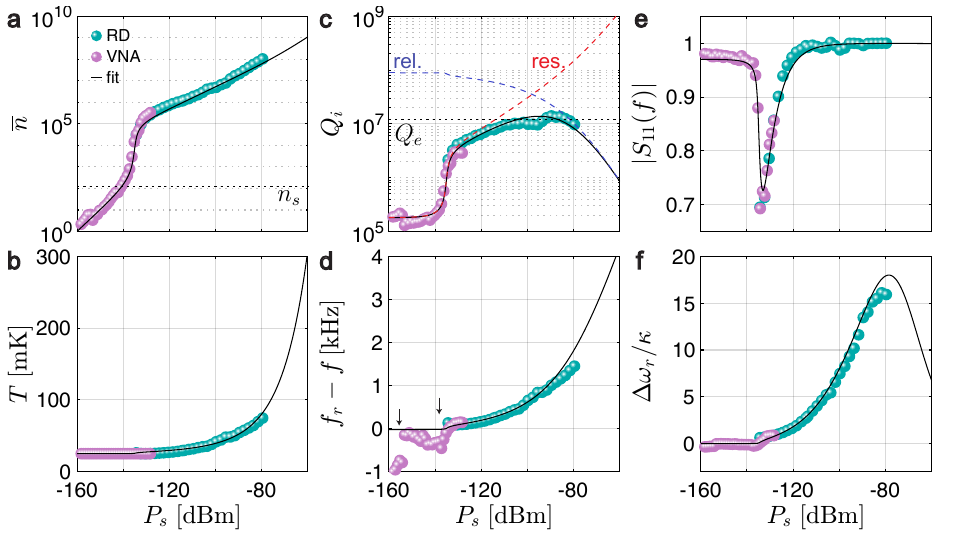}
\caption{Power dependence of the $502.06$~MHz PCR probed at a fixed frequency $f=502.0657$~MHz and with a refrigerator temperature $T_0=25$~mK. (a) The average phonon occupancy $\nbar$, (b) effective TLS temperature $T$, (c) internal quality factor $Q_i$, (d) resonator--probe-tone detuning $f_r-f$, (e) magnitude of the reflection coefficient $|S_{11}(f)|$, (f) resonance frequency shift in number of linewidths, $\Delta \omega_r(T,T_0)/\kappa(T)=2\pi(f_r(T)-f_r(T_0))/\kappa(T)$, as a function of the probe power $P_s$. Teal and purple dots are data extracted from respectively ringdown (RD) and vector network analyzer (VNA) measurements. The black solid lines are fits to the numerical model. The dashed red and blue lines in (c) show the contributions from respectively $Q_\textrm{res}$ and $Q_\textrm{rel}$, corresponding to TLS resonant and relaxation damping processes, and the black dotted line indicates the value of $Q_e$. The two arrows in (d) indicate possible frequency jumps due to gradual saturation with $P_s$ of individual TLSs strongly-coupled to the resonator. The model parameters are summarized in Table~\ref{tableFitParams}.}
\label{fig3}
\end{figure*}

The preceding analysis demonstrates that the nonlinear reactive-dissipative response of the PCR arises from the complex frequency pull from a readout-power-heated system of TLSs. While our model successfully reproduces the measured power-dependent resonance shift, the associated rise in TLS temperature is inferred from fitting rather than directly measured. Furthermore, estimating the dissipated microwave power relies on a model for the power-dependent quality factor. However, single-tone spectroscopy falls short in the strongly nonlinear regime, where the absence of a closed-form model for the distorted lineshape hinders accurate extraction of the quality factor. Although our fitting targets the resonance shift directly, the model’s ability to capture the dissipative response is therefore supported only indirectly, through how well it reproduces the measured resonance depth.
Additional limitations in the extraction of $Q$ from scattering data are discussed in Appendix~\ref{app:frPssolver}.
Time-domain ring-down measurements provide an alternative approach to characterizing these nonlinear resonators. In Fig.~\ref{fig3}, we revisit previously reported ring-down data from a different PCR with $f_r=502.06$~MHz, which had only been partially modeled in Ref.~\cite{Emser2024}. We now show that our TLS nonlinearity model enables a self-consistent, global fit to the entire ring-down dataset, spanning nearly eight orders of magnitude in phonon occupancy. 

We determine the quality factor from the mechanical decay time, obtained by fitting an exponential model to the magnitude-averaged measured ring-down signal. With the external quality factor $Q_e=1.19\times 10^7$ known from spectral measurements in the linear regime, the internal quality factor can be disentangled. The resonance frequency at a given $P_s$ is inferred from the beating at the detuning frequency between the probe tone and the reflected transient that occurs as the PCR rings up. Using a previously established $f_r(T)$ calibration curve from temperature sweeps in the linear regime, an effective TLS temperature is then deduced by inverting the frequency–temperature relation -- effectively using the PCR resonance as a thermometer for the coupled TLS system. Knowing both $f_r$ and $Q_i$, the intracavity phonon number can be determined with Eq.~\ref{Pdnbar}. Finally, the reflection coefficient at the probe frequency, $|S_{11}(f)|$ can be deduced as $|A_s/(A_s-A_r)|$ where for a given $f$, $A_s$ is the measured steady-state amplitude of the demodulated pulse and $A_r$, the amplitude of the ring-down decay when power is turned off \cite{Emser2024}. 

Unlike the spectral measurements in Fig.~\ref{fig2} where the probe tone is swept across resonance, here the probe frequency is fixed and initially set to match the resonance frequency in the limit of vanishing power. Since their relative detuning can only grow as the resonance is shifted, no hysteretic switching ever occurs in any of the resonance parameters, contrary to the swept frequency response, and all the quantities remain smooth functions of the probe power. 
As $P_s$ is increased from $-160$~dBm, the intracavity phonon number $\nbar$ initially grows linearly with $P_s$ (Fig.~\ref{fig3}(a)) and the resonance frequency does not shift much compared to its linewidth. In this linear regime, however, we do observe small glitches in $f_r$ (black arrows in Fig.~\ref{fig3}(d)), which we interpret as resulting from the gradual cancellation of individual dispersive shifts from a few strongly-coupled TLSs as $\nbar$ is increased, similarly to the jump-down in frequency in Dataset~\#1.
Around $P_s\approx-140$~dBm, $\nbar$ approaches the critical value for TLS saturation, $n_s\approx 130$, and the internal quality factor increases sharply with power, following $Q_i\propto\sqrt{1+\nbar/n_s}$ (Fig.~\ref{fig3}(c)). 
Since $\nbar\sim Q_i^2 P_s$ in this regime, the rapid rise in $Q_i$ results in a self-accelerating increase in $\nbar$ and in the associated dissipated power.
Between $-140$ and $-135$~dBm, $\nbar$ grows by two orders of magnitude, triggering significant device heating and the onset of reactive nonlinearity as $\nbar$ reaches $n_h\approx 10^4$ (Appendix~\ref{app:AppGth}).
As $P_s$ continues to rise, the resonance frequency begins to shift away from the fixed probe frequency due to heating. Fig.~\ref{fig3}(d) shows the resulting monotonic increase in $f_r-f$.
At $P_s\approx -130$~dBm ($\nbar\approx 1.2\times 10^5$), the probe-resonator detuning exceeds the resonator linewidth (Fig.~\ref{fig3}(f)), slowing the growth of $\nbar$, which now scales as $\nbar\propto P_s/(f-f_r)^2\sim P_s^{1/2}$. In this regime, $\nbar$ becomes limited by detuning rather than by the internal quality factor. Although at a slower rate, $Q_i$ still keeps increasing due to further TLS saturation and critical coupling is eventually reached at $P_s\approx-105$~dBm.
The increasing detuning with $P_s$ is also evident in the behavior of the reflection coefficient $|S_{11}(f)|$ shown in Fig.~\ref{fig3}(e). At vanishing probe power, the PCR is under-coupled, leading to a shallow resonance with $1-|S_{11}|=2Q/Q_e\approx 0.02$. As $Q_i$ rises, the resonance rapidly grows deeper, reaching a minimum in $|S_{11}|$ near $P_s\approx -134$~dBm where $Q_i\approx Q_e/2$, close to critical coupling. Beyond $P_s=-130$~dBm, the PCR frequency shifts by more than its linewidth, making the probe increasingly off-resonant such that $|S_{11}|\to 1$. 
At around $P_s=-95$~dBm, the device temperature reaches approximately $50$~mK -- double its base temperature -- and $Q_i$ begins to decline due to increased TLS relaxation damping, now amplified by the elevated temperature.
These results clearly identify device heating as the primary mechanism limiting the maximum achievable quality factor to around $10^7$: although the device starts at a base temperature of $T_0=25$~mK where relaxation damping is negligible, readout-power-induced heating can elevate the temperature to the point where TLS relaxation damping becomes the dominant loss mechanism at high powers. 

\begin{figure*}[!t]
\centering
\includegraphics[width=0.85\linewidth]{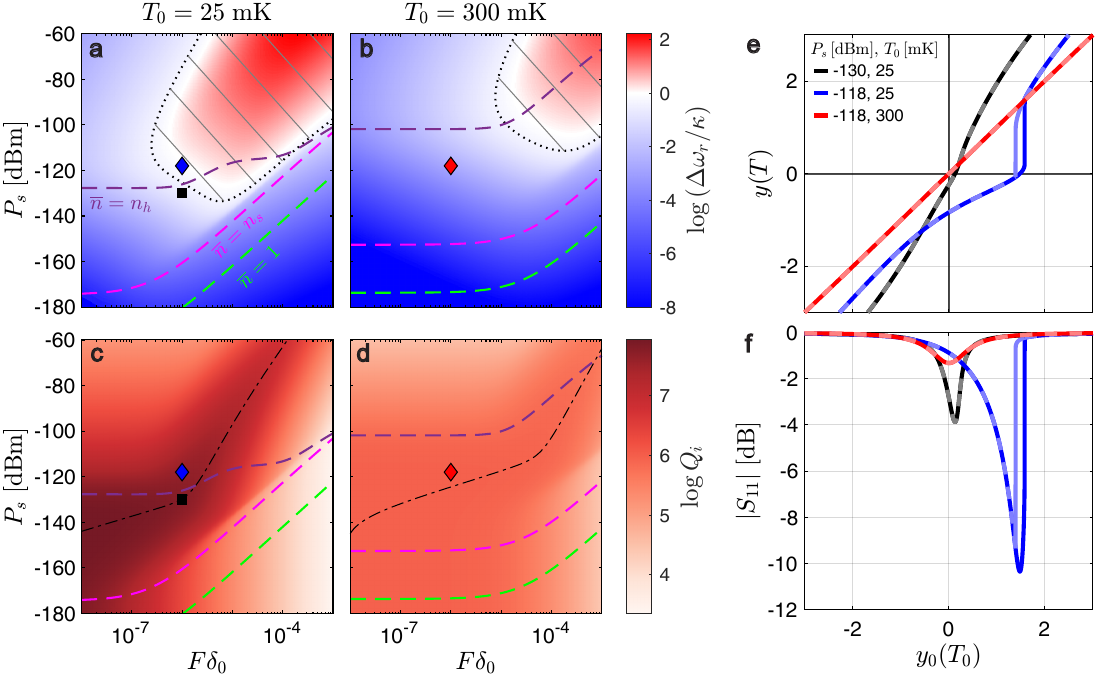}
\caption{Simulated impact of TLS loss tangent on the thermal nonlinearity strength. (a,b) The resonance frequency shift in number of linewidths $\Delta\omega_r(T,T_0)/\kappa=2\pi(f_r(T)-f_r(T_0))/\kappa(T)=-y$ and (c,d) the internal quality factor $Q_i$ as a function of the probe power $P_s$ at the sample and the TLS loss tangent $F\delta_0$ for two illustrative values of base temperature $T_0=25$~mK (a,c) and $300$~mK (b,d). In (a,b) the black dotted line shows the $y=y_c$ contour that delimits the resonance bistability region at high $P_s$. The three colored dashed lines represent contours of achieved phonon occupancies, $\nbar=1,n_s,n_h$ (respectively green, magenta and purple) and demarcate four qualitatively different regimes (see Supplemental Material S4). In (c,d) the black dash-dotted line tracks the maximum value of achieved $Q_i$ in the $P_s\mbox{-}F\delta_0$ space. (e) The achieved fractional detuning $y(T)=Q(T) x(T)$ and (f) the corresponding magnitude of the reflection coefficient $|S_{11}|$ as a function of the applied fractional detuning $y_0(T_0)=Q(T_0)x_0(T_0)$ for three illustrative points shown with the black, blue and red markers in (a-d) corresponding to $F\delta_0=10^{-6}$ and $P_s=-130, -118, -118$~dBm respectively. The responses for upward/downward linear sweeps of $y_0$ are distinguished using dark/light colors.}
\label{fig4}
\end{figure*}

\section{Reactive nonlinearity phase diagram}\label{sec:phaseDiag}

Because it directly governs the growth of $Q_i$ with $\nbar$ and the amount of dissipated power, the product $F\delta_0$ serves as a meaningful figure of merit for evaluating the impact of TLSs on resonator performance. Here, we aim to establish an upper bound on $F\delta_0$ that ensures linear operation of the PCRs.
Using the model parameters extracted from the ring-down data in Fig.~\ref{fig3}, we simulate the power-dependent response of a typical thin-film quartz PCR over a range of TLS loss tangents spanning $10^{-8}\leq F\delta_0\leq 10^{-3}$. In Supplemental Material S4, we derive a critical value for the fractional detuning $\Delta\omega_r/\kappa$ that marks the onset of hysteretic switching, which allows us to construct a ``phase diagram'' for the TLS-induced reactive nonlinearity.
In Fig.~\ref{fig4}(a-b), we highlight the upper-right region of the $P_s-F\delta_0$ parameter space where bistability occurs. Notably, this analysis shows that at a base temperature of $T_0=25$~mK, our resonators with typical values $F\delta_0\sim 10^{-5}$ lie right at the threshold of this bistable region, where bistability occurs for probe powers as low as $-135$~dBm, making the reactive nonlinearity particularly strong in these devices.
If $F\delta_0$ were either an order of magnitude higher or lower, bistability would only emerge for higher probe powers, $P_s\gtrsim-120$~dBm. Crucially, our analysis suggests that if fabrication improvements could reduce $F\delta_0$ to around $10^{-7}$, then at $T_0=25$~mK the PCR would remain in the linear/weakly-nonlinear regime across the entire power range, as the resonance frequency shift would never exceed the resonator linewidth -- effectively suppressing the onset of bistability altogether. A higher external coupling rate could also reduce the overall sensitivity to readout power and push the bistability region to higher power \footnote{See Supplemental Material S5}.

In Fig.~\ref{fig4}(f) we present the simulated scattering parameter $|S_{11}|$ as a function of probe tone frequency, expressed in terms of the applied fractional detuning at $T_0$, defined as $y_0(T_0)=Q(T_0)x_0(T_0)$. The plot illustrates three representative cases with $F\delta_0=10^{-6}$. At $T_0=25$~mK and probe power $P_s=-130$~dBm (black square in Fig.~\ref{fig4}(a-c)), the simulation indicates an internal quality factor $Q_i\approx 5\times 10^7$ and an average phonon number $\nbar\approx 2\times 10^6$. The resonance response is predominantly Lorentzian and the system shows only a weak nonlinearity, as evidenced by the nearly linear relationship $y(y_0)$ shown in Fig.\ref{fig4}(e).
Increasing the probe power to $-118$~dBm shifts the operating point into the bistable region of the ``phase diagram'' (blue diamond), resulting in a strongly asymmetric and hysteretic $|S_{11}|$ response. 
Despite a 4.5 times increase in phonon number to $\nbar\approx 1.1\times 10^7$, $Q_i$ drops to approximately $4\times 10^7$ due to enhanced relaxation damping from the TLSs. This slight reduction in $Q_i$ leads to a deeper resonance, as the system becomes closer to critical coupling ($Q_e\approx 1\times 10^7$). 
Finally, at the same probe power but elevated temperature $T_0 = 300$~mK (red diamond), the model predicts a complete suppression of the reactive nonlinearity: the resonance reverts to a Lorentzian shape with $y = y_0$. However, the increased temperature enhances relaxation damping from off-resonant TLSs, reducing the internal quality factor to $9 \times 10^5$ at a phonon occupancy of $\nbar \approx 3.6 \times 10^5$. Thus, while a higher operating temperature can mitigate nonlinear effects, it also limits the maximum achievable mechanical $Q$ -- highlighting a fundamental trade-off between thermal stability and performance.

\section{Discussion and conclusions}\label{sec:Disc}

In this work our analysis focused solely on the steady-state nonlinear response of the resonator, neglecting time-dependent dynamics associated with TLS heating. Interestingly, over the wide range of probe powers studied here, the ringdown decay following power turn-off remained apparently exponential, with no clear sign of a mechanical-$Q$ drop as the stored energy dissipates. Likewise, the transient beating observed during ring-up showed no indication of frequency chirping, implying that the TLS temperature and probe-resonator detuning equilibrate faster than the resonator field, allowing both to be treated as quasi-stationary (see Appendix~\ref{sec:cavityDynamics}). 

These observations point to distinct thermal time scales: rapid heating upon probe turn-on ($\lesssim 10$~{\textmu}s), and much slower cooling after power removal ($\sim 10$~ms). 
Such asymmetry between heating and cooling time scales may stem from the temperature dependence of the dissipated power and was well described in the context of quasiparticle heating \cite{Thompson2013}. At low probe power, where the resonator is under-coupled and resonant-TLS damping dominates, the dissipated power $P_d\approx 4(Q_\textrm{res}(T)/Q_e)P_s$ depends only weakly on temperature near $25$~mK, since $\coth{(h f_r/2k_B T)}\approx 1$. At high power, however, strong detuning and relaxation-TLS damping render the dissipation strongly temperature dependent, $P_d \approx P_s / (Q_e Q_\mathrm{rel}(T) x(T)^2) \sim T^d$, leading to slower cooling dynamics as the TLS system heats, in qualitative agreement with our ringdown data. The observed thermal relaxation time $\tau$ then reflects the competition between the intrinsic TLS thermalization time $\tau_c$, limited at low power by phonon transport through the PCR tethers ($\tau_c\propto 1/G_{th}(T)$, see Appendix~\ref{app:GthRC}), and the resonator-induced heating, with $\tau^{-1}=\tau_c^{-1}-\tau_h^{-1}$ and $\tau_h^{-1}\propto\dvstar{P_d(T)}{T}$. While at low power $\dvstar{P_d(T)}{T}\approx 0$ and $\tau\approx \tau_c$, at high power $\tau_h$ can approach $\tau_c$, yielding an apparent cooling time $\tau\gg\tau_c$ that no longer reflects the intrinsic TLS dynamics -- a situation akin to electrothermal feedback in transition-edge sensors \cite{Irwin1995}. 

More generally, the interplay between different time scales -- including the resonator decay time, thermal relaxation time, and readout sweep rate -- may produce rich dynamical behavior beyond our steady-state TLS model, such as relaxation oscillations \cite{Diallo2015, Haque2022} and mode-locking \cite{Weng2015}, with the temperature-induced frequency shift acting as a ``built-in'' feedback mechanism that can stabilize the resonance. These dynamical effects have been extensively observed in high-$Q$ optical whispering-gallery mode resonators, in which thermal nonlinearities cause comparable blueshifts and dynamical behavior, due to the combined thermal expansion of the resonator material and the thermo-refractive effect  \cite{Carmon2004, Schmidt2008}. 
As a final remark, we note that although the reported thermal nonlinearity is detrimental to high-$Q$ performance -- limiting energy buildup and slowing TLS saturation -- the resulting hysteretic switching may offer opportunities for sensing applications or for realizing a ``TLS-based parametric amplifier,'' as hinted in Ref.~\cite{Thomas2022}. 
In both cases, a thorough characterization of the heating dynamics and the response time of the TLS-induced nonlinearity will be essential.

The data that support the findings of this study are
available from the corresponding author upon reasonable
request.

\begin{acknowledgments}
This material is based upon work supported by the Air Force Office of Scientific Research and the Office of Naval Research under award number FA9550-23-1-0333. We acknowledge support from the Office of the Secretary of Defense via the Vannevar Bush Faculty Fellowship, Award No. N00014-20-1-2833 and N000142512111. We thank Zurich Instruments and Edward J. Kluender for their aid in the setup of the SHFQC+ Qubit Controller for ringdown measurements. We are grateful to Sarang Mittal, Kazemi Adachi and Pablo Aramburu Sanchez for fruitful discussions related to this work. 
\end{acknowledgments}


\appendix

\counterwithin{figure}{section}
\counterwithin{table}{section}
\renewcommand{\thefigure}{\thesection\arabic{figure}}
\renewcommand{\thetable}{\thesection\arabic{table}}

\section{Model for the TLS-induced nonlinearity}\label{app:TLSnonlin}

\subsection{TLS reactive nonlinearity}\label{app:TLSreac}

In this appendix, we derive a first-order model for the TLS-induced reactive nonlinearity arising from the temperature-dependent resonance frequency shift in Eq.~\ref{dfTLS}. This yields an expression for the reduced detuning in Eq.~\ref{S11self}, $x(f, P_d)$, valid in the limit of weak device heating. 
A key point to emphasize is that the detuning $x$ is not directly controllable. Although the readout frequency $f$ is externally set, the resonance frequency $f_r(P_d)$ shifts with dissipated power in the presence of a reactive nonlinearity, making the detuning power-dependent. We therefore distinguish the \textit{realized} detuning, $x=(f-f_r(P_d))/f_r(P_d)$, from the \textit{applied} detuning, $x_0=(f-f_r(0))/f_r(0)$, where $f_r(0)$ is the ``bare'' resonance frequency in the limit of vanishing readout power, $P_s$, such that $P_d\rightarrow 0$.

We determine $x(f, P_d)$ by linearizing all relevant quantities around the operating temperature $T_0$. We assume that the TLS system is initially thermalized at $T_0$ and denote $f_r(T_0)$ the resulting resonator frequency. From Eq.~\ref{dfTLS}, we define a ``temperature coefficient of frequency'' (TCF) \footnote{We use here a fractional form where TCF is expressed in parts per degree, to keep with the usual definition for SAW devices, \cite{Bechmann1962, Chung2004}.}, which quantifies how the resonance frequency shifts, $f_r(T_0)\to f_r(T_0)+\delta f_r$, in response to a small increase, $\delta T\ll T_0$, of the TLS temperature, $T_0\to T_0+\delta T$:
\begin{multline}\label{TCF0}
    TCF(T_0)\equiv f_r(T_0)^{-1}\frac{d}{d T}\Delta f_r(T)\Bigr|_{T_0} \approx \frac{F\delta_0^\textrm{reac}}{\pi}\Biggl[\frac{1}{T_0}\\-\frac{h f_r}{2\pi k_B T_0^2}\Im \biggl\{\Psi'\bigl(\frac{1}{2}+\frac{h f_r}{2\pi i k_B T_0}\Bigr)\biggr\}\Biggr],
\end{multline}
Here, $\Psi'$ stands for the first derivative of the complex digamma function (also known as the trigamma or polygamma function of order 1) and $f_r^{(0)}$ was approximated by $f_r(T_0)$, which refers to the resonator's frequency at the operating temperature $T_0$, in the limit of vanishing probe power -- ideally matching the refrigerator temperature. As a probe tone with reduced detuning $x$ is swept across the resonance, power $P_d(x)$ is dissipated in the resonator, leading to a steady-state temperature increase $\delta T(x,T_0)$. In the regime where $\delta T\ll T_0$, this temperature rise can be expressed in terms of an effective thermal resistance $R_\textrm{th}(T_0)$ as
\begin{equation}\label{dT}
    \delta T(x,T_0) = P_d(x)\; R_\textrm{th}(T_0). 
\end{equation} 
For phononic crystal resonators, $R_\textrm{th}(T_0)$ is primarily determined by the lattice thermal resistance of the clamping structures supporting the defect site, which serve as the thermal pathway to the surrounding cold substrate. 
Here we assume that the temperature directly settles to its steady-state value, neglecting transients associated with the finite heat capacity of the defect site. A discussion of the conditions under which this approximation is valid is provided in Appendix~\ref{app:GthRC}. 
The temperature increase directly translates into a frequency shift for the resonator, which can be expressed as:
\begin{align}\label{fTCF}
    \frac{\delta f_r(x, T_0)}{f_r(T_0)} = TCF(T_0)\; P_d(x)\; R_\textrm{th}(T_0).
\end{align}
Since temperature $T$ characterizes the internal process driving the nonlinearity, we simplify the notation by writing $\delta f_r(T)$ instead of $\delta f_r(x, T_0)$, as $T$ is implicitly defined by $T(x, T_0)$. The realized reduced detuning $x(T)$ at temperature $T$ is then given by:
\begin{align}\label{xT}
    x(T) &= \frac{f - f_r(T_0)-\delta f_r(T)}{f_r(T_0)+\delta f_r(T)}\nonumber\\
    &\underset{\;\;\mathclap{\delta f_r\ll f_r}}{\approx} x_0(T_0) - \delta x(T),
\end{align}
where we identified the zeroth-order term as the applied detuning $x_0(T_0)$ and the first-order correction $\delta x(T)$:
\begin{equation}\label{xT0}
    x_0(T_0)=\frac{f - f_r(T_0)}{f_r(T_0)},\quad\delta x(T)=\frac{\delta f_r(T)}{f_r(T_0)}.
\end{equation}
The dissipated power $P_d(x)$ given in Eq.~\ref{Pdnbar} can be recast as the product of a \textit{detuning} efficiency $\chi_d$ and a \textit{coupling} efficiency $\chi_c$ \cite{Zmuidzinas2012}:
\begin{equation}
    P_d(x) = \chi_c\,\chi_d(x)\,P_s, \label{Pdprod}
\end{equation}
where
\vspace{-0.7cm}
\begin{align}[left=\empheqlbrace]
    &\chi_c = \frac{4 Q_e Q_i}{(Q_e+Q_i)^2}=\frac{4 Q^2}{Q_e Q_i},\label{chiC}\\
    &\chi_d(x) = \frac{1}{1+4 Q^2 x^2}.\label{chiD}
\end{align}
Combining Eqs.~\ref{fTCF}, \ref{xT} and \ref{Pdprod} yields an implicit equation for $x$, which we can express in a normalized form by introducing the realized fractional detuning measured in total linewidths $y=Q x$:
\begin{align}\label{yT}
    &y(T) = y_0(T_0) + \frac{a(T_0)}{1+4y(T)^2}\\
    &\textrm{with}\quad a(T_0)\equiv -R_\textrm{th}(T_0)TCF(T_0)\frac{4Q^3}{Q_e Q_i}P_s.\nonumber
\end{align}
A similar cubic equation was derived by Swenson \textit{et al.} to model the kinetic inductance nonlinearity in superconducting resonators \cite{Swenson2013}. The Duffing-like dynamics encoded in this equation as well as analytical approximations to its solutions are discussed in Supplemental Material S3. $y$ and $y_0$ refer to the probe tone's detuning measured in total linewidths relative to respectively the power-shifted resonance and the unshifted resonance in the limit of vanishing readout power. When $P_s$ vanishes, $a\to 0$ and the resonance is unshifted, $y=y_0$. The parameter $a$ therefore encodes the strength of the reactive nonlinearity. Its magnitude at $T_0$ is controlled by $TCF(T_0)$, which depends strongly on temperature, as illustrated in Fig.~\ref{fig1}(e). The TCF changes sign at a crossover temperature $T_c$ given by approximately \footnote{This approximation for the crossover temperature $T_c$ can be derived using a known inequality constraining the derivative of the digamma function \cite{Chen2005}, $(x+1/2)/x^2\leq \Psi'(x)\leq (x+1)/x^2$ for $x>0$. Solving $TCF(T_c)=0$ numerically yields $k_B T_c/h f_r\approx 0.4408$, which is within $~2\%$ of the approximate result from Eq.~\ref{TcTCF}.}:
\begin{equation}\label{TcTCF}
    \frac{k_B T_c}{h f_r}=\frac{\sqrt{2}}{\pi}\approx 0.45
\end{equation}

As a result, the nonlinearity can be either softening when $T_0<T_c$ ($TCF<0$, $a>0$), or hardening for $T_0>T_c$ ($TCF>0$, $a<0$).
For a resonator at $f_r=500$~MHz, the temperature crossover happens at $T_c\approx 11$~mK, right at the lower limit of the temperature range accessible with standard dilution refrigerator technology. The measurements reported here are therefore mostly in the high-temperature regime, $T_0\gg T_c$, where the TCF and nonlinearity parameter can be approximated by
\begin{equation}\label{TCFhighT}
    TCF(T_0)\underset{T_0\gg T_c}{\sim}\frac{F\delta_0^\textrm{reac}}{\pi T_0}
\end{equation}
\begin{equation}
    a(T_0) \underset{T_0\gg T_c}{\sim} -\frac{F\delta_0^\textrm{reac}}{\pi}\frac{R_\textrm{th}(T_0)}{T_0}\frac{4Q^3}{Q_e Q_i}P_s.
\end{equation}
This expression for $a(T_0)$ highlights that the nonlinear behavior of the resonator is thermally driven ($1/T_0$ dependence) and originates from the coupling of the resonator to TLSs ($\propto F\delta_0$). Worse device thermalization (higher $R_\textrm{th}$) and more dissipated power increase $a$ as both effects result in a stronger $\delta T$ and consequently, a larger resonance frequency shift.

Similarly, we can obtain a low-temperature approximation, valid when $T_0\ll T_c$.
Inserting $\Re\{ \Psi(1/2+x/(2\pi i))\}\underset{x\to\infty}{\sim} \ln{(x/2\pi)}-\pi^2/6x^2+\mathcal{O}(1/x^4)$ into Eq.~\ref{dfTLS} and differentiating with respect to $T$ yields a linear scaling with $T_0$: 
\begin{equation}\label{TCFlowT}
    TCF(T_0) \underset{T_0\ll T_c}{\approx} -\frac{\pi}{3}F\delta_0^\textrm{reac}\left(\frac{k_B}{h f_r(T_0)}\right)^{\mathclap{2}} T_0
\end{equation}
\begin{equation}
    a(T_0)\underset{T_0\ll T_c}{\approx} \frac{4\pi}{3}F\delta_0^\textrm{reac} \frac{Q^3}{Q_e Q_i}\left(\frac{k_B}{h f_r(T_0)}\right)^{\mathclap{2}}T_0\,R_\textrm{th}(T_0) P_s.
\end{equation}

The previous analysis assumes a linearization of all quantities around the operating temperature $T_0$, $x(T)\approx x_0(T_0)-\delta x(T_0)$ and $T = T_0 + \delta T(x,T_0)$. As a result, the TCF and the nonlinearity parameter $a(T_0)$ are constants. 
While this approximation is convenient and enables simple analytical results, it breaks down when the resonance is lossy and the dissipated readout power causes significant heating, $\delta T\geq T_0$. In practice, one expects $a[T(y)]$ to vary with the applied detuning, yielding more complex nonlinear dynamics than described in this appendix. In that case, the equation for the realized detuning $y$ is no longer a simple cubic polynomial and has to be solved numerically. This will be the focus of Appendix~\ref{app:TLSIterativeSolver}.

\subsection{TLS dissipative nonlinearity}\label{app:TLSdissip}

Having solved the $x(f, P_d)$ dependence in $S_{11}$, we now proceed with characterizing the dissipative component of the nonlinearity arising from the $Q(P_d)$ dependence inherited from the coupling to TLSs. In Eq.~\ref{Qimodel}, we broke down the internal quality factor into three contributions: a power- and temperature-dependent part $Q_\textrm{res}(\nbar,T)$ capturing resonant-TLS damping, a power-independent but temperature-dependent part $Q_\textrm{rel}(T)$ modeling relaxation-TLS damping, and a fixed part $Q_\textrm{bkg}$ to account for any residual power- and temperature-independent internal loss source.

The resonant-TLS damping term, $Q_\textrm{res}$, is a function of the stored energy in the resonator, measured in terms of $\nbar$, which itself depends on the mechanical linewidth and \textit{a fortiori} $Q_\textrm{res}$. This power-dependent loss therefore drives an additional dissipative nonlinearity that is present even for well-thermalized resonators where the reactive nonlinearity discussed in Appendix~\ref{app:TLSreac} is negligible.
This effect was modeled in details in Ref.~\cite{Thomas2020}, starting from the usual $Q_\textrm{res}(T,\nbar)$ parametrization from the STM, Eq.~\ref{QTLS}, with the assumption that the temperature $T$ of the TLS system is fixed, so that the mechanical loss due to TLSs is determined solely by the resonator's stored mechanical energy $h f_r\nbar$. Here, we generalize the results from Ref.~\cite{Thomas2020} using a refined parametrization for $Q_\textrm{res}(T,\nbar)$ that offers improved flexibility for fitting experimental data \cite{Crowley2023, Emser2024}:
\begin{equation}\label{Qtlsparam}
    Q_\textrm{res}^{-1}(T,\nbar) = \frac{Q_\textrm{res,min}^{-1}(T)}{\sqrt{1+\Bigl(\frac{\nbar}{n_s}\Bigr)^\beta \tanh{\left(\frac{\hbar\omega_r}{2k_B T}\right)} } },
\end{equation}
The exponent $0\leq\beta\leq 1$ is an additional phenomenological parameter to describe nonuniform TLS saturation that may arise from the spatially-varying strain distribution over the resonator's mode volume \cite{Wang2009} and $Q_\textrm{res,min}^{-1}(T)=F\delta_0^\textrm{diss}\tanh{(\hbar\omega_r/2k_B T)}$ quantifies the maximum mechanical loss due to resonant TLSs. We also make the temperature dependence of the critical phonon number $n_s$ explicit by recognizing that $n_s^{-1}\propto T_1\propto\tanh{(\hbar\omega_r/2k_B T)}$, where $T_1$ is the average lifetime of the TLS ensemble in thermal equilibrium.

Following Ref.~\cite{Thomas2020}, we express the total quality factor in terms of a TLS saturation parameter $\alpha$ as
\begin{equation}\label{Qtalpha}
    Q = \frac{Q_\textrm{min}}{1-r\alpha}
\end{equation}
with
\begingroup
\setlength{\jot}{2pt} 
\hspace*{-0.5em}
\begin{align}[left=\empheqlbrace]
    &Q_\textrm{min}^{-1} = Q_e^{-1}+Q_\textrm{bkg}^{-1} + Q_\textrm{res,min}^{-1}(T) + Q_\textrm{rel}^{-1}(T) \label{QminEq}\\
    &Q_\textrm{max}^{-1} = Q_e^{-1}+Q_\textrm{bkg}^{-1} + Q_\textrm{rel}^{-1}(T) \label{QmaxEq}\\
    &r =(Q_\textrm{max}-Q_\textrm{min})/Q_\textrm{max} \label{rEq}\\
    &\alpha = 1-1/\sqrt{1+(\nbar/n_s)^\beta \tanh{(\hbar\omega_r/2k_B T)}}.\label{alphEq}
\end{align}
\endgroup
By definition, $0\leq r,\alpha\leq 1$. $Q_\textrm{min}$ and $Q_\textrm{max}$ are the smallest and largest values that $Q$ can take, which corresponds respectively to the two limits where near-resonant TLS are either fully polarized ($\alpha=0$) or fully saturated ($\alpha=1$).  
Assuming $T=T_0$ in Eq.~\ref{Qtalpha}, then both $r$ and $Q_\textrm{min}$ are fixed as they depend only on fixed parameters, $Q_e$, $Q_\textrm{bkg}$ and $Q_\textrm{res,min}$, so that finding the steady-state behavior amounts to solving for $\alpha$ given a readout power $P_s$. 
Expressing $\nbar$ in terms of $\alpha$ from Eq.~\ref{alphEq}, substituting into Eq.~\ref{Pdnbar} and solving for $\alpha$, we obtain $\alpha = f(\alpha)$, where
\begin{equation}\label{falpha}
    f(\alpha) = 1-\frac{(1-r\alpha)^\beta}{\sqrt{(1-r\alpha)^{2\beta}+\chi_d^\beta(\alpha)\xi^\beta\tanh{\bigl(\frac{\hbar\omega_r}{2k_B T}\bigr)}}}
\end{equation}
\bigskip
\vspace{-1.5cm}\begin{equation}\label{chiDtls}
     \textrm{and}\quad0\leq \chi_d(\alpha) = \frac{(1-r\alpha)^2}{(1-r\alpha)^2+(2Q_\textrm{min}x)^2}\leq 1.   
\end{equation}
The detuning efficiency $\chi_d$ from Eq.~\ref{Pdprod} is now expressed in terms of $\alpha$ and $x$, and $\xi$ is a normalized readout power that defines the critical power $P_c$ for TLS saturation:
\begin{equation}\label{xitls}
    \xi=4 Q_\textrm{min}^2 P_s/(Q_e\hbar\omega_r^2 n_s)=P_s/P_c.
\end{equation}
The advantage of this formulation into a fixed-point problem $\alpha=f(\alpha)$ is that $\alpha$ can be easily found numerically, starting from an initial guess and then repeatedly computing $f(\alpha)$. For the case $\beta=1$, Thomas \textit{et al.} showed that this fixed-point equation has one unique solution satisfying $0\leq\alpha\leq 1$ and that the iterative sequence $\alpha_{n+1}=f(\alpha_n)$ always converges to this solution in the limit $n\to\infty$, provided that it starts from $\alpha=0_+$ \cite{Thomas2020}. 

In practice, if the device temperature variations are neglected, $Q_\textrm{min}$ and $Q_\textrm{max}$ are fixed and do not vary over the course of a sweep. Setting $Q_\textrm{min}$ and $Q_\textrm{max}$ determines $r$, which fixes the range over which $Q$ can vary. The parameter that controls the TLS saturation is $\xi$ and it is controlled by the applied readout power $P_s$. With $r$ and $\xi$ fixed, one can compute $f(\alpha)$ starting from a small enough value of $\alpha$. After a few iterations, the converged value of $\alpha$ can then be plugged into Eq.~\ref{Qtalpha} to yield a self-consistent value for $Q$.

In the presence of readout-power heating, the TLS temperature also acquires a power dependence, $T(\nbar)$, and varies as the probe tone is swept across the resonance. Two additional effects, whose magnitude depends on the amount of device heating, may therefore affect the strength of the dissipative nonlinear response: (1) the temperature dependence of $Q_\textrm{res,min}$ which was neglected so far and results in a further saturation of the resonant TLSs and (2) the competing contribution $Q_\textrm{rel}$ from relaxation-TLS damping. Because $Q_\textrm{rel}^{-1}\propto T(\nbar)^d$ and $\nbar$ peaks on resonance, the maximum of $Q$ no longer coincides with the maximum in dissipated power at $x=0$ in the presence of device heating. Relaxation damping therefore reduces the amount of power dissipated on resonance, which also suppresses the strength of the reactive response. This further highlights the need to treat on equal footing both the reactive and dissipative response arising from the coupling to TLSs.
In Appendix~\ref{app:genDuffing}, we show how this can be done in the limit of small heating where all quantities can be linearized around $T_0$, but in the general case relevant where $\delta T\sim T_0$, which is the situation described in this work, the problem needs to be solved numerically. 

\subsection{Generalized Duffing model}\label{app:genDuffing}

Here we generalize the first-order ``Duffing'' type model from \ref{app:TLSreac} to account for mixed reactive and dissipative nonlinear behavior. We follow the formalism developed in Ref.~\cite{Thomas2022} to model the quasiparticle-driven kinetic inductance nonlinearity and adapt it to the case of TLS nonlinearity.
First, we identify the quantity $\delta T$ descriptive of the TLS heating process as the state variable that drives the nonlinearity and linearize all quantities around the operating point $T_0$, $T\approx T_0+\delta T$. 
According to Eq.~\ref{TCF0}, the resonator's fractional frequency shift is then simply $\delta f_r/f_r=TCF(T_0)\delta T$. 
The change in internal loss contains an additional contribution due to the $\nbar$ dependence:
\begin{multline}
    Q_i^{-1}[\nbar(T_0+\delta T),T_0+\delta T]=Q_i^{-1}(\nbar, T_0)\\+\delta T\frac{\delta Q_i^{-1}(\nbar, T_0)}{\delta T}+\delta T \frac{\delta\nbar(T_0)}{\delta T}\frac{\delta Q_i^{-1}(\nbar,T_0)}{\delta \nbar}\\+\mathcal{O}(\delta T^2)
\end{multline}
The three terms are obtained from Eq.~\ref{Qtlsparam} and its derivatives with respect to $T$ and $\nbar$. As we seek here a model applicable to low powers and temperatures, the relaxation damping contribution to $Q_i$ can be neglected. Writing $\epsilon=\hbar\omega_r/(2k_BT_0)$, we obtain in the limit of vanishing power, $\nbar\to0$ (for which $\delta T\to 0$), and for the particular case of $\beta=1$:
\begin{align}[left=\empheqlbrace]
    &\frac{\delta Q_i^{-1}}{\delta T}=-\frac{F\delta_0^\textrm{diss}}{T_0}\epsilon\sech^2{(\epsilon)} \label{dQidT}\\
    &\frac{\delta Q_i^{-1}}{\delta \nbar}=-\frac{F\delta_0^\textrm{diss}}{2 n_s}\tanh^2{(\epsilon)}.\label{dQidn}
\end{align}
From Eqs.~\ref{dT} and \ref{Pdnbar}, one can relate the change of phonon occupancy to $\delta T$ as $\delta\nbar/\delta T=1/(F\delta_0^\textrm{diss}R_{th}(T_0)\hbar\omega_r^2\tanh{\epsilon})$.
To first order in $\delta T$, the complex-valued resonator shift can then be expressed as
\begin{equation}\label{cpxShift}
    \frac{\delta f_r}{f_r}+i\frac{\delta Q_i^{-1}}{2}=e^{i\varphi}\frac{\delta T}{T_*}.
\end{equation}
The phase angle $\varphi=\tan^{-1}(T_d/T_r)-\pi/2\leq 0$, governed by the two temperature scales $T_r=1/TCF(T_0)$ and $T_d^{-1}=1/T_{d0}+1/T_{d1}$, controls the ratio of reactive to dissipative response, and $T_d$ is defined in terms of the following two temperatures:  \begin{align}[left=\empheqlbrace]
    &T_{d0}=16 n_s R_\textrm{th}(T_0) \frac{(k_B T_0)^2}{\hbar}\frac{\epsilon^2}{\tanh{(\epsilon)}}\\
    &T_{d1}=\frac{2T_0}{F\delta_0^\textrm{diss}}\frac{1}{\epsilon \sech^2{(\epsilon)}}.
\end{align}
Expressing the thermal resistance in terms of $N_{ch}$ quanta of thermal conductance, $R_{th}=3h/(N_{ch}\pi^2k_B^2T_0)$ \AppGth, the ratio of these two temperature scales reads $T_{d0}/T_{d1}=48n_s/(N_{ch}\pi)F\delta_0^\textrm{diss}f(\epsilon)\approx 10^{-3}f(\epsilon)$, where $f(\epsilon)=\epsilon^3\sech^2{\epsilon}/\tanh{\epsilon}\leq1/\sqrt{2}$. The temperature dependence of $Q_i^{-1}$, \ref{dQidT}, can therefore be neglected and $T_d\approx T_{d0}=4R_{th}(T_0)\hbar\omega_r^2 n_s/\tanh{(\epsilon)}$ for all practical purposes.

Finally, Eq.~\ref{cpxShift} is supplemented with a dynamical equation for $\delta T$, relating the temperature increase to the time-averaged stored energy in the resonator expressed in terms of $\nbar$:
\begin{equation}\label{dTdyn}
    \frac{d \delta T}{d t}=-\frac{\delta T}{\tau}+\frac{T_*}{\tau}\frac{\nbar}{n_*},
\end{equation}
where $\tau=R_\textrm{th}C_\textrm{th}$ is the response time associated to the device heating dynamics (Appendix~\ref{app:GthRC}) and the relevant temperature and energy scales of the nonlinearity are given by
\begin{align}[left=\empheqlbrace]
    &T_*=\frac{T_d}{\sqrt{1+(T_d/T_r)^2}}\label{nonlinScT} \\
        &n_*=\frac{Q_i(T_0)T_*}{\hbar\omega_r^2R_{th}(T_0)}\label{nonlinScn}.
\end{align}
The steady-state solution of Eq.~\ref{dTdyn} coincides with Eq.~\ref{dT}, $\delta T\to T_*\nbar/n_*=R_\textrm{th} P_d$. This first-order description remains valid as long as $\delta T\ll T_0$, which is equivalent to the condition $\nbar\ll n_* (T_0/T_*)=(1+\gamma)n_h$ derived in Appendix~\ref{app:AppGth}. 
When $T_d\gg T_r$, $T_*\to T_r$, $\varphi\to0$ and the nonlinearity is mainly reactive. Conversely, when $T_d\ll T_r$, $T_*\to T_d$ and $\varphi\to-\pi/2$ and the dissipative character of the nonlinearity dominates. $T_*$ can therefore be seen as the relevant temperature scale for the nonlinearity, interpolating between reactive and dissipative limits. 

Remarkably, the operating-point equation for this mixed reactive-dissipative nonlinearity can still be cast into the same form as Eq.~\ref{yT}, at the price of replacing the applied and realized fractional detunings $y_0$ and $y$ by generalized quantities $k_0$ and $k$ \cite{Thomas2022}: 
\begin{align}[left=\empheqlbrace]
    &k_0=\frac{\Im(z)}{2\Re(z)} \label{k0Swenson}\\
    &k=k_0-\frac{a}{1+4k^2},\label{kSwenson}
\end{align}
where $z=(1+2 i y_0)e^{-i\varphi}$ and $a=a_*/\Re(z)^3$. The input power $P_s$ determines the degree of the nonlinearity $a_*=4Q(T_0)^3 P_s/(\hbar\omega_r^2(T_0)Q_e n_*)$ at the operating point $T_0$, similarly to $\xi$ in Eq.~\ref{falpha}. Given the applied fractional detuning $y_0=Q(T_0)(\omega-\omega_r(T_0))/\omega_r(T_0)$, the generalized fractional shift $k$ obtained by solving Eq.~\ref{kSwenson} then defines $p=(1+2i k)\Re(z)e^{i\varphi}$, from which $S_{11}=1-2Q(T_0)/(pQ_e)$ is deduced.

When the nonlinearity has both reactive and dissipative characters, the threshold power for bifurcation is increased by a factor $\mathcal{F}=1/(8 \cos^3(\pi/3-\varphi))>1$ compared to the purely reactive case \cite{Thomas2022}. For a purely reactive nonlinearity ($\varphi=0$), one recovers $k_0=y_0$ and $a=\mathcal{F}a_c=a_c=4/(3\sqrt{3})$ \AppSwenson. Since $\mathcal{F}\to\infty$ as $\varphi\to-\pi/6$, the bifurcation regime becomes inaccessible when the amount of dissipative nonlinearity is such that $-\pi/2\leq\varphi<-\pi/6$. This inequality on $\varphi$ translates into the condition $T_d/T_r\geq\sqrt{3}$ for switching to occur. 
Using the high-temperature approximation of the TCF valid at $50$~mK, Eq.~\ref{TCFhighT}, this can be rewritten as a condition on the TLS loss tangent:
\begin{equation}
    F\delta_0^\textrm{reac} \geq \frac{\pi^2}{8\sqrt{3}}N_{ch}\frac{n^2_{th}}{n_s}\tanh{\frac{1}{2 n_{th}}},
\end{equation}
where $n_{th}=k_B T_0/(h f_r)$ is the high-temperature approximation for the average number of thermal phonons in the resonator. With $N_{ch}=0.4,\; n_s=100,\; T_0=50$~mK and $f_r=520$~MHz, this inequality evaluates to $F\delta_0^\textrm{reac}\geq2.8\times 10^{-3}$ with $n_{th}\approx 2.0$. In the $50$~mK data presented in Fig.~\ref{fig1}, however, $F\delta_0\sim 10^{-5}$, which is two orders of magnitude lower than this bound, and yet the bifurcation regime is reached. Indeed, in the power regime where switching occurs, the assumption $\delta T\ll T_0$ is no longer verified and consequently this first-order generalized Duffing model cannot be applied. To fit the experimental data, instead of using Swenson's equation, we therefore resort to a non-perturbative approach and solve the full set of coupled equations numerically (Appendix~\ref{app:TLSIterativeSolver}). At $T_0=25$~mK, the $TCF\approx 150$~ppm/K value from Fig.~\ref{fig1}(e) results in $T_r\approx10^4$~K, while $T_d\approx 10^2$~K. The ratio $T_d/T_r\approx 10^{-2}$ confirms that at base temperature, the TLS nonlinearity is mainly dissipative as long as $\nbar\ll n_s$ and $\delta T\ll T_0$. In this regime, Eqs.~\ref{nonlinScT}-\ref{nonlinScn} simplify to $T_*\approx T_{d0}$ and $n_*\approx4n_s/(F\delta_0^\textrm{diss}\tanh^2{\epsilon})\approx1.8\times 10^8$, which allows one to estimate the critical phonon number for TLS heating, $n_h\approx 1.2\times 10^4$. This value agrees well with Fig.~\ref{fig3}.
Finally, given that $Q_i$ was linearized around $\nbar=0$, we note that this first-order theory is only valid in the restricted range $\nbar\ll n_s\approx 100$, which also ensures $\delta T\ll T_0$ since $n_s< n_h$. Consequently, it cannot be used to model the full dataset in Fig.~\ref{fig2}, where $\nbar$ varies by six orders of magnitude. Since $Q_i$ is unbounded below, attempts to apply this model for $\nbar> 2n_s/\tanh{\epsilon}$ would result in an unphysical $Q_i<0$.

\subsection{Iterative TLS nonlinearity solver}\label{app:TLSIterativeSolver}

\begin{figure*}[!ht]
\centering
\includegraphics[width=1.0\linewidth]{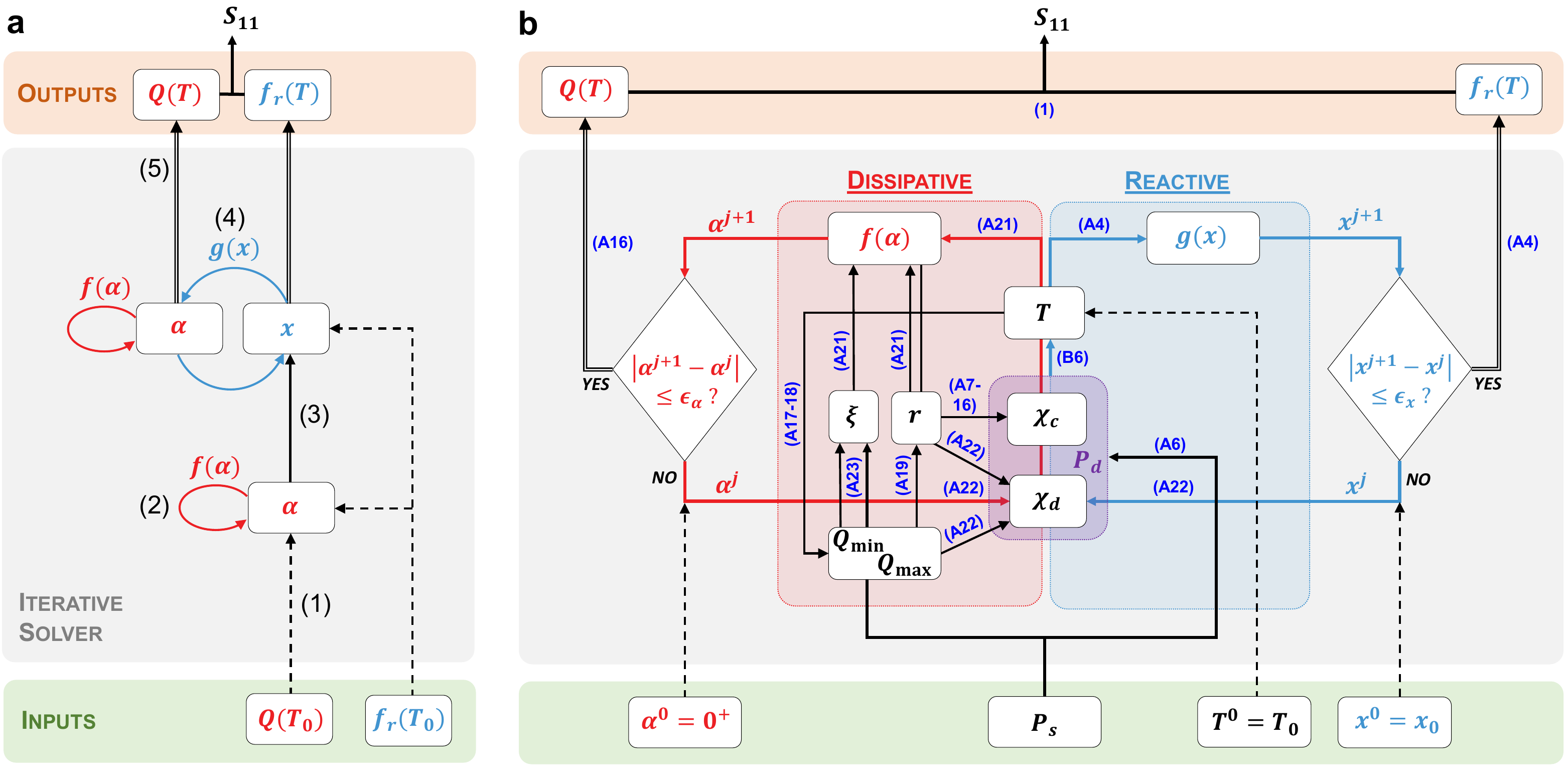}
\caption{Iterative solver for the TLS nonlinearity. (a) Overview of the numerical method: starting from an initial temperature $T_0$, the TLS saturation parameter 
$\alpha$ is updated iteratively to determine the total quality factor $Q(T_0)$ corresponding to an input power $P_s$. Given $Q(T_0)$, the resulting dissipated power is used to estimate the TLS temperature $T$ and corresponding detuning $x(T)$, which is refined through a second iteration loop. This process continues until convergence of $\alpha$ and $x$, yielding final values of $Q(T)$, $f_r(T)$, and $S_{11}$. (b) Detailed implementation of the $\alpha$- and $x$-iterations, shown as red and blue loops, respectively. Initialization and termination steps are indicated by dashed and double-lined arrows. Labels of the relevant equations for each step are shown in dark blue.}
\label{figIterativeSolver}
\end{figure*}

In Appendix~\ref{app:genDuffing}, we derived a generalized Duffing model by linearizing $Q$ and $f_r$ around $T_0$, which is valid only in the limited regime $\nbar\ll n_s<n_h$, where $\delta T\ll T_0$. To go beyond this first-order approximation, we now employ a numerical approach that fully accounts for the nonlinear dependence of $Q$ and $f_r$ on both $\nbar$ and $T$, as described by Eqs.~\ref{dfTLS}-\ref{Qimodel}. 
Our goal is to simulate the reflection coefficient $S_{11}$ over an array of $N$ linearly spaced probe frequencies $\{f_n\}_{n\in[1,N]}$ centered around the unshifted resonance frequency, such that $f_1<f_r(T_0)<f_N$. For \textit{upward} sweeps, the $n$\textsuperscript{th} probe frequency $f_n$ is given by $f_n = f_1 + \frac{n-1}{N-1}(f_N-f_1)$, while \textit{downward} sweeps are obtained by exchanging the indices $1$ and $N$. 
In the low-power limit, the TLS ensemble is assumed to thermalize at the applied refrigerator temperature $T_0$ and the corresponding reduced detuning at frequency step $n$ is $(x_0)_n=(f_n-f_r(T_0))/f_r(T_0)$ with $f_r(T_0)$ from Eq.~\ref{dfTLS}. For each $(x_0)_n$ and given a probe power $P_s$, we seek to obtain the self-consistent realized detuning $x_n$ (or equivalently $(f_r)_n=f_r(T_n)$ with $T_n$ the achieved TLS temperature at step $n$) and total quality factor $Q_n=Q(T_n)$ using the following iterative method illustrated in Fig.~\ref{figIterativeSolver}(a):
\begin{enumerate}
  \item \textbf{Initialization}: the TLS ensemble temperature is initially set to the refrigerator temperature $T_0$.
  \item \textbf{$\alpha$-iteration}: the quality factor $Q(T_0)$ achieved for a given input power $P_s$ is computed from the iterative update $\alpha^{j+1}=f(\alpha^j)$, where $\alpha$ is the TLS saturation parameter defined in Appendix~\ref{app:TLSdissip}.
  \item \textbf{$\alpha$-$x$ propagation}: the dissipated power, determined from the resulting $Q(T_0)$ and resonance frequency $f_r(T_0)$ is used to estimate the effective TLS temperature $T$, which is then mapped to the corresponding detuning $x(T)$.
  \item \textbf{$x$-iteration \& $\alpha$-update}: the detuning is refined iteratively via $x^{j+1}=g(x^j)$. At each step $j$, the current $x^j$ value corresponding to $T^j$ is used to recompute $Q^j$ by repeating the $\alpha$-iteration. From the updated $Q^j$, a new detuning $x^{j+1}$ is obtained. 
  \item \textbf{Termination}: once convergence is reached, the final values of $\alpha^j$ and $x^j$ yield $Q(T)$ and $f_r(T)$, from which $S_{11}$ is calculated.
\end{enumerate}

In Fig.~\ref{figIterativeSolver}(b), we detail the implementation of the $\alpha$ and $x$-iterations used to compute the self-consistent $Q(T)$ and $f_r(T)$. We denote by $X_n^j$ the value of quantity $X$ at frequency step $n$ and iteration $j$. At each frequency step $n$, we initialize the TLS temperature to the applied refrigerator temperature $T_n^{j=0}=T_0$, and compute the corresponding resonance frequency $(f_r)_n^0=f_r(T_n^0)$ using Eq.~\ref{dfTLS}. This amounts to initializing the realized detuning to the applied value $x_n^0=(x_0)_n$. 
\begin{itemize}
    \item \textbf{$\alpha$-iteration}: To obtain the quality factor $Q_n$, we first evaluate with Eqs.~\ref{QminEq}-\ref{QmaxEq} its minimal and maximal values at $T_0$, $Q_\textrm{min}(T_0)$ and $Q_\textrm{max}(T_0)$, and combine them into the normalized ratio $r$ (Eq.~\ref{rEq}). From $Q_\textrm{min}(T_0)$ and $P_s$, we calculate the dimensionless probe power $\xi$ (Eq.~\ref{xitls}) and map $x_n^0$ to a detuning efficiency $(\chi_d)_n^0$ via Eq.~\ref{chiDtls}. Starting from $\alpha_n^0=0^+$, we then compute the iterative sequence $\alpha_n^{j+1}=f(\alpha_n^j)$ by repeatedly applying Eq.~\ref{falpha} and updating $(\chi_d)_n^j$ with the new $\alpha_n^j$ value until the convergence criterion $|\alpha_n^{j+1}-\alpha_n^j|\leq\epsilon_\alpha$ is satisfied. This iterative procedure is depicted as the red loop in Fig.~\ref{figIterativeSolver}(b)). From the converged value $\alpha_n^J$ and using $r$ and $Q_\textrm{min}(T_0)$, the realized quality factor $Q_n$ is then obtained via Eq.~\ref{Qtalpha}. 
    \item \textbf{$x$-iteration} We solve for $x_n$ using a second fixed-point iteration, illustrated as the blue loop in Fig.~\ref{figIterativeSolver}(b). From the converged $Q_n$, we compute the detuning efficiency $(\chi_d)_n$ (Eq.~\ref{chiDtls}) and coupling efficiency $(\chi_c)_n$ (Eq.~\ref{chiC}). Their product, multiplied by the input power $P_s$, yields the dissipated power $(P_d)_n$ (Eq.~\ref{Pdprod}). We then determine the TLS temperature $T_n$ corresponding to $(P_d)_n$ using Eq.~\ref{TvsPd}, compute the corresponding resonance frequency $(f_r)_n$ (Eq.~\ref{dfTLS}), and translate it into a realized detuning $x_n = (f_n-(f_r)_n)/(f_r)_n$. The detuning is refined iteratively via $x_n^{j+1}=g(x_n^j)$, where at each iteration $j$, the function $g$ recomputes $Q_n^j$ from $x_n^j$ through an $\alpha$-iteration, and updates the detuning to $x_n^{j+1}$ based on the new $Q_n^j$. The iteration continues until the convergence criterion $|x_n^{j+1}-x_n^j|\leq \epsilon_x$ is met. To model our high-Q resonance in the high-power regime, we require $\epsilon_\alpha\sim10^{-5}$ and $\epsilon_x\sim 10^{-8}$.
\end{itemize}

While our TLS nonlinearity solver is conveniently cast as a fixed-point iteration, convergence is not always guaranteed especially at high input powers. Thomas et al.~\cite{Thomas2020} showed that the dissipative response, governed by the iteration $\alpha^{j+1}=f(\alpha^j)$, converges reliably to a unique solution. In contrast, the reactive response can become multi-valued at high power, producing hysteresis and discontinuities. In this hysteretic regime, convergence of the $x$-iteration to the physical solution requires suitable preconditioning.
Several convergence acceleration strategies exist for fixed-point schemes, varying in complexity and applicability~\cite{Kelly1995, Ortega1970}. The usual Newton-Raphson method, which achieves quadratic convergence using the derivative $g'(x)$, is not suitable here due to the possible discontinuity of $g$. Steffensen’s method~\cite{Johnson1968}, a derivative-free approximation to Newton’s method, also fails under such non-smooth conditions.
A more robust class of techniques relies on combining the history of past residuals to construct improved update directions. The simplest is linear mixing, which updates the input via $x^{j+1}=(1-\beta)x^j+\beta g(x^j)$. A small damping factor $0\leq\beta\leq 1$ ensures stability, but slows convergence; larger $\beta$
accelerates updates but risks divergence. While linear mixing works even when derivatives are ill-defined, it may become impractically slow for poorly conditioned problems.
To improve robustness and efficiency, we employ Anderson acceleration~\cite{Anderson1965, Walker2011}, a more aggressive convergence acceleration method well-suited for nonlinear, high-dimensional problems. It requires only a single function evaluation per iteration and no derivatives, solving a small least-squares problem to optimally combine previous iterates. Though its convergence remains linear in general, we found that for our particular application, Anderson acceleration delivered a substantial practical speedup compared to simple linear mixing.

\section{Readout-power heating model}\label{app:ROheating}

\subsection{Phonon bottleneck: a qualitative picture of the TLS heating}\label{app:phBottle}

\begin{figure}[!ht]
\centering
\includegraphics[width=1.0\linewidth]{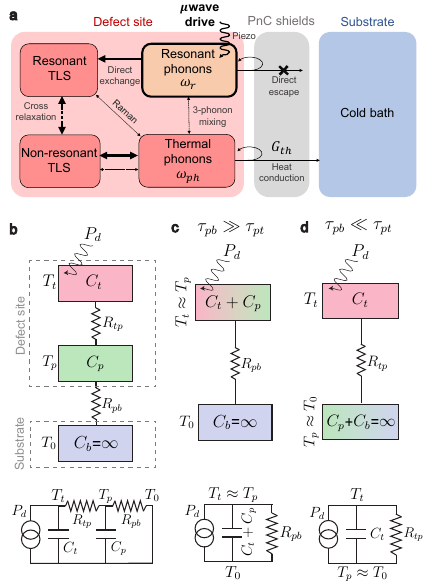}
\caption{Readout-power heating of the TLS ensemble. (a) Conceptual sketch of a PCR resonator as a cavity (red) hosting resonant phonons coupled to an ensemble of resonant/non-resonant TLSs and thermal phonons. The cavity is connected to the substrate acting as a cold bath (blue) through acoustically shielded beams (gray). (b) The system can be modeled as a three-node thermal conductance model ($t=\textrm{TLS}$, $p=\textrm{phonons}$, $b=\textrm{bath}$), which reduces to a two-temperature problem in the limits $\tau_{pb}\gg\tau_{pt}$ (c) and $\tau_{pb}\ll\tau_{pt}$ (d). Equivalent lumped-element circuits are sketched in each limit.}
\label{figRC}
\end{figure}

Here we aim to provide a qualitative picture for the microwave-power heating of the TLS ensemble. In Fig.~\ref{figRC}(a) we provide a conceptual sketch of a PCR resonator illustrating how the resonant phonon mode hosted by the central block (``defect site'') couples to TLS and thermal phonons as well as the substrate cold bath.
The defect-mode phonons excited by the microwave probe tone may be absorbed by resonant TLSs, effectively heating up the TLS ensemble at the resonator's frequency. Because of the acoustic bandgap engineered in the resonator clamping structure, this narrow population of excited TLSs cannot efficiently relax by re-emitting resonant phonons, nor can it spread spectrally to off-resonant TLSs in absence of interactions. Therefore, in this idealized system, broadband heating of the TLS ensemble should in principle not occur under a resonant microwave drive. Nonetheless, a second-order scattering process exists, whereby an excited TLS may interact simultaneously with two off-resonant lattice phonon modes whose difference frequency matches its energy \cite{Silbey1990}. In such a Raman-type relaxation, a wide band of the phonon spectrum becomes available to facilitate TLS relaxation, in particular thermal phonons with frequencies below and also above the phononic bandgap, since $\Delta\omega_b<k_B T\sim\omega_r$.
The phononic shields still constitute a good cavity for thermal phonons near the bandgap edge, which can make multiple passes within the defect region and locally equilibrate with the TLS system before eventually radiating through the clamping structure and into the cold substrate bath \cite{MacCabe2020}. In addition, long-range interactions between TLSs are responsible for spectral diffusion of their energy level splitting \cite{Laikhtman1986, Burin2004, Faoro2015, Burin2018}, which somewhat circumvents the ``phonon bottleneck'' by allowing the phonon energy absorbed from one resonant TLS to spread out to other non-resonant TLSs, thus effectively heating up a broad ensemble of TLSs. 
If this process of interaction-induced relaxation within the TLS ensemble and their simultaneous thermalization with the hot phonons of the defect site happens faster than the ballistic escape of the latter through the phononic shields, the ensemble formed by the TLSs and the thermal phonons of the central block can be considered in internal equilibrium and this system as a whole then relaxes to the substrate's cold bath.

\subsection{Three-node thermal conductance model}\label{app:GthRC}

The thermalization dynamics of the system sketched in Fig.~\ref{figRC}(a) may be too complicated to model because of the complex nonlinear interactions between its constitutive parts. Nonetheless, many features can be reproduced by a phenomenological thermal conductance model involving only three temperature nodes. Our readout-power heating model describes the central block of a PCR as a hot cavity coupled through the acoustically-shielded tethers to a cold bath/reservoir provided by the substrate (Fig.~\ref{figRC}(b)). The cavity is modeled as a hot TLS ensemble with heat capacity $C_t$ and temperature $T_t$ interacting with a population of thermal phonons described by their heat capacity $C_p$ and temperature $T_p$. The substrate has fixed temperature $T_0$ and is assumed to have infinite heat capacity $C_b\to\infty$. The total system can be described by its equivalent ``lumped-element'' circuit (Fig.~\ref{figRC}(b)) comprising three temperature nodes $T_i$ with thermal capacitance $C_i$ connected by their respective thermal resistances $R_{ij}=R_{ji}$, where $i,j\in\{t=\textrm{TLS}$, $p=\textrm{phonons}$, $b=\textrm{bath}\}$ and the bath node is grounded ($T_b\equiv T_0$) because of its infinite heat capacity.
The heating dynamics of this three-node thermal structure is captured by the following set of coupled differential equations, which describes conservation of the heat current at the ``t'' and ``p'' nodes:
\begin{align}[left=\empheqlbrace]\label{CTR1}
     C_t \dv{T_t}{t} &= P_d -\frac{T_t-T_p}{R_{tp}}\\
     C_p \dv{T_p}{t} &= -\frac{T_p-T_0}{R_{pb}}-\frac{T_p-T_t}{R_{tp}}\label{CTR2},
\end{align}
where $P_d$ is the microwave power dissipated in the TLS ensemble, $R_{tp}$ is the thermal resistance connecting the TLS and thermal phonons subsystems and $R_{pb}$ models the thermal resistance of the PCR tethers to the substrate. For simplicity, the thermal resistances are assumed constant, independent of temperature, which is valid only in the limit where temperature differences are small, $|T_i-T_j|\ll T_i,T_j$. This allows one to express the heat current between any two nodes in an ``Ohm's law'' fashion as the difference of the nodes' temperature divided by the thermal resistance connecting the two. The heating dynamics is governed by three time constants, $\tau_{tp}=R_{tp}C_t$, $\tau_{pt}=R_{tp}C_p$ and $\tau_{pb}=R_{pb}C_p$. In the steady-state, when $t\gg\tau_{tp},\tau_{pt}\tau_{pb}$, no current flows through the capacitors and we have simply:
\begin{equation}\label{TinfRPd}
    P_d=\frac{T_t^\infty-T_p^\infty}{R_{tp}}=\frac{T_p^\infty-T_0}{R_{pb}}
\end{equation}
such that $T_t^\infty =(R_{tp}+R_{pb})P_d+T_0$.

At low temperature, TLSs are efficient phonon scatterers (the scattering rate of phonons at frequency $\omega_{ph}$ scales like $\Gamma_{ph}\propto\tanh{(\hbar\omega_{ph}/(2k_B T))}$ \cite{Phillips1987}) and we expect to be in the limit $\tau_{pt}\ll\tau_{pb}$ which, dividing this by $C_p$, is equivalent to $R_{tp}\ll R_{pb}$. The thermal structure can then be approximated by a two-node version, as shown in Fig.~\ref{figRC}(c), where the TLS ensemble quickly equilibrates with the thermal phonons ($T_t\approx T_p$) over a time $\tau_{pt}$ and the heat conduction is limited by $R_{pb}$, \textit{i.e.} by the phonon transport across the PCR tethers. The temperature of this joint system $T\equiv T_t=T_p$ then evolves according to $d T/d t=-(T-T^\infty)/\tau_h$, where the steady-state temperature of the cavity, $T^\infty=T_0+R_{pb}P_d$, is reached over a time $\tau_h=(C_t+C_p)R_{pb}\gg \tau_{pt}$.
However, once the TLS system is hot, the bottleneck in the heat transport is no longer the ballistic escape of phonons through the tethers, but the internal energy transfer between the TLS system and the phonons in the central block. We then have the opposite limit where $\tau_{pt}\gg\tau_{pb}$ (\textit{i.e.} $R_{tp}\gg R_{pb}$) and the circuit reduces to the one shown in Fig.~\ref{figRC}(d). In that limit, $T_p\approx T_0$ and Eq.~\ref{CTR1} simplifies to $d T/d t\approx-(T-T^\infty)/\tau_c$ where $T=T_t$ reaches $T^\infty=T_0+R_{tp}P_d$ within a time $\tau_c=C_t R_{tp}\gg\tau_h$. The cooling dynamics of the TLS ensemble therefore happens over a longer time scale compared to the heating. 

\subsection{Two-node thermal conductance model beyond the low-power limit}\label{app:AppGth}

Previously we showed that the thermalization of the TLS ensemble can be modeled as that of a hot cavity coupled to a cold reservoir -- the reservoir being either the cold substrate or the thermal phonons in the central block depending on the initial bath temperature value $T_0$. Here, we seek to improve over the previous two-node thermal conductance model which was valid only in the limit of low readout-power heating: we now assume that the thermal resistance between cavity and reservoir bears a temperature dependence. Eq.~\ref{CTR1} should then be replaced by:
\begin{equation}\label{CTR3}
    C_{th} \dv{T}{t} = P_d -\int_{T_0}^{T} G_{th}(T) dT.
\end{equation}
In the low-temperature limit where $\tau_{pt}\ll\tau_{pb}$, the cavity temperature $T$ should be understood as an effective temperature for the joint system comprised of the TLSs and thermal phonons of the defect site, its heat capacity contains both TLS and phonon contributions, $C_{th}=C_t+C_p$, and $G_{th}=1/R_{pb}$ models the thermal conductance of the PCR tethers. 
The left-hand side of Eq.~\ref{CTR3} represents the time derivative of the thermal energy stored in the cavity, $\dot{E}_{st}=C_{th} \dvstar{T}{t}$, while the two terms on the right-hand site describe respectively the rate of thermal energy generation, $\dot{E}_{in}$, which corresponds to the dissipated microwave power $P_d$, and the rate at which heat escapes from the cavity to the cold reservoir, $-\dot{E}_{out}$. In the steady-state, $\dvstar{T}{t}\approx 0$ and $\dot{E}_{in}=\dot{E}_{out}$. The power flow into the cavity due to microwave absorption by the TLSs equals the power flow from the hot cavity into the cold reservoir and the cavity thermalizes at an effective temperature $T$, given by:
\begin{equation}\label{PdGth}
    P_d = \int_{T_0}^{T} G_{th}(T) dT.
\end{equation}
For moderate microwave power such that $P_d\ll G_{th}(T_0) T_0$, we can neglect the temperature variation of $G_{th}$ and Eq.~\ref{PdGth} reduces to the ``Ohm's law'' formulation of the previous section, $P_d = (T - T_0)G_{th}(T_0)$, which predicts a linear dependence of the cavity temperature $T$ with the dissipated power $P_d$. Otherwise, when $P_d\gtrsim G_{th}(T_0) T_0$, the temperature dependence of $G_{th}(T)$ can no longer be neglected and Eq.~\ref{PdGth} should be used. 
At low temperature where phonon transport is ballistic, the lattice thermal conductance is expected to scale as a power law of the thermal phonon's temperature, $G_{th}\propto T^\gamma$, with the temperature exponent $\gamma$ depending on the dimensionality of the phonon gas. We find that for the specific tether geometry of our PCR resonators, a constant $\gamma=2.83$ models well the thermal conductance up to $\sim 200$~mK, which allows us to parametrize the thermal conductance as $G_{th}=G_{th}(T_0)(T/T_0)^\gamma$ \AppGth. 
Substituting this formula into Eq.~\ref{PdGth} yields the following expression for the cavity temperature as a function of the dissipated power:
\begin{equation}\label{TvsPd}
    T = T_0\Bigl(1 + \frac{1+\gamma}{T_0 G_{th}(T_0)}P_d\Bigr)^\frac{1}{1+\gamma}. 
\end{equation}
In the limit of strong heating, $T$ no longer scales linearly with power, but shows a weaker, sublinear $\sim P_d^{1/(1+\gamma)}$ dependence. This expression is quite general and can model a wide range of device heating phenomena, \textit{e.g.} the case $\gamma=4$ models well Joule heating of electrons in thin metal films, where the dissipated power flows into the substrate phonons via a thermal resistance $G_{e-ph}\propto T^4$ mediated by electron-phonon scattering with a rate $\tau_{e-ph}^{-1}\propto T^3$ \cite{Wellstood1994}. 

Injecting Eq.~\ref{Pdnbar}, one obtains the cavity's temperature $T$ as a function of the intracavity phonon number $\nbar$:
\begin{equation}\label{Trnbar}
    T = T_0\Bigl(1+\frac{\nbar}{n_h}\Bigr)^\frac{1}{1+\gamma},\quad n_h=\frac{\pi N_{ch}Q_i(\nbar)}{6(\gamma + 1)}n_{th}^2,
\end{equation}
where $n_{th}=k_B T_0/(h f_r)$ is the high-temperature approximation for the average number of thermal phonons, and we expressed $G_{th}(T_0)=N_{ch}g_0$ as $N_{ch}$ quanta of thermal conductance $g_0=\pi^2 k_B^2 T_0/(3 h)$. As an important caveat, the typical phonon occupancy $n_h$ at the onset of readout-power heating is expressed here in an implicit way as it still depends on the circulating power $\nbar$. Replacing $Q_i(\nbar)$ by $Q_i(n_s)$ may however provide a reasonable order-of-magnitude estimate, since TLS saturation when $\nbar\sim n_s$ results in a faster increase of $\nbar$ with the applied power $P_s$ and therefore acts as a precursor to readout-power heating.

\subsection{Coupled equations for the cavity phonons \& TLS }\label{sec:cavityDynamics}

The differential equation \ref{CTR3} captures the time evolution of the TLS temperature $T(t)$ due to the dissipated power $P_d(t)=\hbar\omega_r\bigl(T(t)\bigr)\kappa_i\bigl(T(t),n(t)\bigr)n(t)$. We need to supplement it with an equation for the cavity phonons, whose frequency and loss rate also depend on $T$.
In a frame rotating at the drive frequency $\omega$, the Heisenberg equation of motion for the cavity field $a(t)$ reads \cite{Aspelmeyer2014}:
\begin{equation}\label{adot}
    \dv{a(t)}{t} = -\Bigl[i\Delta\bigl(T(t)\bigr)+\frac{\kappa\bigl(T(t),n(t)\bigr)}{2}\Bigr]a(t)+\sqrt{\kappa_e}a_{in}(t),
\end{equation}
where $a_{in}(t)=\sqrt{P_s/\hbar\omega }e^{-i\omega t}$ models the incoming drive tone with power $P_s$ and $\Delta\bigl(T(t)\bigr)=\omega-\omega_r\bigl(T(t)\bigr)$ is the detuning. The scattering coefficient $S_{11}(t)$, defined as the ratio of the reflected field $a_{out}(t)$ to the incoming one, $a_{in}(t)$, is deduced from the usual input-output relation, $a_{out}(t)=a_{in}(t)-\sqrt{\kappa_e}a(t)$:
\begin{equation}\label{S11aoutain}
    S_{11}(t)=\frac{a_{out}(t)}{a_{in}(t)}=1-\sqrt{\kappa_e}\frac{a(t)}{a_{in}(t)}.
\end{equation}
The evolution equation for the phonon number can be obtained by differentiating the product $n=a^\dagger a$ with respect to $t$ and inserting the equations for $a(t)$ and $a^\dagger(t)$. The detuning terms cancel exactly (frequency shifts do not change energy) and one is left with:
\begin{equation}\label{ndot}
    \dv{n(t)}{t}=-\kappa\bigl(T(t),n(t)\bigr)n+2\sqrt{\kappa_e}\Re{\{a_{in}^{*}(t)a(t)\}}
\end{equation}
The first term describes phonon loss (internal+external) and the second term, $\Phi_{in}=2\sqrt{\kappa_e}\Re{\{a_{in}^{*}a\}}$, represents the injected phonon flux from the drive. It depends on the cavity field $a$, because power transfer into a coherently driven cavity is an interference effect: the rate at which energy enters depends on both the amplitude and phase of the intracavity field already present. The separation of time scales further simplifies the equations. 

\subsubsection{Fast cavity dynamics: $\kappa^{-1}\ll\tau_{th}$}\label{sec:fastCavity}
If cavity dynamics are fast compared to the thermal time constant $\tau_{th}=C_{th}/G_{th}$, the cavity field tracks the instantaneous steady state of Eq.~\ref{adot}:
\begin{equation}\label{ainfty}
    a_\infty(t)\approx \frac{\sqrt{\kappa_e}}{i\Delta\bigl(T(t)\bigr)+\kappa\bigl(T(t),n(t)\bigr)/2}a_{in}(t)
\end{equation}
This allows adiabatic elimination of $a(t)$. 
Inserting $a_\infty$ into \ref{S11aoutain}, we recover the usual expression for the reflection coefficient:
\begin{equation}\label{S11Sim}
    S_{11}(t)=1-\frac{\kappa_e}{i\Delta\bigl(T(t)\bigr) +\kappa\bigl(T(t),n(t)\bigr)/2}
\end{equation}
Substituting \ref{ainfty} into the expression for the incoming phonon flux $\Phi_{in}$ and plugging the latter in \ref{ndot} gives $\dot{n}(t)=-\kappa(t) \bigl(n(t)-\nbar(t)\bigr)$ with the expected expression for the quasi-stationary phonon number $\nbar$, 
\begin{equation}
    \nbar(t)=\frac{\kappa_e}{\Delta\bigl(T(t)\bigr)^2+(\kappa\bigl(T(t),n(t)\bigr)/2)^2}\frac{P_s}{\hbar\omega}.
\end{equation}
We are left with a closed-form system of two differential equations for the coupled $T-n$ dynamics :
\begin{align}[left=\hspace*{-0.3cm}\empheqlbrace]\label{tnRungeKuttaT}
    C_{th}\dv{T}{t}&=h f_r(T)\kappa_i(T,n) n - \frac{G_{th}(T_0)}{\tilde{\gamma} T_0^{\tilde{\gamma}-1}}(T^{\tilde{\gamma}}-T_0^{\tilde{\gamma}})\\
    \dv{n}{t}&=-\kappa(T,n)\bigl(n-\nbar(T,n)\bigr)\label{tnRungeKuttaN},
\end{align}
where we introduced $\tilde{\gamma}=\gamma+1$ to lighten notations. These coupled dynamical equations can be solved efficiently using standard numerical methods such as a Runge-Kutta scheme. This provides an alternative approach to computing the swept-frequency response in the nonlinear regime by explicitly simulating the time evolution of $T$ and $n$ at each frequency point. For a given frequency $f_i$, the system is driven with a power $P_s$ for a duration $t_{meas}$, yielding $T_i(t_{meas})$ and $n_i(t_{meas})$, and then $S_{11}(t_{meas})$ using Eq~\ref{S11Sim}.
The simulation is initialized at a frequency $f_1$ sufficiently detuned from resonance with $T_1(0)=T_0$ and $n_1(0)=0$. Subsequent frequency points are computed iteratively from \ref{tnRungeKuttaT}-\ref{tnRungeKuttaN} using as initial conditions the converged values obtained at the previous step, $T_{i-1}(t_{meas})$, $n_{i-1}(t_{meas})$. 

This procedure naturally accounts for sweep directionality and hysteretic effects, and allows to incorporate the finite response time of the measurement chain.
In a vector network analyzer measurement, the IF receiver requires a finite settling time after each frequency step. The IF chain behaves as a low-pass filter with time constant $\tau_{IF}\approx k/BW$, where $k\approx 1-3$ depends on the filter order and instrument implementation. This effect can be simulated by convolving $S_{11}(t)$ with a filtering kernel, $S_{11}(t_{meas})=\int_0^{t_{meas}}S_{11}(t)\tilde{h}_{IF}(t_{meas}-t)dt$ with $\tilde{h}_{IF}(t)=h_{IF}(t)/(\int_0^{t_{meas}}h_{IF}(t')\,dt')$ and $h_{IF}(t)=\exp(-t/\tau)$. Accordingly, steady-state amplitude and phase are reached for $t_{meas}\gtrsim 3\tau_{IF}$.
The $S_{11}$ curves in Fig~\ref{fig1}(b) were measured using a vector network analyzer with an IF bandwidth $BW=200$~Hz, ensuring that the steady-state response was always reached, since the quality factor remained below $Q<k\pi f_r/BW\approx 8.2\times 10^6$ at all probe powers (see Appendix~\ref{app:frPssolver}).

\subsubsection{Slow cavity dynamics: $\kappa^{-1}\gg\tau_{th}$}\label{sec:slowCavity}
In the opposite limit where $T$ responds much faster than the cavity field, we can treat $T(t)$ as quasi-stationary with respect to $a(t)$, $\dot{T}\approx 0$, and so $T\approx T_\infty(|a|^2)$ with
\begin{equation}
    T_\infty(t) = T_0\left(1+\frac{\tilde{\gamma} hf_r(T_\infty)}{T_0 G_{th}(T_0)}\kappa_i(T_\infty, |a(t)|^2)\right)^\frac{1}{1+\gamma}
\end{equation}

Here the temperature equilibrates ``instantly'', and the cavity evolves slowly in the landscape shaped by the quasi-instantaneous thermal response. Plugging $T_\infty$ into the cavity equation \ref{adot}, we obtain a single nonlinear differential equation for $a(t)$: $\dot{a}=\bigl[-i\Delta\bigl(T_\infty(|a|^2)\bigr)-\kappa\bigl(T_\infty(|a|^2),|a|^2\bigr)/2\bigr]a+\sqrt{\kappa_e}a_{in}(t)$.
In the limit of small temperature increase and small drive, we can linearize the frequency shift and internal linewidth around the bath temperature, $\Delta(T)\approx \Delta_0(T_0)+\alpha(T-T_0)$ and $\kappa_{i}(T)\approx\kappa_i(T_0)-\beta(T-T_0)$. The effective equation for the cavity field then becomes $\dot{a}(t)=-i\bigl(\Tilde{\Delta}(T_0)+K|a(t)|^2\bigr)\,a(t)+\sqrt{\kappa_e}a_{in}(t)$ with the complex detuning $\Tilde{\Delta}(T_0)=\Delta_0(T_0)-i \kappa_i(T_0)/2$ and complex Kerr constant:
\begin{equation}
    K=\Bigl(\alpha-i\frac{\beta}{2}\Bigr) \frac{h f_r(T_0)\kappa_i(T_0)}{G_{th}(T_0)}
\end{equation}
This is the equation for a Kerr oscillator with a mixed reactive/dissipative nonlinearity originating from the quasi-instantaneous thermal response. Within this linearized treatment, the internal loss rate $\kappa_i(T)$ is not bounded from below, which can lead to unphysical solutions, including negative internal quality factors at sufficiently high readout power.
Consequently, this linearized formulation is valid only in the limit of vanishing readout power and, in the present case, cannot be used to quantitatively model the data shown in Fig.~\ref{fig1}(b).

\begin{figure*}[!ht]
\centering
\includegraphics[width=1.0\linewidth]{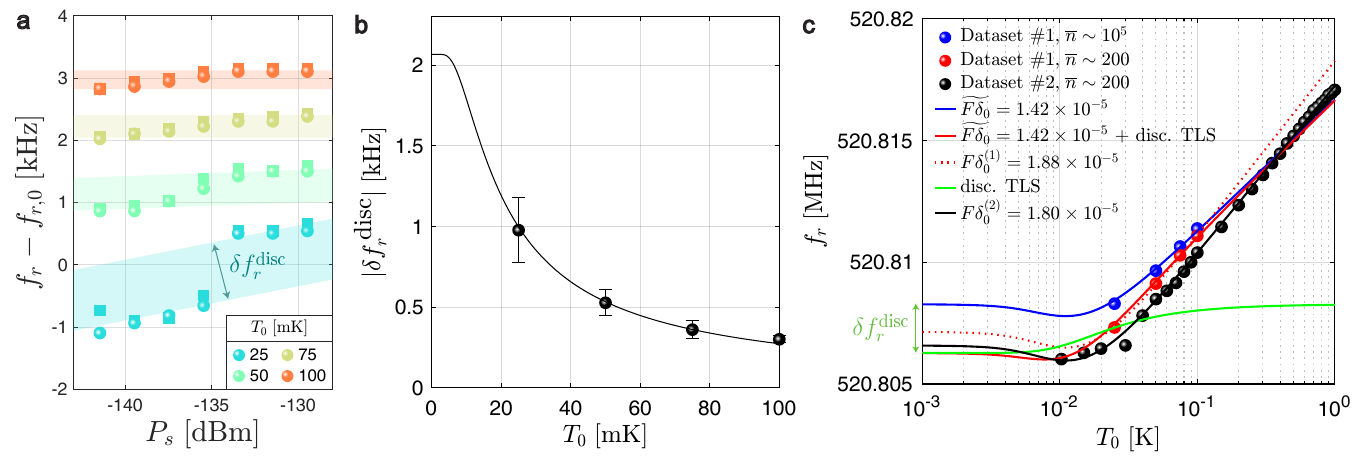}
\caption{Dispersive shift from a discrete TLS. (a) An enlargement of the measured $f_r(P_s)$ in Fig.~\ref{fig2}(a) at low power. The frequency step around $P_s=-135$~dBm is consistent with saturation of a discrete TLS above the PCR resonator frequency. The tangent method is used to extract the step size $\delta f_r^\textrm{disc}$ at each operating temperature $T_0$ by disentangling it from the smooth $f_r$ increase with $P_s$. (b) The magnitude of the frequency step $\delta f_r^\textrm{disc}$ as a function of $T_0$. The solid black line is a fit to $\chi_0(T_0)$ from Eq.~\ref{shift1TLS}. (c) Comparison of $f_r(T_0)$ measured in different cooldowns and with different probe strengths. Solid lines are fits accounting for the contribution from a continuum of TLSs (Eq.~\ref{dfTLS}) and a strongly coupled discrete TLS (Eq.~\ref{shift1TLS}).}
\label{figTLS}
\end{figure*}

\section{Power-dependent resonance fit}\label{app:modelFrPs}

\subsection{Signature of a discrete TLS}\label{app:discreteTLS}

In Fig.~\ref{fig2}(a), the frequency step in the $f_r(P_s)$ curve at low power can be reproduced by including an additional discrete TLS alongside the continuum of TLSs already described by the loss tangent 
$F\delta_0^\textrm{reac}$. To account for the upward step as $P_s$ is raised above $-135$~dBm, the TLS transition frequency must satisfy $\omega_{tls}>\omega_r$, so that its dispersive shift on the resonator is negative. As we show below, a minimal model capturing this frequency step requires only two parameters: the TLS frequency $\omega_{tls}$ and its coupling rate $g$ to the resonator.

In full generality, the average resonator shift $\delta\omega_r$ and spectral broadening $\kappa_r$ due to a single TLS with transition frequency $\omega_{tls}$ and coupling strength $g$ are given by the real and imaginary parts of the complex susceptibility \cite{Capelle2020, Emser2024}:
\begin{equation}
    \delta\omega_r+i\frac{\kappa_r}{2}=-g^2\langle \sigma_z \rangle\left[ \frac{1}{\omega_{tls}-\omega_r+i\Gamma_2}+\frac{1}{\omega_{tls}+\omega_r-i\Gamma_2}\right]\label{shift_1TLSth_wrkap}
\end{equation}
When the TLS is in thermal equilibrium with a bath at temperature $T$, $\langle\sigma_z\rangle=\langle\sigma_z\rangle_\textrm{th}=\tanh{(\hbar\omega_{tls}/(2k_B T))}$. 
In presence of drive phonons from the resonator, $\langle\sigma_z\rangle$ is replaced by:
\begin{equation}\label{sigDr}
    \langle\sigma_z\rangle_\textrm{dr} = \langle\sigma_z\rangle_\textrm{th}\frac{1+\Bigl(\frac{\delta\omega}{\Gamma_2}\Bigr)^2}{1+\Bigl(\frac{\delta\omega}{\Gamma_2}\Bigr)^2+\Bigl(\frac{\Omega}{\sqrt{\Gamma_1\Gamma_2}}\Bigr)^2},
\end{equation}
where $\Gamma_1=1/T_1$ and $\Gamma_2=1/T_2$ are the TLS transverse and longitudinal relaxation rates, and we introduced the detuning $\delta\omega =\omega_{tls}-\omega_r(\nbar,T)$ for ease of notation \cite{Emser2024}. Expressing the on-resonance Rabi frequency in terms of $\nbar$, $\Omega=2g\sqrt{\nbar+1}\approx 2g\sqrt{\nbar}$, and neglecting dephasing ($\Gamma_2=\Gamma_1/2$), one can recast Eq.~\ref{shift_1TLSth_wrkap} into
\begin{widetext}
    \begin{align}[left=\empheqlbrace]\hspace{0.3cm}\label{shift1TLSdr}
     \delta\omega_r &=-\frac{g^2\tanh{\bigl(\frac{\hbar\omega_{tls}}{2k_B T}\bigr)}}{\Gamma_2^2+\delta\omega^2+2g^2\nbar}\Bigl[\delta\omega +(\delta\omega+2\omega_r)\frac{\Gamma_2^2+\delta\omega^2}{\Gamma_2^2+(\delta\omega+2\omega_r)^2}\Bigr]\\
     \kappa_r &= \frac{2g^2\Gamma_2\tanh{\bigl(\frac{\hbar\omega_{tls}}{2k_B T}\bigr)}}{\Gamma_2^2+\delta\omega^2+2g^2\nbar}\Bigl[1-\frac{\Gamma_2^2+\delta\omega^2}{\Gamma_2^2+(\delta\omega+2\omega_r)^2}\Bigr].
    \end{align}   
\end{widetext}
Within the rotating-wave approximation, $\omega_{tls}+\omega_r\gg\omega_{tls}-\omega_r$ and the second term in the square brackets, known as the Bloch-Siegert shift, can be neglected. We also neglect the TLS linewidth $\Gamma_2\ll\delta\omega$ and, in the dispersive limit $g\ll \delta\omega$, recover to first order in $\nbar/n_c$ the expression for the cavity pull obtained from exact diagonalization of the Jaynes-Cummings Hamiltonian \cite{Boissonneault2010, Tosi2024}:

\begin{align}\label{shift1TLS}
    \delta\omega_r(\nbar,T) &\approx -\frac{g^2}{\delta\omega(1+2\nbar(\frac{g}{\delta\omega})^2)}\tanh{\left(\frac{\hbar\omega_{tls}}{2k_B T}\right)}\nonumber\\&
    \underset{g\ll\delta\omega}{\approx} \frac{\chi_0(T)}{\sqrt{1+\nbar/n_c}}    
\end{align}
$
\quad\textrm{where}\;\left\{
    \begin{array}{ll}
      \chi_0(T)=-(g^2/\delta\omega)\tanh{(\hbar\omega_{tls}/2k_B T)}\\
      n_c=(\delta\omega/2g)^2
    \end{array}
  \right.
$

\begin{figure*}[t]
\centering
\includegraphics[width=1.0\linewidth]{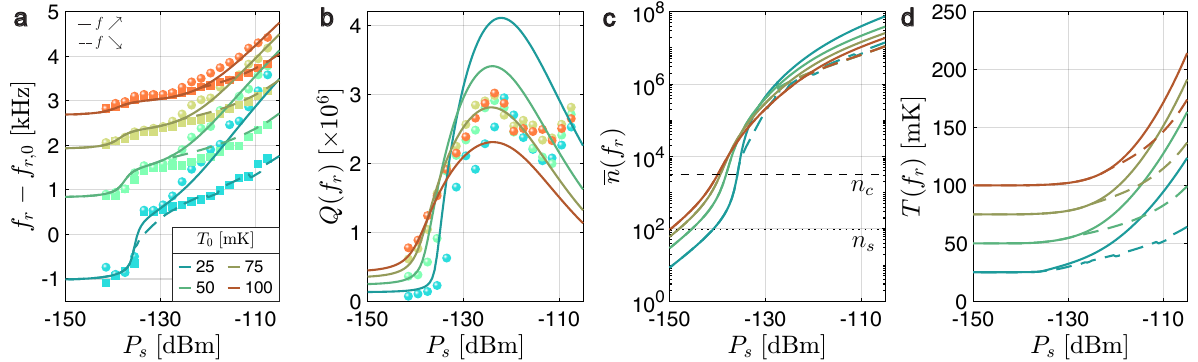}
\caption{Fitting the power-dependent resonance frequency (extended data). (a) Same as Fig.~\ref{fig2}(a). (b) Total quality factor at resonance, $Q(f_r)$, inferred from $|S_{11}(f_r)|$. (c) Mean phonon number at resonance, $\nbar(f_r)$, and (d) TLS temperature at resonance, $T(f_r)$, versus probe power $P_s$.
Disk (square) markers show measured $f_r$ and $Q(f_r)$ for upward (downward) frequency sweeps extracted from the $|S_{11}(f)|$ curves in Fig.~\ref{fig1}(b). Only the upward response is shown in (b). Solid (dashed) lines in (a–d) are theory curves for upward (downward) sweeps from the numerical model described in Appendix~\ref{app:TLSIterativeSolver}, corresponding to the best joint fit of the data in (a–b). Horizontal dotted and dashed lines in (c) mark the critical phonon numbers for saturation of background TLS ($n_s$) and discrete TLS ($n_c$). Model parameters are listed in Table~\ref{tableFitParams}.}
\label{figSim}
\end{figure*}

At low power, the extra TLS is mainly in its ground state, resulting in a maximal downward shift of the resonator frequency. As the refrigerator temperature $T_0$ is raised, the excited state population of the TLS increases, which reduces the magnitude of its dispersive shift on the resonator and therefore the size of the frequency step in Fig.~\ref{fig2}(a). Although the TLS population is initially unaffected by the resonator's phonons because of the large detuning $\delta\omega\gg g$, the mean phonon number $\nbar$ quickly grows with $Q_i$ as $P_s$ is increased above the critical power for TLS saturation. At this point, the effective Rabi frequency for the discrete TLS becomes comparable to its detuning to the resonator (equivalently $\nbar\sim n_c$) and the dispersive shift $\delta\omega_r$ drops with $\nbar$ according to Eq.~\ref{shift1TLS}. 

This minimal two-parameter model is shown to capture well the power and temperature dependence of $f_r$ at low $P_s$. When $\nbar\gg n_c$, $\delta\omega_r\to 0$, while at low enough $P_s$ such that $\nbar\ll n_c$, $\delta\omega_r\to \chi_0(T_0)$. Therefore, the magnitude of the frequency step is given by $\delta f_r^\textrm{disc} = \chi_0(T_0)/2\pi$ and fitting its temperature dependence allows to extract $\omega_{tls}$ and $g$. 
We illustrate this procedure in Fig.~\ref{figTLS}(a), where for each $T_0$ value, the tangent method is used to extract $\delta f_r^\textrm{disc}$ from $f_r(P_s)$. In Fig.~\ref{figTLS}(b), we plot the extracted $\delta f_r^\textrm{disc}$ as a function of $T_0$ and fit it with $\chi(T_0)$. We extract $\omega_{tls}/2\pi=546$~MHz and a coupling strength $g/2\pi=230$~kHz. At this point, fitting the frequency step as a function of $P_s$ is not yet possible, as it requires one to know the achieved $\nbar(P_s)$.

\subsection{TLS loss tangent extraction}\label{app:extractTLSFd}

The standard technique for extracting the TLS loss tangent involves performing a temperature sweep and fitting the measured $f_r(T_0)$ curve using Eq.~\ref{dfTLS}. This expression assumes coupling to a continuum of TLSs uniformly distributed in frequency and therefore does not depend on $\nbar$, since the overall shift results mainly from dispersively coupled TLSs. 
In regimes where shifts from individual TLSs can be resolved, Eq.~\ref{dfTLS} may still capture well the temperature-dependent response but yields an \textit{apparent} TLS loss tangent that depends on the actual strength of the probe tone. 
We illustrate this effect in Fig.~\ref{figTLS}(c) where we compare the temperature-dependent resonance frequency measured in three different conditions. 
The blue and red data points are from Dataset~\#1 and correspond to vertical slices at $P_s=-129$~dBm ($\nbar\sim 10^5$) and $-141$ ($\nbar\sim 10^2$) in Fig.~\ref{figTLS}(a). The best fit to Eq.~\ref{dfTLS} yields respectively $\widetilde{F\delta_0}=(1.42\pm0.23)\times 10^{-5}$ and  $F\delta_0^{(1)}=(1.88\pm 0.10)\times 10^{-5}$. The weaker value at $P_s=-129$~dBm reflects an apparent density where the contribution from the discrete TLS has been normalized out, since it is saturated at that power and no longer pulls on the resonator. 
The extracted loss tangent at $P_s=-141$~dBm agrees with the more accurate estimate from Dataset~\#2, $F\delta_0^{(2)}=(1.80\pm 0.02)\times 10^{-5}$, inferred from a finer temperature sweep up to $T_0=1$~K with a similar probe strength, $\nbar\sim 10^2$.
As a reference, we re-plot Dataset~\#2 from Fig.~\ref{fig1}(e) to highlight the different resonance frequency in the $T_0\to 0$ limit. Although the average TLS density and coupling remain essentially unchanged, the value of $f_r(T_0)$ depends on the actual realization of the TLS ensemble, which may vary slightly from cooldown to cooldown. 

\subsection{Swept-frequency response fit}

The theory curves in Fig.~\ref{fig2}(a) are generated using the numerical model detailed in Appendix.~\ref{app:TLSIterativeSolver}, with parameters summarized in Table.~\ref{tableFitParams}. The values of $\kappa_e$ and $F\delta_0^\textrm{diss}$ are extracted by fitting the low-power reflection coefficient. The parameters $f_{r,0}$, $F\delta_0^\textrm{reac}$ and the phonon bath dimensionality $d$ are obtained from the temperature sweep, Dataset \#2 (Appendix~\ref{app:extractTLSFd}). We set $Q_\textrm{bkg}=60\times 10^6$, based on the estimated radiation leakage through the phononic shields reported in our prior work \cite{Emser2024}. The thermal conductance temperature exponent, $\gamma=2.83$, is determined through simulation (see Supplemental Material S1). The discrete TLS frequency and coupling strength $\omega_{tls}$ and $g$ are extracted by fitting the height of the low-power resonance frequency step (Appendix~\ref{app:discreteTLS}). Out of the 13 parameters, only four remain free in the fit: $\beta$, $n_s$, $G_{th}(T_0)/g_0(T_0)$ and $Q_\textrm{rel}(0.5\text{K})$. The best fit yields a $\beta$ value close to 1 indicating that nonuniform TLS saturation is not required to reproduce the experimental data. Accordingly, we fix $\beta=1$ in the model.

From the extracted $\omega_{tls}$ and $g$, we can compute the critical phonon number $n_c$ above which the dispersive shift $\chi_0$ from the discrete TLS vanishes. We find $n_c\approx(\omega_{tls}-2\pi f_r(T_0))^2/(2g)^2\approx 3100$, which agrees with Fig.~\ref{figSim}(c) showing $\nbar\approx n_c$ at approximately the same probe power where the frequency step occurs (\textit{e.g.} $P_s\approx-135$~dBm for the $25$~mK series). 

\begin{table*}[t]
\centering
\begin{tabular}{@{}lllll@{}}
\toprule
\toprule 
\textbf{Parameter} & \textbf{Description}& \textbf{Eq.} & \textbf{Fig.~\ref{fig2}} & \textbf{Fig.~\ref{fig3}} \\
\midrule
$\kappa_e/2\pi\;\textrm{[Hz]}$ & External coupling rate of the resonator & \ref{S11self} & 69.9 & 42.1 \\
$f_{r,0}\;\textrm{[MHz]}$ & Resonator frequency at $T=0$ & \ref{dfTLS} & 520.808275 & 502.06550 \\
$F\delta_0^\textrm{reac}$ & Intrinsic TLS loss tangent at $T=0$ (reactive response) & \ref{dfTLS} &  $1.42\times 10^{-5}$ & $1.14\times 10^{-5}$ \\
$F\delta_0^\textrm{diss}$ & Intrinsic TLS loss tangent at $T=0$ (dissipative response) & \ref{QTLS} & $1.61\times 10^{-5}$ & $1.23\times 10^{-5}$\\
$n_s$ & Critical phonon number for TLS saturation & \ref{QTLS} & 94.2 & 128 \\
$\beta$ & Phenomenological parameter for nonuniform TLS saturation & \ref{Qtlsparam} & 1.0 & 1.0 \\
$Q_\textrm{bkg}$ & Intrinsic (background) internal quality factor & \ref{Qimodel} & $60\times 10^6$ & $60\times 10^6$ \\
$Q_\textrm{rel}(0.5\textrm{K})$ & TLS relaxation-damping quality factor at $0.5$~K & \ref{Qimodel} & $0.33\times 10^6$ & $0.36\times 10^6$ \\
$d$ & TLS relaxation-damping temperature exponent & \ref{Qimodel} & 1.69 & 1.84 \\
$G_{th}(T_0)/g_0(T_0)$ & Number of effective thermal conductance channels & \ref{TvsPd} & 0.37 & 0.58 \\
$\gamma$ & Thermal conductance temperature exponent & \ref{TvsPd} & 2.83 & 2.83 \\
$g/2\pi\;\textrm{[kHz]}$ & Coupling strength between the resonator and the discrete TLS & \ref{shift1TLS} & 230 & / \\
$\omega_{tls}/2\pi\;\textrm{[MHz]}$ & Discrete TLS frequency & \ref{shift1TLS} & 546 & / \\
\bottomrule
\bottomrule 
\end{tabular}
\caption{Fitting parameters for the frequency and time-domain resonator data.}
\label{tableFitParams}
\end{table*}

\subsection{Quality factor extraction from resonance depth}\label{app:frPssolver}

\begin{figure}[t]
\centering
\includegraphics[width=1.0\linewidth]{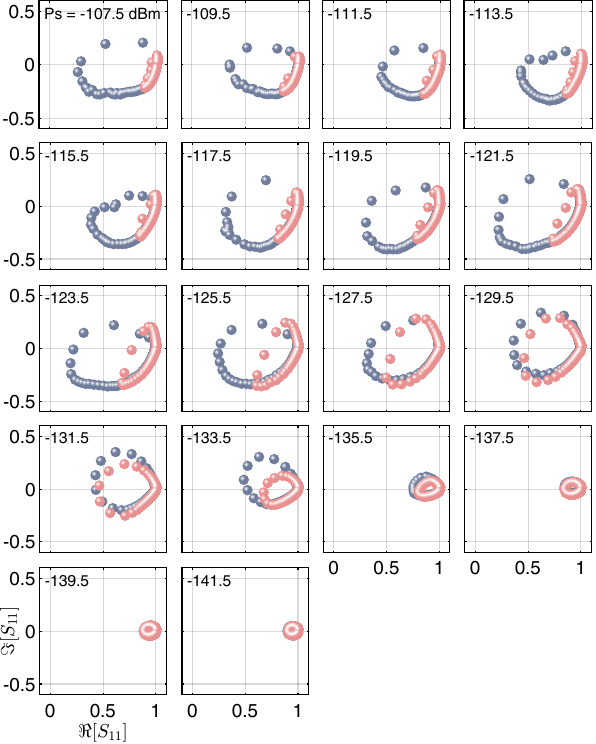}
\caption{Argand plot of the measured $S_{11}(f)$ (Dataset \#1) at $T_0 = 50$~mK, illustrating the distortion of the resonance curve at high power. The probe power, $P_s$, is indicated in the top-left corner of each plot, with upward and downward frequency sweeps shown in blue and red, respectively.}
\label{figResArgand}
\end{figure}

The standard circle-fitting method for extracting $Q$ (see \textit{e.g.} Ref.~\cite{Baity2023}) cannot be applied to the characterization of quartz PCRs in their nonlinear regime. At vanishing probe power, where the resonator response remains linear, the complex reflection coefficient $S_{11}(f)$ traces the familiar circle in the complex plane. However, as illustrated in Fig.~\ref{figResArgand}, for probe powers as low as $P_s\approx -137$~dBm, the response is already not circular any more, but distorted into a tear-drop–like shape due to quality-factor enhancement from saturation of resonant TLSs \cite{Thomas2020}, with the distortion increasing further at higher powers. The resonance curve becomes hysteretic and, above $-125$~dBm, is only partially sampled due to the discontinuous jump in $|S_{11}|$, precluding any circle fit. 

In this regime, one may instead attempt to estimate $Q$ from the resonance depth in $|S_{11}(f)|$. Specifically, for an upward sweep of the probe frequency $f$, $|S_{11}(f)|$ reaches its minimum value, $s\equiv|S_{11}(f_r)|=1-2Q/Q_e$, when $f$ reaches $f_r$ ($x(f_r)=0$), and this equation can be inverted to obtain $Q$. However, contrary to circle fitting that exploits both amplitude and phase information over the full $S_{11}(f)$ response, this approach relies only on $s$, a single scalar quantity, rendering the extracted $Q$ highly sensitive to noise and calibration errors. While this approach is viable at low power, where the resonators are strongly undercoupled ($Q_e/Q_i\approx 40$), it becomes increasingly unreliable as the system approaches critical coupling. In particular, the relative uncertainty in $Q$ increases as $\sigma_Q/Q=(1+Q_e/Q_i)\sigma_s/2=\sigma_s/(1-s)$, which diverges when $Q=Q_e/2$ at $s=1$. For this reason, we resort instead to time-domain ringdown measurements to accurately extract $Q$ as a function of $P_s$ (Section~\ref{sec:Ringdowns}). Ringdown measurements, however, probe the system at a fixed frequency, leading to a different level of readout-induced heating at a given probe power compared to swept-frequency measurements, and consequently a different measured response.

For completeness, Fig.~\ref{figSim} presents, in addition to the extracted resonance frequency, the quality factor inferred from the same spectroscopy measurements (Dataset~\#1) using the resonance-depth method. Also shown are the inferred mean phonon number and the effective TLS temperature at resonance.
In practice, $Q$ is obtained using a slightly modified expression for the reflection coefficient based on the diameter-correction method (DCM) \cite{Khalil2012, Chen2022}:
\begin{equation}
    S_{11}(f)=1-\frac{(2Q/Q_e)(1+j\tan{\phi})}{1+ 2jQ(f-f_r)/f_r},
\end{equation}
where the rotation angle $\phi$ is introduced to fit the small resonance asymmetry that may originate from an impedance mismatch between the feedline and the sample.
For finite $|\phi|\ll2\pi$, the minimum in $|S_{11}(f)|$ is no longer achieved when $f=f_r$, but is shifted slightly to $f\approx f_r(1+\frac{\phi}{4Q(1-Q/Q_e)})$. 
The expression to map the resonance depth $s$ to $Q$ is then generalized to 
\begin{equation}
    Q^{\pm}=\frac{Q_e}{2}\cos^2{(\phi)}\left(1\pm\sqrt{1+(s^2-1)\sec^2{(\phi)}}\right),
\end{equation}
where $Q^{-}$ (resp. $Q^{+}$) corresponds to the under-coupled (resp. over-coupled) case. $s$ reaches its minimum, $s_\textrm{min}=|\sin{\phi}|$, when $Q$ equals $Q_c=Q_e\cos^2{(\phi)}/2$. From the measured data, we infer a small rotation angle of $\phi=-0.18$~rad, corresponding to a maximum observable resonance depth of $S_{11}^\textrm{min}=20\log{s_\textrm{min}}\approx -15$~dB and $Q_c\approx 3.6\times 10^6 $. This agrees with the $25$~mK data in Fig.~\ref{fig1}(b), where the measured resonance depth attains about $15$~dB at $P_s\approx -123$~dBm but never exceeds it. 

The large \textit{apparent} discrepancy at high power between the extracted quality factor and the model in Fig.~\ref{figSim}(b) may result from two effects: (1) increasing uncertainty in the inferred $Q$ as it approaches $Q_c$, and (2) actual spectral broadening due to resonance-frequency fluctuations induced by a growing population of thermally occupied, far-detuned TLSs \cite{Maksymowych2025}. Such frequency noise is not included in our model, which only accounts for mechanical damping from TLSs but neglects dephasing effects. Therefore, while our model captures the power-dependent frequency shift and reproduces fairly well the asymmetric lineshape at a fixed probe power, it cannot provide a complete global fit of the complex $S_{11}$ response across all powers and temperatures.

\vfill

\twocolumngrid

\bibliography{biblio}

\end{document}


\title{\fontsize{13}{15}\selectfont Supplemental Material for:\\
TLS-induced thermal nonlinearity in a micro-mechanical resonator}

\author{C. Metzger}
\email{cyril.metzger@yale.edu}
\affiliation{Department of Physics, Yale University, New Haven, Connecticut 06511, USA}
\affiliation{JILA, National Institute of Standards and Technology and the University of Colorado, Boulder, Colorado 80309, USA}
\affiliation{Department of Physics, University of Colorado, Boulder, Colorado 80309, USA}

\author{A. L. Emser}
\author{B. C. Rose}
\affiliation{JILA, National Institute of Standards and Technology and the University of Colorado, Boulder, Colorado 80309, USA}
\affiliation{Department of Physics, University of Colorado, Boulder, Colorado 80309, USA}

\author{K. W. Lehnert}
\affiliation{Department of Physics, Yale University, New Haven, Connecticut 06511, USA}
\affiliation{JILA, National Institute of Standards and Technology and the University of Colorado, Boulder, Colorado 80309, USA}
\affiliation{Department of Physics, University of Colorado, Boulder, Colorado 80309, USA}

\date{\today}

\maketitle

\begin{table*}[!t]
\centering
\caption{Table of symbols.}
{\scriptsize
\begin{tabular}{@{}ll@{}}
\toprule
\toprule 
\textbf{Symbol} & \textbf{Description} \\
\midrule
$f_r=\omega_r/2\pi$ & PCR resonance frequency \\
$f_{r,0}$ & PCR resonance frequency at $T=0$ and $\nbar=0$ \\
$f_r^{(0)}$ & Bare PCR resonance frequency in the absence of TLS \\
$f_b$ & Central frequency of the acoustic bandgap \\
$\Delta f_b$ & Spectral width of the acoustic bandgap \\
$\omega_{tls}/2\pi$ & Transition frequency of the discrete TLS \\
$g/2\pi$ & Coupling rate between the discrete TLS and the PCR  \\
$f$ & Microwave probe tone frequency \\
$P_s$ & Microwave probe power applied to the sample \\
$P_d$ & Microwave power dissipated in the PCR \\
$Q_i$ & Internal quality factor of the PCR \\
$Q_e$ & External quality factor of the PCR \\
$Q$ & Total (loaded) quality factor of the PCR \\
$Q_\textrm{min}$ & Total quality factor at $\nbar=0$ \\
$Q_\textrm{max}$ & Total quality factor when all near-resonant TLSs are saturated \\
$Q_\textrm{res}$ & Internal quality factor limited by resonant TLSs \\
$Q_\textrm{res,min}$ & Internal quality factor limited by resonant TLSs at $\nbar=0$\\
$Q_\textrm{rel}$ & Internal quality factor limited by relaxation TLSs \\
$Q_\textrm{bkg}$ & Internal quality factor limited by intrinsic mechanical loss only (no TLS) \\
$\beta$ & Phenomenological $\nbar$ exponent accounting for nonuniform TLS saturation \\
$d$ & TLS relaxation-damping temperature exponent \\
$\alpha$ & TLS saturation parameter (0 if fully polarized, 1 if fully saturated) \\
$r$ & Normalized difference between the min and max values of $Q$ \\
$\chi_d$ & Detuning efficiency between the microwave probe and the PCR \\
$\chi_c$ & Coupling efficiency between the microwave probe and the PCR \\
$\kappa_i/2\pi$ & Internal loss rate of the PCR \\
$\kappa_i^{(0)}/2\pi$ & Intrinsic internal loss rate in the absence of TLSs \\
$\kappa_{i,\textrm{res}}/2\pi$ & Internal loss rate due to resonant TLSs \\
$\kappa_e/2\pi$ & External coupling rate of the PCR \\
$\kappa/2\pi$ & Total loss rate of the PCR \\
$\delta_0$ & Average loss tangent at $T=0$ \\
$\delta_0^\textrm{diss}$ & Average loss tangent at $T=0$ due to near-resonant TLSs \\
$\delta_0^\textrm{reac}$ & Average loss tangent at $T=0$ due to far-detuned TLSs \\
$F$ & Filling fraction of TLSs in the resonator material \\
$T_0$ & Base-plate temperature of the refrigerator \\
$T$ & Effective temperature of the TLS ensemble \\
$T_c$ & TLS hardening-softening crossover temperature \\
$T_d$ & Temperature scale for the TLS dissipative nonlinearity \\
$T_r$ & Temperature scale for the TLS reactive nonlinearity \\
$T_*$ & Temperature scale for the mixed reactive and dissipative TLS nonlinearity \\
$T_t$ & Defect-site TLS temperature in the thermal conductance model \\
$T_p$ & Defect-site thermal phonons temperature in the thermal conductance model \\
$TCF$ & Temperature coefficient of frequency \\
$a$ & TLS reactive nonlinearity strength in Swenson's model \\
$a_c$ & Critical value of $a$ for the onset of the hysteretic regime \\
$\varphi$ & Nonlinearity tangent controlling the ratio of dissipative to reactive response \\
$\nbar$ & Average number of phonons in the PCR \\
$n_{th}$ & Average number of thermal phonons \\
$\nbar_r$ & Average number of phonons when probing the PCR on resonance \\
$n_s$ & Critical phonon number for near-resonant TLS saturation \\
$n_h$ & Phonon number scale for the onset of TLS heating \\
$n_c$ & Critical phonon number for the discrete TLS saturation \\
$n_*$ & Phonon number scale for the mixed reactive and dissipative TLS nonlinearity \\
$S_{11}$ & Reflection coefficient of the microwave probe off the PCR \\
$\phi$ & Rotation angle accounting for the resonance asymmetry in the measured $S_{11}$ \\
$x$ & Realized detuning between the microwave probe and the PCR \\
$x_0$ & Applied detuning between the microwave probe and the PCR \\
$y$ & Realized detuning $x$ expressed in number of linewidths \\
$y_0$ & Applied detuning $x_0$ expressed in number of linewidths \\
$R_{th}=1/G_{th}$ & Thermal resistance between the PCR defect site and the substrate's cold bath \\
$C_{th}$ & Heat capacity associated to the PCR defect site \\
$N_{ch}$ & Number of effective thermal conductance channels \\
$\gamma$ & Thermal conductance temperature exponent \\
\bottomrule
\bottomrule 
\end{tabular}
}
\label{tableSyms}
\end{table*}

\makeatletter
\long\def\@makecaption#1#2{%
  \vskip\abovecaptionskip
  \sbox\@tempboxa{\footnotesize\textbf{#1:} #2}%
  \ifdim \wd\@tempboxa >\hsize
    \begin{minipage}{\hsize}%
      \scriptsize      
      \setstretch{1.0}   
      \justifying         
      \noindent\textbf{#1.} #2\par
    \end{minipage}%
  \else
    \begin{minipage}{\hsize}%
      \scriptsize
      \setstretch{1.0}
      \justifying
      \noindent\textbf{#1.} #2\par
    \end{minipage}%
  \fi
  \vskip\belowcaptionskip
}
\makeatother

\newcommand{\dv}[2]{\frac{\textrm{d}#1}{\textrm{d}#2}} 
\newcommand{\dvstar}[2]{\textrm{d}#1/\textrm{d}#2}

\begingroup
\setcounter{tocdepth}{1}
\tableofcontents
\endgroup

\clearpage

\setcounter{section}{0}
\renewcommand{\thesection}{S\arabic{section}}

\setcounter{figure}{0}
\renewcommand{\thefigure}{S\arabic{figure}}

\setcounter{equation}{0}
\renewcommand{\theequation}{S\arabic{equation}}

\setcounter{table}{0}
\renewcommand{\thetable}{S\arabic{table}}

%
\begingroup
\fontsize{10}{12}\selectfont

\section{Scalar model for the thermal conductance $G_{th}$ of the PCR tethers}\label{app:AppGthGamma}

In this section, we motivate the choice of a power law with temperature for the thermal conductance of the PCR tethers, $G_{th}\propto T^\gamma$, and discuss the right value to choose for the temperature exponent $\gamma$, based on a scalar model for the thermal conductance of a rectangular beam. 
The system we aim to model consists of a small block of single-crystal quartz (the ``cavity'') suspended from the bulk substrate (the ``reservoir'') by two thin beams that are patterned spatially in a periodic fashion so as to generate a one-dimensional bandgap in their phononic dispersion (see Fig.~1(a) for a picture of a representative device). The thermal conductance through the delocalized phonon modes of such beams was computed in Ref.~\cite{Cleland2001} by Cleland et al. using a scalar phonon model and we summarize here the main results.

To start with, we initially neglect the periodic modulation of the beam width that generates the phononic bandgap and consider first heat conduction through a simple rectangular beam of length $L$, width $w$ and thickness $t<w$, as conceptually sketched in Fig.~\ref{figGth}(a). 
The eigenspectrum of longitudinal phonon modes in the beam is given by $\omega_{klm}(k)=\sqrt{c^2k^2+\omega_{lm}^2}$, where $c$ is the phonon velocity, $k$ is the longitudinal wave vector along the beam and the two indices, $l$ and $m$ label the transverse wave vectors $k_l=l\pi/ w$, $k_m=m\pi/ t$ with $l,m\in\mathbb{N}$. The cutoff frequency $\omega_{lm}$ of each band can be expressed as $\omega_{lm}=\pi c\sqrt{(l/w)^2+(m/t)^2}$ and we plotted the lowest-lying modes in Fig.~\ref{figGth}(b).
Using the Landauer formula for one-dimensional transport, the thermal conductance of the beam is then obtained by summing the contribution of all the phonon modes in the beam assuming that they are populated with an equilibrium thermal distribution given by the temperature of the hot cavity \cite{Angelescu1998, Rego1998}:
\begin{equation}\label{Gth}
    G_{th} = \frac{k_B^2 T}{h}\sum_{l,m}\int_{x_{lm}}^{\infty}\mathcal{T}_{lm}(x)\frac{x^2e^{x}}{(e^{x}-1)^2}d x,
\end{equation}
where the sum runs over the thermally populated bands of delocalized phonon modes in the beam labeled by $l,m$. $x_{lm}=\hbar\omega_{lm}/(k_B T)$ denotes the reduced cutoff frequency of band $(l,m)$ and $\mathcal{T}_{lm}(x)$ its transmissivity at reduced frequency $x=\hbar\omega/(k_B T)$.

\begin{figure}[!ht]
\centering
\includegraphics[width=0.7\linewidth]{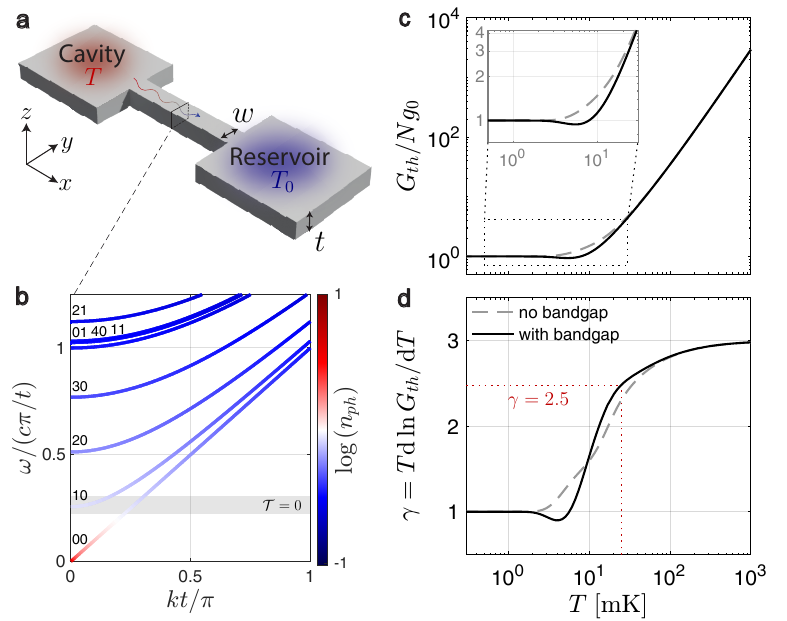}
\caption{Modeling the thermal conductance through the PCR tethers. (a) A phonon cavity at elevated temperature $T$ is in thermal contact with a cold reservoir at temperature $T_0$ through a narrow bridge. (b) Dispersion relation of longitudinal phonons in the bridge, modeled as a rectangular beam of width $w=3.9$~{\textmu}m and thickness $t=1$~{\textmu}m,  assuming a wave velocity $c=4$~km/s. Each band is labeled by its two mode indices $(l,m)$ and the line color encodes the average number of phonons in the mode assuming thermal equilibrium at $T_0=25$~mK, $\log{(n_{ph}(\omega))}$ (truncated below $1$ since $n_{ph}$ diverges at $\omega=0$). (c) Thermal conductance of the beam $G_{th}$ as a function of the cavity temperature $T$, as computed from the scalar model in presence (solid black) or not (dashed gray) of a $\Delta f_b=170$~MHz wide phonon bandgap centered at $f_b =530$~MHz. (d) Exponent $\gamma$ of the temperature dependence of the thermal conductance showing the dimensional crossover from 1D to 3D. At $T_0=25$~mK, $\gamma=2.5$.}
\label{figGth}
\end{figure}

At the lowest temperature and longest thermal-phonon wavelengths, only massless modes of the beam are populated. The scalar model considered here only supports one such mode with a null cutoff frequency: the lowest band $(l,m)=(0,0)$. Considering this mode only and assuming that its transmissivity is constant with frequency, $\mathcal{T}_{00}(x)=\mathcal{T}_{00}$, the thermal conductance evaluates to $G_{th}(T)=\mathcal{T}_{00}g_0\propto T$, where $g_0(T)=\pi^2k_B^2T/(3h)$ is the universal quantum of thermal conductance. This simplified model only assumes longitudinal phonon modes, but an elastic beam can also support one torsional and two flexural modes, so 4 massless modes in total \cite{Nishiguchi1997, Tanaka2005}. This means that at the lowest temperature, the PCR tethers should behave as one-dimensional waveguides for the phonons and their thermal conductance is expected to approach the quantized value $4g_0(T)$ in the limit of perfect phonon transmissivity to the bulk substrate. The defect site of the PCR being supported by two such tethers, we expect $G_{th}(T\to0)=8g_0(T)$, a value that is independent of the beam geometry $L,w,t$ and that is proportional to $T$. However, as Fig.~\ref{figGth}(b) shows, higher modes with finite cutoff frequency become populated at finite temperature and also contribute to the heat transport. 
One can derive an expression for the crossover temperature $T_{1D}$ at which the beam ceases to behave as a one-dimensional waveguide for phonons. This occurs when the dominant phonon wavelength becomes smaller than the beam width (or equivalently when the thermal wavevector $k_\textrm{th}=k_B T/(\hbar c)>\pi/w$), \textit{i.e.} when $T> T_{1D}=\pi\hbar c/(k_B w)$. Assuming $c=4$~km/s and a beam width $w=4$~{\textmu}m, one finds $T_{1D}\approx 24$~mK. As $T$ is increased above  $T_{1D}$, one therefore expects a crossover to $2D$ and then to $3D$ as $T\gg \pi\hbar c/(k_B t)$, in which limit the Debye law $G_{th}\propto T^3$ is eventually recovered \cite{Kuhn2004, Wang2007}.

Eq.~\ref{Gth} can be rewritten as $G_{th}/g_0(T)=(3/\pi^2)\int_0^\infty dx\, N(x)\,x^2 e^x/(e^x -1)^2 $, where $N(x)$ counts the number of modes at reduced frequency $x$. It provides a general recipe to compute the thermal conductance of the beam knowing $N(x)$. One can choose an appropriate UV cutoff to replace the infinite bound in the integral. For it to converge properly, we find that this cutoff needs to be at least $x_{max}\approx 10$. This means that to compute $G_{th}(T)$ up to $T=0.1$~K, one needs to simulate the phonon bandstructure over a broad range of frequencies up to at least $k_B x_{max}/h\times 0.1~K\approx 20$~GHz. This would take a prohibitively long time using a traditional finite-element solver like \textsc{Comsol}\textsuperscript{\textregistered}. Therefore, instead of simulating the bandstructure for the exact geometry, we use the simplified model of a rectangular beam for which the cutoff frequencies of the longitudinal phonon modes are known analytically. The associated thermal conductance is computed numerically and shown as a function of the cavity temperature $T$ in Fig.~\ref{figGth}(c) (dashed gray line). Its logarithmic derivative with respect to temperature provides the temperature exponent $\gamma$, shown in Fig.~\ref{figGth}(d), which varies smoothly from $1$ to $3$ as $T$ is increased, evidencing the dimensional cross-over of the phonon gas in the beam. From this numerical simulation, we therefore expect $2.5\leq \gamma\leq 2.9$ over the range of temperatures achieved in this study, $25 \leq T\leq 200$~mK, with $\gamma\approx 2.5$ at base temperature $T_0=25$~mK.

Finally, to model the effect of the phononic bandgap arising from the spatial modulation of the beam width, we assume a phonon mode transmissivity $\mathcal{T}(f)=0$ if $|f-f_b|<\Delta f_b$ or $1$ otherwise, where $f_b=530$~MHz is the simulated central frequency of the bandgap and $\Delta f_b=170$~MHz its size. The corresponding $G_{th}(T)$ and $\gamma(T)$ are shown in solid black lines in Fig.~\ref{figGth}(c-d). In agreement with \cite{Cleland2001}, we observe a suppression of $G_{th}$ at temperatures where the dominant phonon wavelength becomes of the order of the phononic-crystal repeat distance $p=c/(2f_b)\approx 3.7$~{\textmu}m. Since $p\approx w$ in our device, this also occurs around $T\approx T_{1D}$. We find that for the range of temperature considered experimentally, $25 \leq T\leq 200$~mK, $G_{th}(T)$ is well fit by $(7.2\times 10^{-10})\times T[K]^{2.83}\;W/K$ and therefore use a constant value $\gamma=2.83$ in our model for the readout-power heating.

\section{Nonlinear shift from TLSs due to hole-burning from an off-resonant drive}\label{app:TLSnonlinshift}

In a single-tone spectroscopy experiment, a resonant drive should in principle have no effect on the overall dispersive shift imparted by the TLSs onto the mechanical resonance, in the case where the TLSs are uniformly distributed in frequency. This is because it saturates the same density of TLSs on either side of the resonance, resulting in a symmetric cancellation of the partial shifts from positively and negatively detuned TLSs. However, an off-resonant drive will modify the population of TLSs in an asymmetric way around the resonance frequency, such that a net dispersive shift is imparted to the cavity. The magnitude of this nonlinear shift was derived by Kirsh \textit{et al.} for tunneling states (TS) \cite{Kirsh2017} and for the simpler case of two-level systems (TLS) with uniform coupling strength to the resonator by Capelle \textit{et al.} \cite{Capelle2020}. 
Here we estimate the magnitude of this shift for our quartz PCR resonators and show that it cannot explain the reactive behavior evidenced in Fig.~1. In other words, we clarify that the reactive nonlinearity reported in this work cannot be accounted for by resonant depolarization of TLSs, but is the result of a broadband heating of a spectrally large population of TLSs due to the dissipated power from the readout.

\begin{figure}[!ht]
\centering
\includegraphics[width=0.65\linewidth]{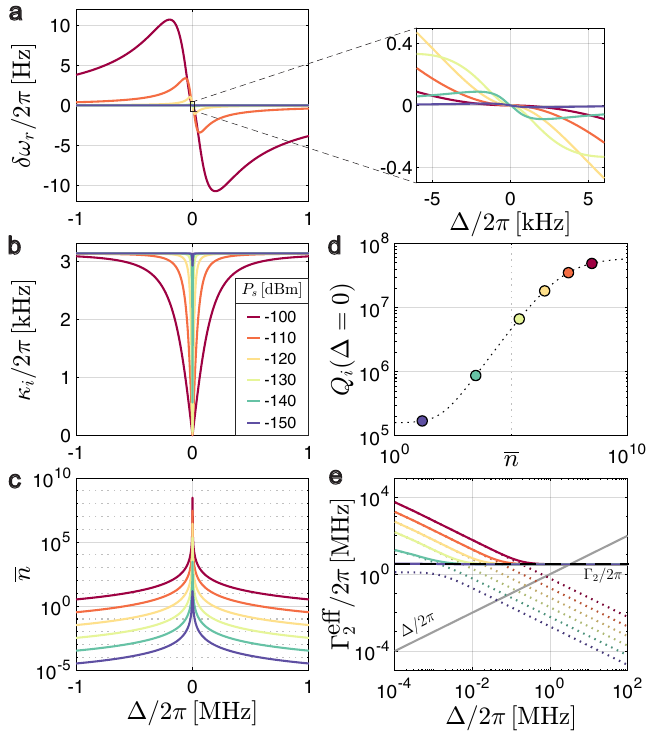}
\caption{Nonlinear response due to hole-burning in the TLS ensemble. (a) The cavity frequency pull $\delta\omega_r$, (b) internal mechanical loss rate $\kappa_i$ and (c) average intracavity phonon number $\nbar$ as a function of the probe-tone detuning $\Delta = \omega-\omega_r$ for various probe power $P_s$. We assume a uniform TLS-cavity coupling strength $g/2\pi=230$~kHz and a critical phonon number for TLS saturation of $n_s=100$. To make contact with Fig.~1, the inset on the top right shows a magnified $\delta\omega_r(\Delta)/2\pi$ over the kHz range, evidencing a negligible sub-Hz shift. The main effect of the probe is a narrowing of the cavity line due to saturation of the TLS ensemble. (d) The internal quality factor $Q_i=\omega_r/\kappa_i$ at $\Delta=0$ as a function of $\nbar$. The black dotted line corresponds to Eq.~\ref{TLSshifteq2smallDet}. (e) The width $\Gamma_2^\textrm{eff}$ of the spectral hole in the TLS ensemble as a function of $\Delta$. The colored dotted lines correspond to the large driving limit where $\Gamma_2^\textrm{eff}\approx\Omega/\sqrt{2}=\sqrt{2}g\nbar(\Delta)$. The black dashed line shows the TLS intrinsic linewidth $\Gamma_2=\sqrt{2g^2n_s}\approx2\pi\times 3.2$~MHz and the solid gray line indicates $\Delta$ for reference. Maximal cavity pull from the TLSs would occur for $\Omega(\Delta)/\sqrt{2}=\Delta\gg\Gamma_2$, which is not realized in the regime considered here.}
\label{figTLSshift}
\end{figure}

To estimate the frequency shift $\delta\omega_r$ and mechanical damping $\kappa_i$ due to hole-burning in the TLS ensemble by a probe tone, we use the following expressions based on Ref.~\cite{Capelle2020} that assume a flat spectral distribution of TLSs and no interaction between individual TLSs:
\begin{align}[left=\empheqlbrace]
    \delta \omega_r &= -\frac{\Gamma_0}{2} \frac{(\Delta/\Gamma_2)(\nbar/n_s)}{\sqrt{1+\nbar/n_s}\Bigl[(\Delta/\Gamma_2)^2 + (1+\sqrt{1+\nbar/n_s})^2\Bigr]}\label{TLSshifteq1}\\
    \kappa_i &= \kappa_i^{(0)}+\Gamma_0\left[1-\frac{\nbar/n_s}{\sqrt{1+\nbar/n_s]}}\frac{1+\sqrt{1+\nbar/n_s}}{(\Delta/\Gamma_2)^2 + (1+\sqrt{1+\nbar/n_s})^2}\right].\label{TLSshifteq2}
\end{align}
Here, $\nbar$ is the average intracavity phonon number, $n_s=\Gamma_2^2/(2g^2)$ the critical phonon number for TLS saturation with $\Gamma_2=\Gamma_1/2$ the intrinsic TLS linewidth, and $\Delta=\omega-\omega_r$ is the probe tone detuning from the cavity frequency. $\kappa_i^{(0)}$ and $\Gamma_0\equiv\omega_r F\delta_0\tanh{(\hbar\omega_r/(2 k_B T))}$ set respectively the background (non-TLS) cavity loss rate and the maximal cavity damping due to TLSs. 

In the small detuning limit where $\Delta\ll\Gamma_2$, Eqs.~\ref{TLSshifteq1}-\ref{TLSshifteq2} reduce to
\begin{align}[left=\empheqlbrace]
    \delta \omega_r &= -\frac{\Gamma_0}{2}\frac{(\nbar/n_s)(\Delta/\Gamma_2)}{\sqrt{1+\nbar/n_s}(1+\sqrt{1+\nbar/n_s})^2}+\mathcal{O}\Bigl(\frac{\Delta}{\Gamma_2}\Bigr)^2\\
    \kappa_i &= \kappa_i^{(0)}+\frac{\Gamma_0}{\sqrt{1+\nbar/n_s}}+\mathcal{O}\Bigl(\frac{\Delta}{\Gamma_2}\Bigr)^2.\label{TLSshifteq2smallDet}
\end{align}
and we recover in Eq.~\ref{TLSshifteq2smallDet} the well-known expression for the power-dependent absorption by a TLS ensemble. 

Crucially, $\delta\omega_r\propto-\Delta$ and vanishes at $\Delta=0$, which is opposite to the situation observed experimentally where the shift is seen to be maximal at resonance. The latter situation corresponds to maximal power being dissipated internally and so points to heating as the mechanism for the nonlinear shift observed in the experiment, instead of resonant depolarization of the TLS ensemble. 
Since both effects can in principle occur, we proceed with estimating the magnitude of the resonant effect. We use the value for the extracted TLS-cavity coupling strength, $g/2\pi=230$~kHz, and assume $n_s=100$ and $\omega_r/\kappa_i^{(0)}=60\times 10^6$, in line with the values extracted from the fit in Fig.~2. This choice of $g$ and $n_s$ corresponds to an intrinsic TLS linewidth of $\Gamma_2=\sqrt{2g^2 n_s}\approx 2\pi\times 3.2$~MHz.
In Fig.~\ref{figTLSshift}(a-b), we show the magnitude of the cavity pull $\delta\omega_r(\Delta)$ and the internal loss rate $\kappa_i(\Delta)$ as a function of the probe-tone detuning $\Delta$ for different values of probe power $P_s$. To enforce self-consistency, both quantities are computed numerically using the following iterative procedure where steps 1-3 are repeated until convergence is reached:
(0) Initialize $\omega_r=\omega_r(T_0)=2\pi\times 500$~MHz and $\kappa_i=\kappa_i^{(0)} + \Gamma_0$.
(1) For each $P_s$ value, compute $\nbar(\Delta)=\kappa_e/[\Delta^2+((\kappa_i+\kappa_e)/2)^2](P_s/\hbar\omega_r)$, where $\kappa_e$ is the external coupling rate to the cavity. 
(2) Feed $\nbar(\Delta)$ into Eqs.~\ref{TLSshifteq1}-\ref{TLSshifteq2} to compute $\delta\omega_r(\Delta)$ and $\kappa_i(\Delta)$.
(3) Update the resonance frequency and detuning, $\omega_r\to\omega_r+\delta\omega_r$ and $\Delta\to\Delta-\delta\omega_r$. 

Fig.~\ref{figTLSshift}(a) shows that the nonlinear cavity shift from the TLS saturation by the probe reaches at most $\sim 10$~Hz at the maximal probe power $P_s=-100$~dBm, which is about two orders of magnitude smaller than the shift evidenced in Fig.~2(a). As seen in the figure's inset, in the $-5<\Delta<5$~kHz window corresponding to the measurement span in Fig.~1(b), the shift becomes as low as sub-Hz. In this small-detuning limit $\Delta\ll\Gamma_2$, the resonant reactive effect can therefore be safely neglected, which supports the analysis reported in the main part of this paper. As evidenced in Fig.~\ref{figTLSshift}(b), the main effect of the probe tone is a narrowing of the mechanical resonance. In Fig.~\ref{figTLSshift}(d) we show the corresponding internal quality factor at resonance, $\Delta=0$, the increase of which with $\nbar$ is well captured by Eq.~\ref{TLSshifteq2smallDet}.

Eq.~\ref{TLSshifteq1} predicts a maximal cavity pull $\delta\omega_r^\textrm{max}=\Gamma_0/4$ when the width $\Gamma_2^\textrm{eff}\equiv\Gamma_2\sqrt{1+\nbar/n_s}$ of the spectral hole burnt in the TLS ensemble by the probe matches the detuning $\Delta$ \cite{Capelle2020}. In the large drive limit $\Omega\gg\Gamma_2$, where $\Omega=2g\sqrt{\nbar}$ is the Rabi frequency, $\Gamma_2^\textrm{eff}=\sqrt{\Gamma_2^2+2g^2\nbar}\approx\Omega/\sqrt{2}$ and the condition for large cavity pull reduces to $\Omega=\sqrt{2}\Delta$. Fig.~\ref{figTLSshift}(e) shows that this regime of large shift is never realized for the range of parameters considered experimentally. This would require one of the colored dotted lines to reach the point where the black dashed line ($\Gamma_2$) and the gray solid line ($\Delta$) cross. This would happen if the probe power exceeded $P_s^\textrm{c}=(\hbar\omega_r/\kappa_e)(\Gamma_2^4/(2g^2))=(\hbar\omega_r/\kappa_e)2g^2 n_s^2$. For $g/2\pi=230$~kHz, $\kappa_e/2\pi=70$~Hz, $n_s=100$ and $T=25$~mK, one obtains $P_s^c\approx-75$~dBm and $\delta\omega_r^\textrm{max}\approx0.78$~kHz. Such a strong probe was not used here, so we can safely disregard the resonant reactive effect in fitting the Fig.~2 data.

\section{Analysis of Swenson's equation}\label{app:AppSwenson}

In this appendix, we analyze the solutions of Swenson's equation for the realized fractional detuning, $y=Q x$, which captures the reactive nonlinear behavior in the limit of weak readout-power heating ($\Delta T\ll T_0$) and instantaneous heating ($C_{th}=0$):
\begin{equation}\label{yT}
    y(T) = y_0(T_0) + \frac{a(T_0)}{1+4y(T)^2}\quad\textrm{with}\quad a(T_0)\equiv -R_\textrm{th}(T_0)TCF(T_0)\frac{4Q^3}{Q_e Q_i}P_s.
\end{equation}
This equation was initially derived by Swenson \textit{et al.} in Ref.~\cite{Swenson2013} to model the softening nonlinearity induced by a power-dependent kinetic inductance. It is known to give rise to a classic Duffing-like oscillator dynamics for the realized detuning \cite{Duffing1921,Kovacic2011}. Throughout this section, we assume a fixed total quality factor $Q$ and therefore neglect any dissipative nonlinearity. In Appendix~A.3, we described how Swenson's equation can be generalized to also incorporate the dissipative nonlinearity up to first order in $\delta T$. 
First, we rewrite Swenson's equation to make explicit its cubic polynomial form:

\begin{equation}\label{yTcubic}
    4 y^3 - 4 y_0\cdot y^2 + y - (y_0+a) = 0.
\end{equation}

\textit{Solutions phenomenology \& transition from linear to hysteretic regime}. This cubic equation can be solved analytically using Cardano's method \cite{Cardano1545}, but the obtained expressions are rather cumbersome. Here, we focus instead on a qualitative description of the solutions obtained numerically. 
In Fig.~\ref{fig_S1}, we show the three cubic roots $y_i,\, i\in\{1,2,3\}$ of Eq.~\ref{yTcubic} for the realized fractional detuning $y$ as a function of the applied fractional detuning $y_0$ for five different values of the Swenson nonlinearity parameter $a$. These three roots are in general complex numbers and only real solutions give rise to stable physical solutions.
The Swenson parameter $a$ encodes the strength of the nonlinear response. Crucially, it is proportional to the probe tone power $P_s$.
At low power when $|a|\ll 1$, we recover the linear regime, the realized detuning corresponds to the applied one, $y\approx y_0$, and the dissipated power $P_d(y_0)\propto\chi_d(y_0)$ is a Lorentzian function of the probe frequency (see Fig.~\ref{fig_S1}, \nth{3} row) centered around the bare resonance. When $a$ is non-zero, the resonance lineshape becomes distorted and the frequency at which the maximum power is dissipated no longer corresponds to the bare resonance frequency: it is blue-shifted when $a<0$ (\textit{hardening} nonlinearity) and red-shifted when $a>0$ (\textit{softening} nonlinearity). For a weak nonlinearity, $|a|<a_c$ where $a_c=4/(3\sqrt{3})\approx 0.77$ (Fig.~\ref{fig_S1}, \nth{2} and \nth{4} rows), the distortion from a pure Lorentzian remains small and all quantities are single-valued. For a strong nonlinearity $|a|>a_c$ (Fig.~\ref{fig_S1}, \nth{1} and \nth{5} rows), the distortion of the response is such that the resonator enters a \textit{bistable regime}: the realized detuning $y$ becomes a multi-valued function of the applied detuning $y_0$, which gives rise to a \textit{hysteretic} response with switching behavior depending on the sweep direction of the probe frequency. This bifurcation physics is well explained in Refs. \cite{Swenson2013, Yurke2006} and additional results from Swenson's model, including a derivation of the threshold $a_c$ for the onset of hysteresis, are available in Appendix~A of Ref.~\cite{Thomas2020}. The two physical solutions, labeled \textit{upward} and \textit{downward} (dotted black lines in Fig.~\ref{fig_S1}), are obtained by selecting for each value of applied detuning $y_0$ the one solution $y_i$ for which $\Im [y_i(y_0)]=0$. In the bistable regime, there is a continuous range of values $y_0\in[y_0^{<,d},y_0^{>,d}]$ for which all three solutions are real (but only two are stable), in which case the physical solution is then determined by the sweep direction: starting with a strongly detuned probe frequency $y_0<y_0^{<,d}$ (upward) or $y_0>y_0^{>,d}$ (downward), only one out of the three solutions satisfies $\Im [y_i(y_0)]=0$, and it will remain the physical solution until it suddenly develops a non-zero imaginary part when $y_0\geq y_0^{>,d}$ (upward) or $y_0\leq y_0^{<,d}$ (downward), value at which the system will spontaneously jump to the other stable real solution.
\medskip

\begin{figure}[!t]
\centering
\includegraphics[width=0.6\linewidth]{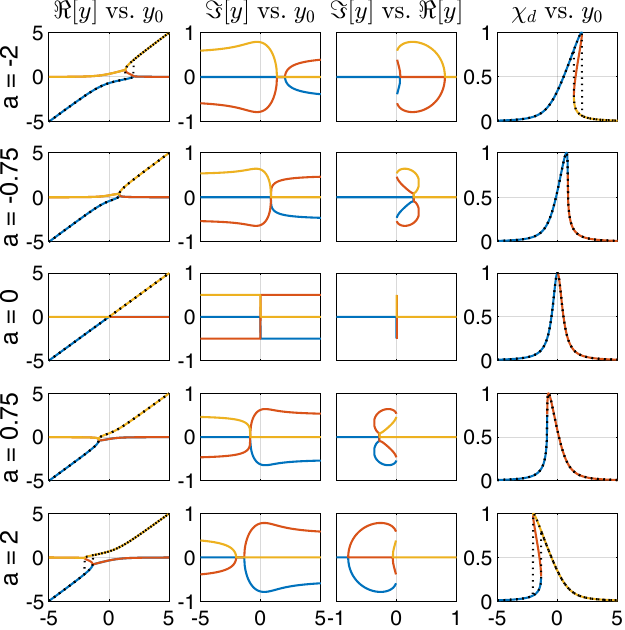}
\caption{Solutions of Swenson's equation as a function of the applied fractional detuning $y_0$. The three cubic roots of Eq.~\ref{yT} are plotted for different values of Swenson's nonlinearity parameter $a=-2:1:2$ in each row, showing the transition from a softening ($a>0$) to a hardening ($a<0$) nonlinearity. \nth{1} and \nth{2} columns: the real ($\Re [y]$) and imaginary ($\Im [y]$) parts of the three cubic roots as a function of $y_0$. \nth{3} column: the three cubic roots in an Argand plot, $\Im [y]$ \textit{vs.} $\Re [y]$. \nth{4} column: the detuning efficiency $\chi_d=1/(1+4y\{y_0\}^2)$ as a function of $y_0$. The three solutions are plotted in blue, yellow and red solid lines. From these solutions, one can build the physical solutions corresponding to upward and downward sweeps shown in dotted black lines by enforcing $\textrm{Im}(y)=0$.}
\label{fig_S1}
\end{figure}

\textit{Location of the switching points \& resonance frequency}. The resonance frequency is defined as the frequency that realizes $y=0$, for which $|S_{11}|$ is then minimal. In the nonlinear regime, when $-a$ exceeds $a_c$, $|S_{11}|$ develops a discontinuity, as evidenced in Fig.~\ref{fig_S2}(a), and $y$ becomes bi-valued, $y\to\{y^{>},y^{<}\}$. 
Zero-crossings of $y$ follow a different behavior depending on the direction in which the probe tone $f$ is swept. 
For an upward frequency sweep, \textit{i.e.} sweeping $y_0$ from negative to positive values (blue curve in Fig.~\ref{fig_S2}(a-d)), $y^{>}$ is initially negative, crosses zero at $y_0^{>,r}\geq 0$ and then jumps up to more positive values at $y_0^{>,d}\geq y_0^{>,r}$, where $y^{>}$ switches from $y^{>,d_-}>0$ to $y^{>,d_+}>y^{>,d_-}$. 
The resonance frequency is easily deduced by setting $y=0$ in Eq.~\ref{yTcubic}, which yields $y_0^{>,r}(T_0)=-a(T_0)$. Injecting this result in Eq.~\ref{yT} gives the probe frequency $f^{>,r}$ at which resonance is realized for an upward sweep:
\begin{equation}\label{frup}
    f^{>,r}(T_0)=f_r(T_0)\Bigl(1-\frac{a(T_0)}{Q}\Bigr).
\end{equation}
Because the nonlinearity strength parameter $a$ scales linearly with the readout power $P_s$, this means that $f^{>,r}(T_0)-f_r(T_0)\propto P_s$, which is characteristic of a Duffing type of nonlinearity.
The jump-up frequency $f^{>,d}$ at which the discontinuity in $y$ and $|S_{11}|$ occurs is found by solving $\dvstar{y_0}{y(y^{>,d})}=0$ for $y^{>,d}$, then $y_0^{>,d}=y^{>,d}-a/(1+4(y^{>,d})^2)$. One can derive the following asymptotic expansions valid for $-a> a_c$:
\begin{empheq}[left={\empheqlbrace}]{alignat=2}
    &y^{>,d} &&= -\frac{1}{8 a}-\frac{1}{64 a^3} + \mathcal{O}(1/a^5)\label{ydfor}\\
    &y_0^{>,d} &&= - a -\frac{1}{16 a}-\frac{1}{256 a^3} + \mathcal{O}(1/a^5),\label{y0dfor}
\end{empheq}
from which follows 
\begin{equation}\label{fdup}
    f^{>,d}(T_0)\underset{-a> a_c}{\approx} f_r(T_0)\Bigl(1-\frac{a(T_0)}{Q}-\frac{1}{16\,Q\,a(T_0)}-\frac{1}{256\,Q\,a(T_0)^3}+...\Bigr).
\end{equation}
For an upward sweep, the frequency at which switching occurs, $f^{>,d}$, follows closely the resonance frequency $f^{>,r}$ and the two coincide in the limit $|a|\gg a_c$, as illustrated in Fig.~\ref{fig_S2}(f). Therefore, the switching frequency in $|S_{11}|$ for an upward sweep provides in general a good approximation of the resonance frequency. 

For a downward frequency sweep, \textit{i.e.} sweeping $y_0$ from positive to negative values (red curve in Fig.~\ref{fig_S2}(a-d)), $y^{<}$ is initially positive, then jumps down from $y^{<,d_+}>0$ to $y^{<,d_-}$ at $y_0^{<,d}>0$ and keeps decreasing to lower negative values. 
When $-a>1$, $y^{<,d_-}<0$ and so, contrary to an upward sweep, the zero-crossing condition $y^{<}=0$ is never met (since the zero-crossing happens over the discontinuity). Accordingly, for $-a> 1$, the probe frequency at which $y^{<}$ is closer to zero and $|S_{11}|$ minimal always corresponds to the frequency at which the discontinuity occurs, \textit{i.e.} $y_0^{<,r}=y_0^{<,d}$. For $-a>a_c$, the following approximations can be derived:
\begin{empheq}[left={\empheqlbrace}]{alignat=2}
    &y^{<,d} &&= \frac{(-a)^\frac{1}{3}}{2^\frac{1}{3}} - \frac{1}{3\,2^\frac{2}{3}\,(-a)^\frac{1}{3}} + \frac{1}{24 a} - \frac{7}{324\,2^\frac{1}{3}\,(-a)^\frac{5}{3}} + \mathcal{O}(1/(-a)^\frac{7}{3})\label{ydrev}\\
    &y_0^{<,d} &&=\frac{3(-a)^\frac{1}{3}}{2^\frac{4}{3}} - \frac{1}{2^\frac{8}{3}(-a)^\frac{1}{3}} + \frac{1}{48\,a} - \frac{7}{864\,2^\frac{1}{3}\,(-a)^\frac{5}{3}} + \mathcal{O}(1/(-a)^\frac{7}{3}),\label{y0drev}
\end{empheq}
and the probe frequency $f^{<,d}$ at which the jump-down occurs for a downward sweep is deduced:
\begin{equation}
    f^{<,d}(T_0)\underset{-a> a_c}{\approx} f_r(T_0)\Bigl(1 + \frac{3(-a)^\frac{1}{3}}{2^\frac{4}{3}\,Q} - \frac{1}{2^\frac{8}{3}(-a)^\frac{1}{3}\,Q} + \frac{1}{48\,a\,Q} +...\Bigr)\quad\textrm{and}\quad f^{<,r}(T_0)=\begin{cases}
      f^{<,d}(T_0), &-a> 1, \\
      f^{>,r}(T_0), & -a\leq 1.
    \end{cases}
\end{equation}

\begin{figure*}[!ht]
\centering
\includegraphics[width=1.0\linewidth]{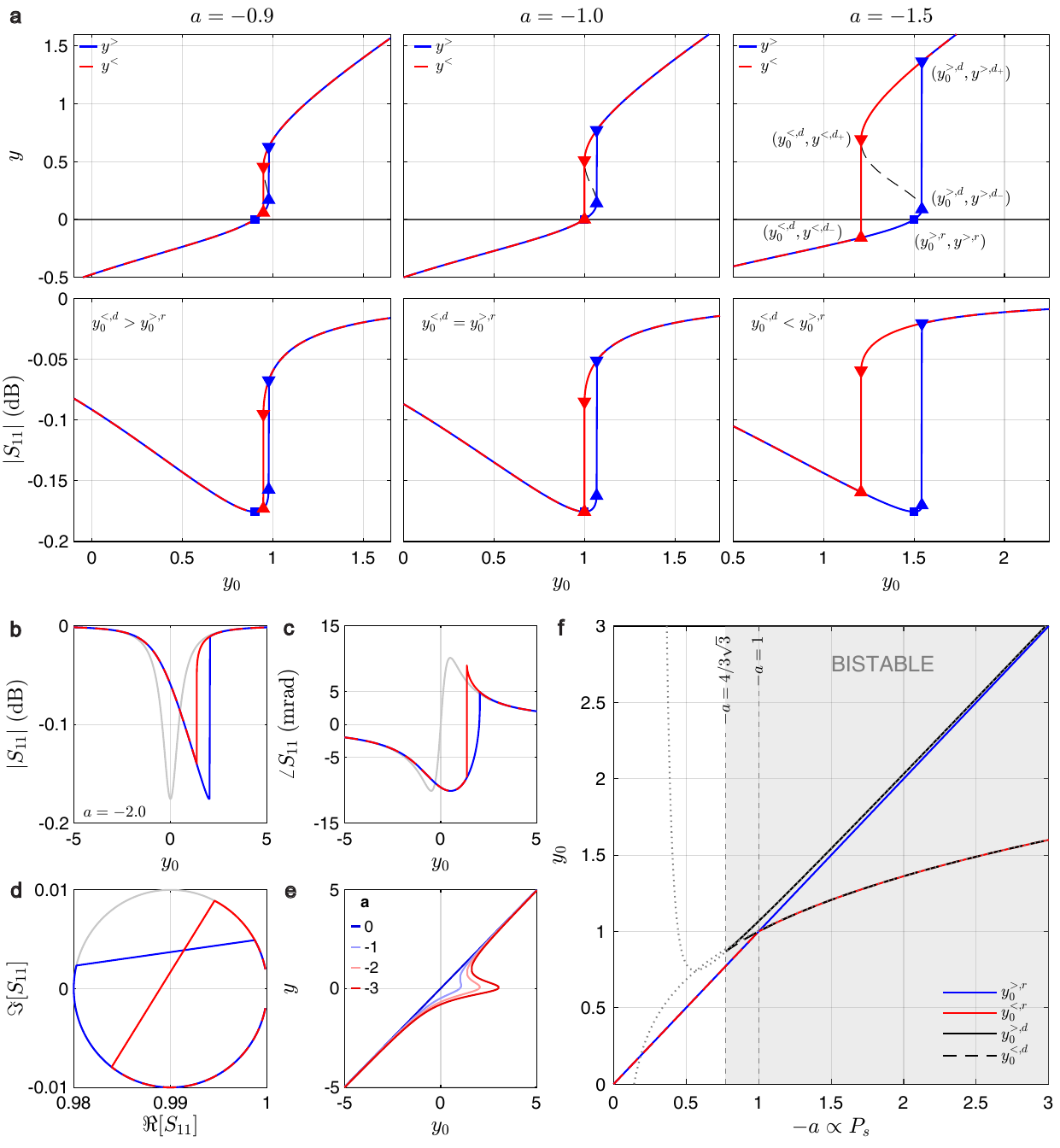}
\caption{Resonance and switching frequencies in Swenson's model. (a) The realized fractional detuning $y$ between the probe tone and the resonance (top row) and the magnitude of the corresponding reflection coefficient $|S_{11}(y)|$ (bottom row) as a function of the applied fractional detuning $y_0$ for three different values of the nonlinearity parameter $a=-0.9, -1.0,-1.5$ for a resonance depth of $\alpha=Q/Q_e=0.01$. The response is shown for both upward (blue lines) and downward (red lines) sweeps of the probe frequency. The thin black dashed line shows the non-physical multi-valued solution. (b-c) The magnitude and phase of $S_{11}$ as a function of $y_0$. The gray solid line depicts the response in the linear case ($a=0$), while the blue and red curves show a typical nonlinear response in the bistable regime ($a=-2$). (d) The Argand plot of $S_{11}$ in (b-c) evidencing the discontinuous jumps characteristic of the bistable nonlinear regime. Both linear and nonlinear responses fall on the same circle in the complex plane, but they sample it differently as $y_0$ is swept. (e) $y$ as a function of $y_0$ for four values of the nonlinearity parameter between $a=0$ and $-3$. When $|a|>a_c=4/(3\sqrt{3})$, $y$ is multi-valued and the response becomes hysteretic. (f) The applied detuning at resonance for both upward ($y_0^{>,r}$, blue line) and downward ($y_0^{<,r}$, red line) sweeps as a function of $a$. These two detunings generally differ from the ones at which switching occurs, $y_0^{>,d}$ and $y_0^{<,d}$ for respectively upward and downward sweeps (shown in black solid and dashed lines). For $-a\geq 1$, $y_0^{<,r}=y_0^{<,d}$ and $y_0^{>,d}\to y_0^{>,r}$ as $a\to-\infty$. The gray solid lines are approximations of $y_0^{>,d}$ and $y_0^{<,d}$ (Eqs.~\ref{y0dfor}, \ref{y0drev}) and the gray shaded area delimiting $-a>a_c$ denotes the bistability region.}
\label{fig_S2}
\end{figure*}

We keep here the notation $r$ to refer to the frequency $f$ at which $|y|$ and $|S_{11}|$ are minimal, although the probe frequency is technically \textit{never} on-resonance for a downward sweep with $-a>1$ since $y\neq 0$.
If $-a\leq 1$, instead, then $y^{<,d_-}\geq 0$ and so, as $y_0$ is further decreased, a zero-crossing in $y$ occurs at $y_0^{<,r}=y_0^{>,r}=-a$.
As shown in Fig.~\ref{fig_S2}(f), Eqs.~\ref{y0dfor} and \ref{y0drev} approximate well $y_0^{>,d}$ and $y_0^{<,d}$ and for $|a|>1$, the first term of the expansions gives the leading behavior, $y_0^{>,d}\sim -a$ and $y_0^{<,d}\sim (-a)^{1/3}$ with $-a\propto P_s$. One recovers the same scaling for the jump-up and jump-down frequencies as the ones for the Duffing oscillator \cite{Brennan2008}. Fig.~\ref{fig_S2}(f) shows the two special values of the nonlinearity strength above which the response becomes discontinuous and bistable ($a=-a_c$) and above which the resonance frequencies for upward and downward sweeps start to differ ($a=-1$).

\section{Influence of TLS density on resonance quality factor and nonlinearity}\label{app:AppQoptim}

In this appendix, we simulate the power behavior of a typical PCR resonance, using the model parameters obtained from fitting the experimental data in Fig.~3. Specifically, we vary the magnitude of $F\delta_0$ and simulate its effect on the power dependence of the resonance quantities: phonon number, quality factor, resonance frequency, etc. The results are displayed in Fig.~\ref{fig6}. $F\delta_0$ is proportional to the TLS density of states and to their squared coupling strength to the resonator \cite{Emser2024} and as such, it constitutes a good figure of merit to parametrize the impact of the TLS bath on the resonator properties, in particular its power dependence.
In these numerical simulations, we assume a base temperature of $T_0=25$~mK and fix the probe frequency to $f=f_r(T_0)$, \textit{i.e.} the probe tone frequency is chosen so as to be on-resonance in the limit of vanishing probe power when readout-power heating is negligible (\textit{i.e.} $y_0=0$). 
In addition, the background loss is discarded ($Q_\textrm{bkg}=\infty$) to keep the focus on the main contribution to mechanical dissipation: TLSs. In Ref.~\cite{Emser2024}, we found that the background (power-independent) loss originated from the radiation leakage through the mirror cells of the phononic crystal, which was estimated to $Q_\textrm{bkg}\sim 60\times 10^6$. In practice, this radiation leakage is suppressed exponentially with the number of mirror cells, $N_\textrm{mirr}$, such that $Q_\textrm{bkg}$ can always be made irrelevant by increasing $N_\textrm{mirr}$ by a few more units. 

\begin{figure*}[!ht]
\centering
\includegraphics[width=1.0\linewidth]{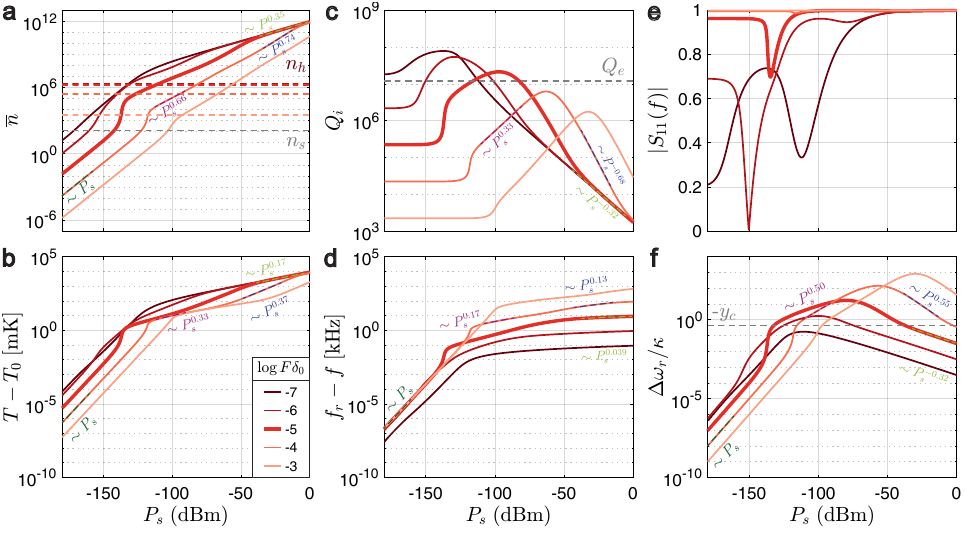}
\caption{Simulated power dependence of resonator characteristics for various TLS loss tangents $F\delta_0$. The probe frequency is set on-resonance at vanishing power, $f=f_r(T_0)$. (a-b) Internal variables: average intracavity phonon number $\nbar$ and effective temperature increase $T-T_0$ as a function of the applied microwave power at device level, $P_s$, for different values of TLS loss tangent $F\delta_0$. The thicker curves correspond to the case $F\delta_0=10^{-5}$, closest to the experimental situation reported in Fig.~3. The dashed gray line indicates the value of $n_s$. $n_h$ varies with $F\delta_0$ and is shown in dashed lines with the same color encoding the value of $F\delta_0$. (c-d) Resonance parameters : internal quality factor $Q_i$ and resonance frequency shift $f_r(T)-f_r(T_0)$ with respect to the resonance frequency without readout power heating. (e) Magnitude of the reflection coefficient $|S_{11}(f)|$ at the probe frequency. (f) Normalized frequency shift in number of resonance linewidths, $\Delta\omega_r(T,T_0)/\kappa(T)=-y=-Q x$. Model parameters: $T_0=25$~mK, $f_{r,0}=502.06550$~MHz, $F\delta_0=1.1395\times 10^{-5}$, $Q_\textrm{bkg}=\infty$, $\kappa_e/2\pi=42.15$~Hz, $n_s=128$, $G_{th}(T_0)/g_0(T_0)=0.58$, $\gamma = 2.83$, $\beta=1.0$, $Q_\textrm{rel}(0.5\textrm{K})=0.36\times 10^6$, $d=1.84$.}
\label{fig6}
\end{figure*}

\begin{table}[!h]%
\footnotesize 
\[%
\setbox\MytempboxA\hbox{\mbox{Var. X}}%
\setbox\MytempboxB\hbox{\mbox{Scaling behavior}}%
\begin{array}{@{}|r|*{4}{c}|}
\hline
\multicolumn{1}{|l}{%
  \edef\myTempA{%
    \number\numexpr\dimexpr\wd\MytempboxA+2\arraycolsep\relax\relax
  }%
  \edef\myTempB{%
    \number
    \numexpr
      \dimexpr\dp\csname @arstrutbox\endcsname+%
              \ht\csname @arstrutbox\endcsname+%
              \arrayrulewidth
      \relax
    \relax
  }%
  \FPpow\myTempC\myTempA{2}%
  \FPpow\myTempD\myTempB{2}%
  \FPadd\myTempC\myTempC\myTempD
  \FProot\myTempC\myTempC{2}
  \FPdiv\myTempD\myTempB\myTempA
  \FParctan\myTempD\myTempD
  \smash{%
    \kern-\arraycolsep
    \rlap{%
      \lower
        \dimexpr
          \dp\csname @arstrutbox\endcsname+\arrayrulewidth
        \relax
        \hbox{%
          \rotatebox[units=-6.283185,origin=br]{\myTempD}{%
            \rule{\myTempC sp}{\arrayrulewidth}%
          }%
        }%
    }%
  }%
}&\multicolumn{4}{c|}{\copy\MytempboxB}\\%
\cline{2-5}%
\copy\MytempboxA& \nbar\ll n_s,n_h & n_s\ll\nbar<n_h & n_s\ll\nbar\sim n_h & \nbar\gg n_s,n_h \\%
\hline
f_r & T & \log{T} & \log{T} & \textrm{const.}\\%
\kappa_i & \textrm{const.} & \kappa_\textrm{res}\propto\nbar^{-\beta/2} & \kappa_\textrm{rel}\propto T^d & \kappa_\textrm{rel}\propto T^d \\%
\nbar & P_s & \frac{P_s}{(f-f_r)^2} & \frac{P_s}{(f-f_r)^2} & \frac{P_s}{\kappa_i^2} \\%
T & P_d & P_d & P_d^\frac{1}{1+\gamma} &  P_d^\frac{1}{1+\gamma} \\%
P_d & P_s & \frac{P_s \kappa_i}{(f-f_r)^2} & \frac{P_s \kappa_i}{(f-f_r)^2} & \frac{P_s}{\kappa_i} \\%
\Delta\omega_r/\kappa & \ll 1 & \gg 1 & \gg 1 & \ll 1 \\%
\hline
\end{array}%
\]%
\caption{Scaling behavior of the resonance parameters\label{table:powerScale}}%
\end{table}

As described in Appendix~A.2, we model TLS relaxation damping as a power law function of temperature, while making the simplifying assumption that $Q_\textrm{rel,0}$ is constant and independent of $F\delta_0$. This effectively neglects any correlation between the density of near-resonant TLSs and that of high-energy, off-resonant TLSs responsible for relaxation damping. 
For a uniform continuum of TLSs, and assuming identical longitudinal and transverse coupling strengths to strain, one would expect $Q_\textrm{rel}^{-1}\propto F\delta_0^2$ \cite{Emser2024}. However, because relaxation damping arises from a broad distribution of thermally populated, off-resonant TLSs, it is plausible that fabrication improvements may affect the density of high-energy TLSs differently from the near-resonant subset characterized by $F\delta_0$. We therefore treat each contribution to the loss independently in our model.

The scaling behavior with power of each quantity can be obtained analytically and is summarized in Table.~\ref{table:powerScale}. Four regimes, characterized by different scaling behaviors for $\nbar$ and $T$, can be distinguished. They correspond to different orderings of $\nbar$ with respect to $n_s$ and $n_h$, the critical phonon numbers for TLS saturation and for the onset of readout-power heating: 
\begin{itemize}
  \item At low enough probe power such that $\nbar\ll n_s,n_h$, the quality factor saturates at $Q_i\approx Q_\textrm{res,min}=1/(F\delta_0\tanh{\{h f_r(T_0)/(2k_B T_0)\})}$ and  the readout power heating remains negligible. Therefore, this case corresponds to the \textit{linear} regime, where all quantities (except $Q_i$) scale like $P_s$. Indeed, with $Q x\ll 1$, the dissipated power varies like $P_d\propto (Q_i/Q_e)P_s\sim P_s$ and $T-T_0\propto P_d \sim P_s$ since $T-T_0\ll T_0$. Similarly, $\nbar\propto(Q_i^2/Q_e)P_s\sim P_s$. The residual frequency shift can be linearized around $T_0$: $\Delta f_r(T, T_0)=f_r(T)-f_r(T_0)\underset{T\gg T_c}{\approx}(F\delta_0/\pi)f_r\ln{(T/T_0)}\underset{T\approx T_0}{\approx} (T-T_0)/T_0 \propto P_s$, which yields $\Delta\omega_r(T,T_0)/\kappa(T)\propto P_s$, in agreement with Fig.~\ref{fig6}(f).

  \item At moderate power above the value for TLS saturation, $n_s\ll\nbar < n_h$, the mechanical dissipation is dominated by TLS resonant damping and decreases with power due to saturation of resonant TLSs, $\kappa_i\propto\nbar^{-\beta/2}$. Because of the rapid increase of $Q_i$ with power, we have now $Q x\gg 1$ and so the phonon number scales instead like $\nbar\propto P_s/x^2\sim P_s^{1-2\eta}$ and $\kappa_i\sim P_s^{-(\beta/2)(1-2\eta)}$, where $0\leq\eta\leq 1$ denotes the weak power scaling of the resonance frequency, $f_r\sim P_s^\eta$. Since the underlying temperature dependence of $f_r$ is a logarithmic one, $\eta$ does not have a fixed value and depends on power. The goal here is to relate the scaling exponents of the other resonance quantities to $\eta$. Since $\nbar<n_h$, the readout-power heating remains small and $T-T_0\propto P_d \propto P_s\kappa_i/x^2\sim P_s^{1-2\eta-\beta(1/2-\eta)}$.

  \item Further increasing power so that $\nbar\sim n_h\gg n_s$, the readout-power heating is such that mechanical dissipation becomes dominated by TLS relaxation damping, $\kappa_i\approx \kappa_\textrm{rel}\propto T^d$, and $T\sim P_d^{1/(1+\gamma)}$ since $T-T_0\gg T_0$. Let's denote $\alpha$ the scaling exponent of the effective temperature with power, $T\propto P_s^\alpha$. Then, the internal loss scales like $\kappa_i\propto P_s^{d\alpha}$. $Q x\gg 1$ is still verified and $\nbar\propto P_s/x^2\sim P_s^{1-2\epsilon}$, where we parametrized $f_r\sim P_s^\epsilon$ and used a different letter to emphasize that $\epsilon\neq\eta$ in this different power regime. Consequently, $T\sim P_s^{(1-2\epsilon+d\alpha)/(1+\gamma)}\equiv P_s^\alpha$. From this self-consistent equation, $\alpha=(1-2\epsilon+d\alpha)/(1+\gamma)$, one deduces $\alpha=(1-2\epsilon)/(1-d+\gamma)$. The scaling exponents of $T, \nbar$ and $Q_i$ can then be all expressed in terms of $d$ and $\gamma$, and are summarized in Table~\ref{table:powerExp}.
  
  \item When $\nbar\gg n_h, n_s$, TLS relaxation damping activated by the device heating makes $Q_i$ low enough that again $Q x\ll 1$, which results in a weaker increase of the dissipated power with $P_s$ and, in turn, slows down the readout-power heating. The phonon number and dissipated power now scale like $\nbar\propto P_s/\kappa_i^2\sim P_s^{1-2 d\alpha}$ and $P_d\propto P_s/\kappa_i$, from which follows $T\sim P_d^{1/(1+\gamma)}\sim P_s^{(1-d\alpha)/(1+\gamma)}\equiv P_s^\alpha$ (we neglect here the weak logarithmic power dependence of $f_r$, see Fig.~\ref{fig6}(d)). From this self-consistent equation, $\alpha=(1-d\alpha)/(1+\gamma)$, one deduces the new temperature exponent $\alpha=1/(1+d+\gamma)$. 
\end{itemize}

\begin{table}[!h]%
\footnotesize 
\[%
\setbox\MytempboxA\hbox{\mbox{Var. X}}%
\setbox\MytempboxB\hbox{\mbox{Scaling exponent $\alpha$}}%
\begin{array}{@{}|r|*{4}{c}|}
\hline
\multicolumn{1}{|l}{%
  \edef\myTempA{%
    \number\numexpr\dimexpr\wd\MytempboxA+2\arraycolsep\relax\relax
  }%
  \edef\myTempB{%
    \number
    \numexpr
      \dimexpr\dp\csname @arstrutbox\endcsname+%
              \ht\csname @arstrutbox\endcsname+%
              \arrayrulewidth
      \relax
    \relax
  }%
  \FPpow\myTempC\myTempA{2}%
  \FPpow\myTempD\myTempB{2}%
  \FPadd\myTempC\myTempC\myTempD
  \FProot\myTempC\myTempC{2}
  \FPdiv\myTempD\myTempB\myTempA
  \FParctan\myTempD\myTempD
  \smash{%
    \kern-\arraycolsep
    \rlap{%
      \lower
        \dimexpr
          \dp\csname @arstrutbox\endcsname+\arrayrulewidth
        \relax
        \hbox{%
          \rotatebox[units=-6.283185,origin=br]{\myTempD}{%
            \rule{\myTempC sp}{\arrayrulewidth}%
          }%
        }%
    }%
  }%
}&\multicolumn{4}{c|}{\copy\MytempboxB}\\%
\cline{2-5}%
\copy\MytempboxA& \nbar\ll n_s,n_h & n_s\ll\nbar<n_h & n_s\ll\nbar\sim n_h & \nbar\gg n_s,n_h \\%
\hline
f_r & 1 & \eta\approx 0.17 &\epsilon\approx 0.13  & \approx 0 \\%
\kappa_i & 0 & -\beta(\frac{1}{2}-\eta)\approx-0.33 &\frac{d(1-2\epsilon)}{1-d+\gamma}\approx 0.68 & \frac{d}{1+d+\gamma}\approx 0.32 \\%
\nbar & 1 & 1-2\eta\approx 0.66 & 1-2\epsilon\approx 0.74 & \frac{1-d+\gamma}{1+d+\gamma}\approx 0.35 \\%
T & 1 & 1-2\eta-\beta(\frac{1}{2}-\eta)\approx 0.33 & \frac{1-2\epsilon}{1-d+\gamma}\approx 0.37 & \frac{1}{1+d+\gamma}\approx 0.17 \\%
\Delta\omega_r/\kappa & 1 & \eta(1-\beta)+\beta/2\approx 0.50 & \frac{\epsilon(1+d+\gamma)-d}{1-d+\gamma}\approx -0.55 & -\frac{d}{1+d+\gamma}\approx -0.32 \\%
\hline
\end{array}%
\]%
\caption{Power scaling of the resonance parameters $X\sim P_s^\alpha$\label{table:powerExp}}%
\end{table}

The fractional detuning $-y$, plotted in Fig.~\ref{fig6}(f), can be used as a figure of merit to gauge the strength of the nonlinear reactive response:
\begin{equation}
    \frac{\Delta\omega_r(T,T_0)}{\kappa(T)}\equiv2\pi\frac{f_r(T)-f_r(T_0)}{\kappa(T)}\underset{T\approx T_0}{\approx}Q\Bigl(\frac{f-f_r(T_0)}{f_r(T_0)}-\frac{f-f_r(T)}{f_r(T)}\Bigr)=y_0-y\underset{\textrm{Eq.}~\ref{yT}}{=}-\frac{a}{1+4y^2}
\end{equation}
In ringdown measurements, we generally enforce $f=f_r(T_0)$, \textit{i.e.} $y_0=0$ so that $\Delta\omega_r/\kappa=-y$.
Using the linearized model for the TLS reactive nonlinearity of Appendix~A.1, which is valid in the limit of small readout-power heating ($\Delta T\ll T_0$), we can derive a critical value $-y_c$ of the fractional detuning $\Delta\omega_r/\kappa$ for the onset of bistability. Solving $-y=-a_c(1+4y^2)$ with $a_c=4/(3\sqrt{3})$ yields $y_c\approx-0.43$. When probing at a fixed frequency such that $y_0=0$, no discontinuity in $S_{11}(y_0=0)$ occurs as the probe power is increased, as shown in Fig.~\ref{fig7}(a). However, if at $y_0=0$, $y<y_c$, then a discontinuous response will be measured if the probe tone frequency is increased such that $y_0$ crosses $y_0^{>,d}$. This was discussed in Appendix~\ref{app:AppSwenson} and is further illustrated in Fig.~\ref{fig7}(a-d).

\section{Effect of the external coupling rate on the nonlinearity strength}
\label{app:effectKappaext}

\begin{figure*}[!ht]
\centering
\includegraphics[width=1.0\linewidth]{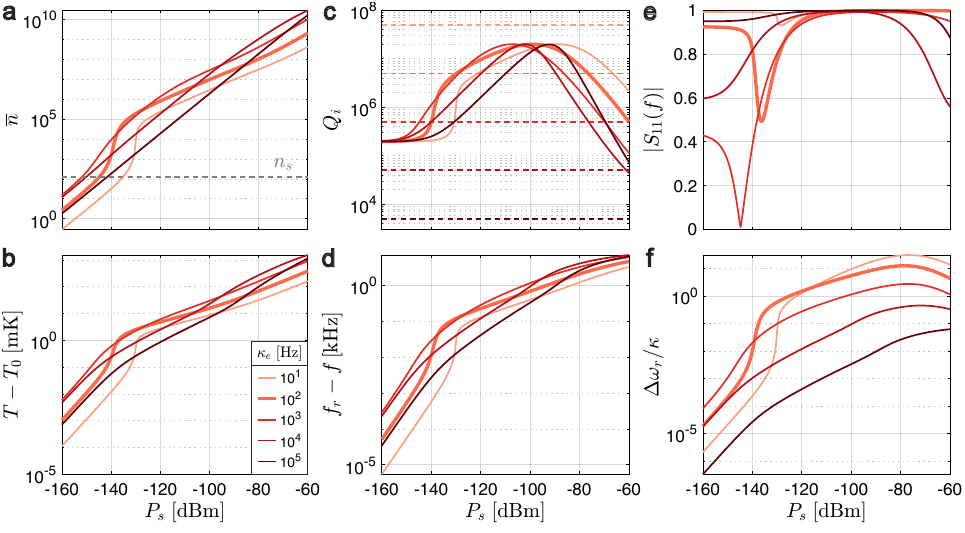} 
\caption{Simulated power dependence of resonator characteristics for various external coupling rates $\kappa_e$. The probe frequency is set on-resonance at vanishing power, $f=f_r(T_0)$. Same parameters and labels as in Fig.~\ref{fig6}. The thicker curves highlight the typical regime for thin-film quartz PCRs, $\kappa_e/2\pi=100$~Hz, and in (c) the value of $Q_e$ is shown in dashed lines with the same color encoding the value of $\kappa_e$.}
\label{figKext}
\end{figure*}

Since the dissipated power and \textit{a fortiori} the resonance frequency shift are maximized when $Q_e=Q_i$, we expect that the degree to which the TLS nonlinearity manifests as the probe power is increased would depend strongly on the coupling quality factor. In Fig.~\ref{figKext}, we simulate the power dependence of the resonance parameters for various external coupling rates spanning the different regimes where the resonator is initially under-coupled or over-coupled. For thin-film quartz PCRs, the weak electromechanical coupling coefficient of quartz ($K^2\approx 0.1\%$) limits the achievable external coupling rate to $\kappa_e\propto K^2\lesssim 2\pi\times 10^2$~Hz and so at base temperature, our PCRs are generically largely under-coupled, $Q_e\gtrsim10^6>Q_i(T_0)$. This is not the case for lithium niobate devices where the $K^2\approx 20\%$ allows for coupling rates several orders of magnitude higher, $\kappa_e/2\pi\sim 10-100$~kHz \cite{Wollack2021}. 

If at vanishing power, the PCR is already over-coupled,  $Q_e\ll Q_i(T_0)$, then the phonon occupancy will initially grow linearly with the probe power, $\nbar\propto Q_e P_s$. Contrary to the under-coupled case, no self-accelerating increase is then expected when $\nbar\sim n_s$ and $Q_i$ shows a smoother rise with $P_s$ from saturation of near-resonant TLSs (Fig.~\ref{figKext}(c)). Increasing the external coupling therefore reduces the overall sensitivity to readout power and pushes the threshold for the onset of hysteretic switching to higher readout power.

\section{TLS reactive nonlinearity: linearized \textit{vs.} non-perturbative model}\label{app:ReacLinVSnonPert}

\begin{figure*}[!ht]
\centering
\includegraphics[width=1.0\linewidth]{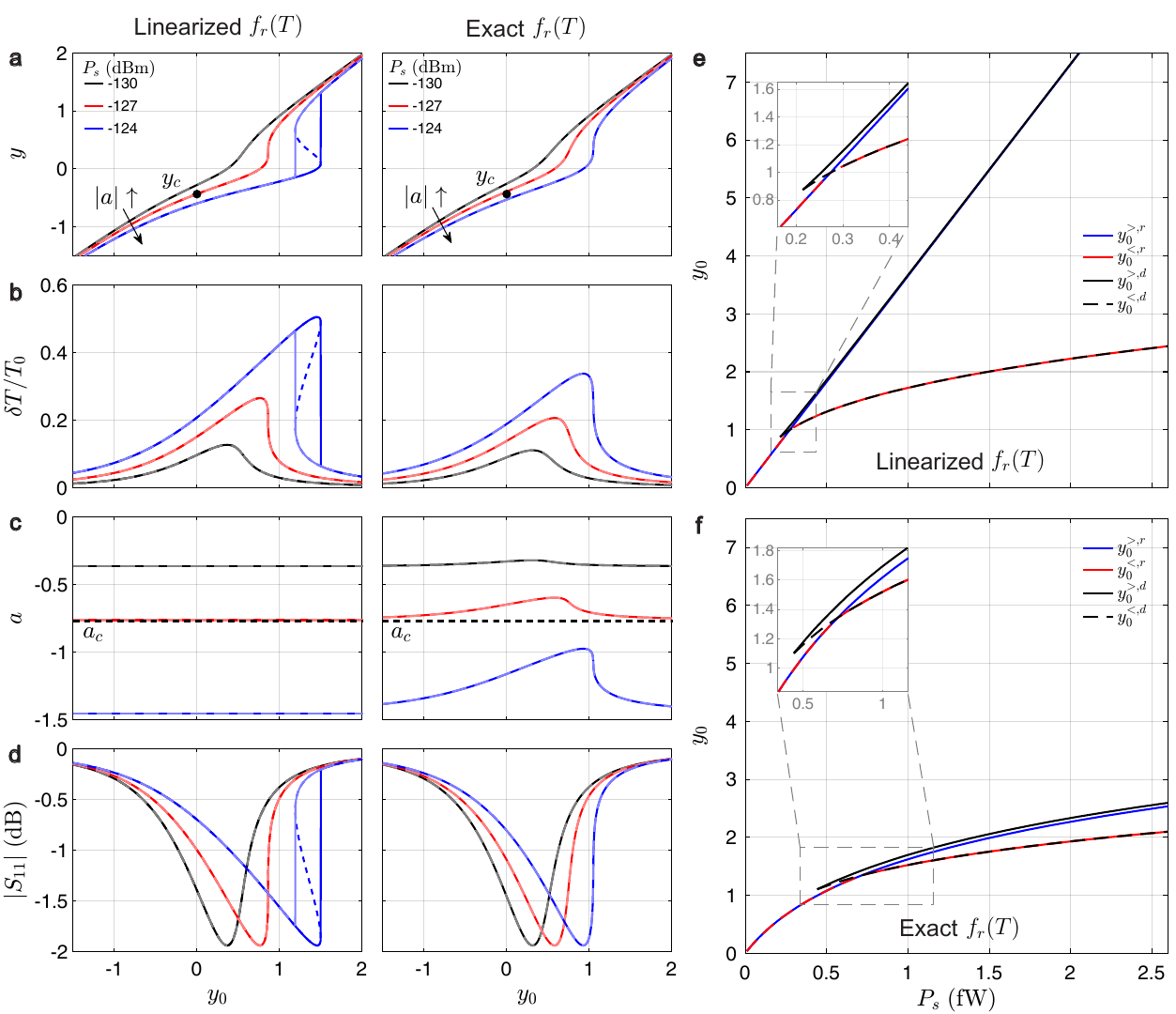}
\caption{TLS reactive nonlinearity: comparison between the linearized and full non-perturbative models. (a) The realized detuning $y$, (b) the relative temperature increase $\delta T/T_0$, (c) Swenson's nonlinearity strength $a$ and (d) the magnitude of the reflection coefficient $|S_{11}|=|1-2\alpha/(1+2 j y)|$ as a function of the applied detuning $y_0$ for three illustrative values of probe power $P_s=-130, -127, -124$~dBm. The critical value of $y$ for the onset of bistability, $y_c\approx-0.43$, is shown with a black dot in (a). For the purpose of this comparison, the power and temperature dependences of $Q$ are neglected and a fixed resonance depth $\alpha=Q/Q_e=0.1$ is assumed. In the bistable regime when $-a>a_c$ ($\approx 0.77$), the response differs for upward and downward  sweeps of $y_0$. To distinguish the two, the downward response is shown with the same color as the upward sweep but with a lighter shade. (e-f) The applied detuning at resonance for both upward ($y_0^{>,r}$, blue line) and downward ($y_0^{<,r}$, red line) sweeps as a function of $P_s$ when a linearized $f_r(T)$ (e) or the exact dependence (f) is assumed. For the latter case, the resonance frequency no longer increases linearly with $P_s$, but shows a weaker $\sim\log(P_s)$ dependence.}
\label{fig7}
\end{figure*}

Here, we compare the reactive response of the resonator predicted by the analytical ``first-order'' model (Swenson's equation, discussed in \ref{app:AppSwenson}) to the exact numerical solution, where the full $f_r(T)$ dependence is taken into account and not just linearized around $T_0$. To allow a fair comparison and focus solely on the reactive part, a fixed quality factor was assumed, corresponding to a resonance depth of $\alpha = Q/Q_e$=0.1. All the other parameters are the same as in Fig.~\ref{fig6}. The resonator response is computed for three illustrative value of probe power $P_s$, chosen such as to represent the weakly nonlinear regime (black), the strong bistable regime (blue) and the critical regime  right before the onset of bistability (red). First, Swenson's nonlinearity parameter corresponding to the applied power is obtained using the following equation:
\begin{equation}\label{aT0}
    a(T_0) = -R_{th}(T_0) TCF(T_0)4 Q_e\alpha^2(1-\alpha)P_s
\end{equation}
Second, Swenson's equation (Eq.~\ref{yT}) is solved for $a(T_0)$, which yields the realized detuning $y$. Finally, the temperature increase and complex reflection coefficient are deduced by plugging $y$ into the following expressions :
\begin{equation}\label{dT0S11alph}
    \frac{\delta T}{T_0}=\frac{4\alpha(1-\alpha)}{1+4 y^2}\frac{R_{th}(T_0)}{T_0}P_s, \qquad S_{11}=1-\frac{2\alpha}{1+2 j y}.
\end{equation}
For the purpose of the comparison, an effective nonlinear strength $a$ is computed for the non-perturbative model by replacing $T_0$ in Eq.~\ref{aT0} with the simulated $T(y_0)$. In Fig.~\ref{fig7}(a-d), the computed responses for the two models agree for the case of $P_s=-130$~dBm (black curves), but differ significantly at higher $P_s$. The case $P_s=-130$~dBm corresponds to the limit of weak probe power, where the device heating remains small such that $\delta T\ll T_0$ (see Fig.~\ref{fig7}(b)) and the linearized model approximates well the resonator response. At higher power, Swenson's linearized model overestimates the strength of the nonlinear response, because it assumes $\delta f_r(T)\propto \delta T$ instead of the actual weaker $\propto \log{T}$ dependence. When the exact $f_r(T)$ dependence is assumed, the critical value for the onset of bifurcation is no longer given by $y_c$ and $a_c$, as illustrated in Fig.~\ref{fig7}(a), where $y(y_0=0)<y_c$ for the $P_s=-124$~dBm case (blue curve) and still the response is not yet multi-valued. The main difference between the two models is that the linearized version predicts a linear increase of the resonance frequency with the applied power, $f_r\propto P_s$, while the exact model shows a much weaker $\sim \log(P_s)$ dependence, as evidenced in Fig.~\ref{fig7}(e-f). In addition, the resonance frequency no longer starts to differ for upward/downward sweeps when $y_0=1$, as discussed in Appendix~\ref{app:AppSwenson}, but at a higher value (see insets in Fig.~\ref{fig7}(e-f)). Finally, in contrast to the linearized model, the switching frequency in the upward sweep, $y_0^{>,d}$, does not converge at high power to $y_0^{>,r}$ anymore, but stays at a fixed offset from it. 

\endgroup

\bibliography{biblio}